\documentclass[a4paper]{article}

\usepackage[utf8]{inputenc}
\usepackage{amsfonts}
\usepackage{mathrsfs}
\usepackage{amsmath}
\usepackage{amssymb}
\usepackage{amsthm}
 \theoremstyle{plain} 
  \newtheorem{theorem}{Theorem}[section]

 \theoremstyle{definition} 
  \newtheorem{definition}{Definition}[section]
 \theoremstyle{remark} 
  \newtheorem{remark}[theorem]{Remark}   
\usepackage{graphicx} 
\usepackage{enumerate}
\usepackage{paralist}
\usepackage{acro}
\DeclareAcronym{CT}{
 short = CT,
 long = computed tomography}
\DeclareAcronym{CNN}{
 short = CNN,
 long = convolutional neural network} 
\DeclareAcronym{LLM}{
 short = LLM,
 long = large language model} 
\DeclareAcronym{EIT}{
 short = EIT,
 long = electrical impedance tomography} 
\DeclareAcronym{SciML}{
 short = SciML,
 long = scientific machine learning} 
\DeclareAcronym{FoE}{
 short = FoE,
 long = fields-of-experts} 
\DeclareAcronym{TV}{
 short = TV,
 long = total variation} 
\DeclareAcronym{QPAT}{
 short = QPAT,
 long = quantitative photoacoustic tomography}
\DeclareAcronym{FNO}{
 short = FNO,
 long = Fourier neural operator} 
\DeclareAcronym{PSNR}{
 short = PSNR,
 long = peak signal-to-noise ratio} 
\DeclareAcronym{IID}{
 short = IID,
 long = independent identically distributed}
\DeclareAcronym{DFP}{
 short = DFP,
 long = Davidon-Fletcher-Powell}
\DeclareAcronym{BFGS}{
 short = BFGS,
 long = Broyden-Fletcher-Goldfarb-Shanno}
\DeclareAcronym{SR1}{
 short = SR1,
 long = symmetric rank-one}
\DeclareAcronym{PSB}{
 short = PSB,
 long = Powell-Symmetric-Broyden} 
\DeclareAcronym{ISTA}{
 short = ISTA,
 long = iterative shrinkage and thresholding algorithm} 
\DeclareAcronym{FISTA}{
 short = FISTA,
 long = fast iterative shrinkage and thresholding algorithm}
\DeclareAcronym{LISTA}{
 short = LISTA,
 long = learned iterative shrinkage and thresholding algorithm} 
\DeclareAcronym{ReLU}{
 short = ReLU,
 long = rectified linear unit}
\DeclareAcronym{LPD}{
 short = LPD,
 long = learned primal-dual}
\DeclareAcronym{ResNet}{
 short = ResNet,
 long = residual neural network}
\DeclareAcronym{RNN}{
 short = RNN,
 long = recurrent neural network}
\DeclareAcronym{FBP}{
 short = FBP,
 long = filtered back-projection}
 \DeclareAcronym{TDV}{
 short = TDV,
 long = total deep variation}
\DeclareAcronym{PINN}{
 short = PINN,
 long = physics-informed neural network}
\DeclareAcronym{PDE}{
 short = PDE,
 long = partial differential equation}
\DeclareAcronym{ODE}{
 short = ODE,
 long = ordinary differential equation}
\DeclareAcronym{PnP}{
 short = PnP,
 long = plug-and-play}
\DeclareAcronym{DEQ}{
 short = DEQ,
 long = deep equilibrium}
\DeclareAcronym{GAN}{
 short = GAN,
 long = generative adversarial network} 
\DeclareAcronym{SGD}{
 short = SGD,
 long = stochastic gradient descent}  
\DeclareAcronym{DeepONet}{
 short = DeepONet,
 long = deep operator network} 
\DeclareAcronym{GNO}{
 short = GNO,
 long = graph neural operator}
\DeclareAcronym{MGNO}{
 short = MGNO,
 long =  multipole graph neural operator}
\usepackage{hyperref}
\hypersetup{%
 unicode=true,    
 pdftoolbar=true,   
 pdfmenubar=true,   
 pdffitwindow=false, 
 pdfstartview={FitH}, 
 pdfnewwindow=true,  
 colorlinks=true,   
 linkcolor=red,    
 citecolor=red,    
 filecolor=magenta,  
 urlcolor=blue    
}
\usepackage{biblatex}
\usepackage{todonotes}

\newcommand{\Real}{\mathbb{R}}

\newcommand{\RecSpace}{X}
\newcommand{\DataSpace}{Y}

\newcommand{\argmin}{\operatorname*{arg\,min}}
\newcommand{\prox}{\mathrm{prox}}
\newcommand{\operator}[1]{\operatorname{\mathcal{#1}}}
\newcommand{\RecOp}{\operator{R}}
\newcommand{\FwdOp}{\operator{A}}
\newcommand{\RegOp}{\operator{S}}
\newcommand{\CostFunc}{\operator{E}}

\newcommand{\grad}{\nabla\!}
\newcommand{\DataNorm}{\operator{D}}
\newcommand{\Id}{\operator{I}}
\newcommand{\signal}{f}
\newcommand{\signaltrue}{\signal^{\ast}}

\newcommand{\signalother}{u}
\newcommand{\signalothernew}{v}
\newcommand{\signalnoise}{\delta\!\signal}
\newcommand{\data}{g}
\newcommand{\dataother}{h}
\newcommand{\datanoise}{e}

\newcommand{\NNOp}{\Lambda}
\newcommand{\NNOpOther}{\Gamma}
\newcommand{\NNparam}{\theta}
\newcommand{\NNparamOther}{\vartheta}
\newcommand{\NNparamSet}{\Theta}

\newcommand{\Loss}{\operator{L}}
\newcommand{\DataLoss}{\Loss_{\DataSpace}}
\newcommand{\SignalLoss}{\Loss_{\RecSpace}}
\newcommand{\UnrollIter}{N}
\newcommand{\Expect}{\operator{\mathbb{E}}}
\newcommand{\random}[1]{\mathsf{#1}}
\newcommand{\est}[1]{\hat{#1}}
\newcommand{\signalrand}{\random{\signal}}

\newcommand{\datarand}{\random{\data}}
\newcommand{\datanoiserand}{\random{\datanoise}}
\newcommand{\PClass}[1]{\mathscr{P}_{#1}}

\newcommand{\Cdot}{\,\cdot\,}

\newcommand{\HessianOp}[1]{\operatorname{Hess}(#1)}
\newcommand{\GateuaxSecond}[1]{\partial^2\! #1}
\newcommand{\LinOp}{\mathscr{L}}

\newcommand{\opB}{\operator{B}}
\newcommand{\opH}{\operator{H}}
\newcommand{\opI}{\operator{I}}
\newcommand{\triangleSignal}{\triangle\!\signal}
\newcommand{\deltaSignal}{\delta\!\signal}
\newcommand{\deltaCost}{\delta\!\operator{Q}}
\newcommand{\matrixForm}[1]{\mathsf{#1}}
\newcommand{\matB}{\matrixForm{B}}
\newcommand{\matH}{\matrixForm{H}}
\newcommand{\matI}{\matrixForm{I}}

\newcommand{\LearnAlg}{T}
\newcommand{\ApproxError}{\operator{E}}
\newcommand{\RecOpOther}{\operator{T}}

\title{Learned iterative networks: An operator learning perspective}
\author{Andreas Hauptmann\thanks{Research Unit of Mathematical Sciences, University of Oulu, Finland and  Department of Computer Science, University College London, United Kingdom.} 
\and Ozan Öktem\thanks{Department of Mathematics, KTH - Royal Institute of Technology, Sweden}}

\begin{document}
\maketitle

\begin{abstract}
    Learned image reconstruction has become a pillar in computational imaging and inverse problems. Among the most successful approaches are learned iterative networks, which are formulated by unrolling classical iterative optimisation algorithms for solving  variational problems. While the underlying algorithm is usually formulated in the functional analytic setting, learned approaches are often viewed as purely discrete. In this survey we present a unified operator view for learned iterative networks. Specifically, we formulate a learned reconstruction operator, defining \emph{how to compute}, and separately the learning problem, which defines  \emph{what to compute}. In this setting we present common approaches and show that many approaches are closely related in their core. 
    We review linear as well as non-linear inverse problems in this framework and present a short numerical study to conclude.
\end{abstract}

\section{Introduction}
Data-driven approaches, and in particular machine learning approaches based on deep neural networks (deep learning) \cite{LeCun:2015aa}, have shown unparalleled capability in performing a variety of tasks.
Initial success stories related to tasks in natural language processing \cite{Afkari-Fahandari:2025aa,Banerjee:2025aa,Zhang:2025aa} and computer vision \cite{Ioannidou:2017aa,Khan:2022aa,Wang:2021aa}, but recent success relates to performing more complex tasks \cite{Li:2025aa,Yin:2024aa,Al-Zoghby:2025aa}.
This has in turn catalysed a surge of interest in applying such data driven learning techniques for performing a variety of computing tasks in science and engineering, like solving ill-posed inverse problems.

As will be outlined in this work, usage of deep learning in such applications often requires invoking \emph{structured learning}. This refers loosely to various techniques for domain adaptation that go beyond choosing domain specific training data and loss function. 
Here, we focus particularly on \emph{learned iterative networks}, which represent a specific form of structured learning that is suited for learning tasks that are given as iterative schemes. 
The idea is to use such an iterative scheme as a blue print for parametrising the class of operators (hypothesis space) that will be used during the learning.
Learned iterative networks are therefore useful for learning a task that is representable by an iterative scheme, like finding a minimiser to some functional or solving an ill-posed inverse problem.

\subsubsection*{Overview}
The overarching viewpoint of this survey is built around an operator learning perspective for solving inverse problems. The inverse problem is given by an operator equation and solving it with a (learned) reconstruction method corresponds to defining and training a \emph{learned reconstruction operator}. 

Defining the learning within such a functional analytic viewpoint yields a unified and structured description of various machine-learning methods for solving inverse problems.
For learned iterative networks, this includes in particular many popular architectures, like learned gradient schemes \cite{Adler:2017aa}, learned proximal networks \cite{gilton2021deep}, and variational networks \cite{hammernik2018learning,Kobler:2017aa}. 
We will also notice that there is a significant conceptual overlap between these architectures when they are viewed as an operator learning problem. 
Additionally, the above approach allows us to decouple the formulation of the learned reconstruction operator from the learning task, where the latter is more concerned about how to train the learned reconstruction operator. 

Thus, this survey will first formulate the operator learning problem in Section~\ref{sec:operatorLearn} and its application to inverse problems with the definition of a learned reconstruction operator. Section~\ref{sec:LearningProbs} then introduces the \emph{learning problem}, i.e., the question of training data and loss functions. 
We continue to discuss the concept of \emph{structured learning} in Section~\ref{sec:MotivationStructuredLearning} as a basis for learning the reconstruction operator by domain adaption to the inverse problem at hand. With these preliminaries we introduce various approaches to construct a learned reconstruction operator before we move on to the core of this chapter and introduce learned iterative networks in Section~\ref{sec:LearnedIterNetwork} along with a short historical account. Section~\ref{sec:LearnedGradient} then concentrates on learned iterative methods based on gradient based optimisation methods, whereas~\ref{sec:LPD} discusses primal-dual architectures. We will then discuss non-linear inverse  problems and higher order methods such as Newton type approaches in Section~\ref{sec:HighOrderNonLin}. To complete this survey chapter, Section~\ref{sec:implementation} concentrates on the practical side with implementation related aspects and numerical examples providing a performance comparison in Section~\ref{sec:Comparison}. Finally, we conclude with a discussion on theoretical results in Section~\ref{sec:TheoryFoundations} for operator learning and how these apply to case of learned reconstruction operators and specifically learned iterative networks.

\section{Operator learning} \label{sec:operatorLearn}
Mathematical models in science and engineering frequently builds on transformations that have continuum function data as input or output.
These sets for inputs or outputs are often formalised as infinite dimensional vector spaces of functions rather than sets of arrays. 
Furthermore, an associated computational task is typically  represented by an operator
\begin{equation}\label{eq:MainOperator} 
 \RecOp \colon \DataSpace \to \RecSpace
\end{equation}
with $\DataSpace$ or $\RecSpace$ denoting the aforementioned function spaces of inputs and outputs.

A prominent example is to solve a differential equation where $\DataSpace$ could be the function space of possible boundary conditions and $\RecSpace$ is the function space of solutions.
Another is to regularise an (ill-posed) inverse problem where $\DataSpace$ is the function space of possible data and $\RecSpace$ is the function space of possible solutions (Section~\ref{sec:InvProb})
Yet another is to compute a minimiser of a parametrised convex cost functional that is defined on some function space (Section~\ref{sec:OptimSolver}). Then, functions in $\DataSpace$ parametrise the convex cost functional and functions in $\RecSpace$ are possible minimisers.

\subsection{Key notions and concepts}\label{sec:KeyNotions}
The mathematical analysis of operators, together with the design and analysis of computational algorithms for their approximate evaluation, typically necessitates the imposition of additional structure. 
Most commonly, the (infinite-dimensional) vector spaces $\DataSpace$ and $\RecSpace$ in \eqref{eq:MainOperator} are assumed to be Banach spaces or Hilbert spaces. 
Within such a framework, functional analysis provides the tools necessary to rigorously define fundamental concepts, like convergence, continuity, and stability. 
These notions are indispensable for both theoretical inquiry and the development and analysis of computational methods.
They are also particularly significant for inverse problems since the central notion of ill-posedness becomes mathematically vacuous in a finite-dimensional setting since all norms are equivalent in finite dimensional vector spaces.

The above stands in contrast to statistical learning, which is rarely formulated and analysed in a infinite dimensional functional analytic setting.
\emph{Operator learning} aims to do this, i.e., it aims to develop a framework for statistical learning and data driven (operator) approximation of tasks whose formalisation involve continuum function data as input or output.
A formal definition of operator learning that is sufficiently broad to encompass the range of possible approaches and learning tasks would necessarily be highly generic, and thus of limited practical use. 
Consequently, we adopt a slightly less formal characterisation that nevertheless seeks to capture the essential traits of operator learning.
In this less formal setting, one can view operator learning as methods within statistical learning for approximating (or discovering) an (unknown) target mapping (\emph{target operator}) of the form in \eqref{eq:MainOperator} from example data (training data) residing in the function spaces $\RecSpace$ or $\DataSpace$.

A key part of operator learning is the choice of a \emph{hypothesis space}, which is the family $\{ \RecOp_{\NNparam} \}_{\NNparam\in\NNparamSet}$ of parametrised mappings $\RecOp_{\NNparam} \colon \DataSpace \to \RecSpace$ that will approximate the target operator.
The parameter set $\NNparamSet$ is typically a finite dimensional vector space and the learning/training corresponds to using data to select the element in the hypothesis space that `best' approximates the target operator $\RecOp$. 
Stated equivalently, one seeks to select the parameter $\est{\NNparam} \in \NNparamSet$ which gives a learned operator  $\RecOp_{\est{\NNparam}}$ that `best' approximates $\RecOp$.
Operator learning is commonly associated with the more specific ambition to extend deep learning to a functional analytic setting.
This means setting up functional analytic variants of deep neural network architectures and learned iterative networks, which is the central theme for this survey, deals with this question when the target operator being approximated is implicitly defined by an iterative scheme.

\subsection{Relevance of learning in functional analytic setting}\label{sec:MotivationFuncAnaly}
An implementation of a computing task that is formalised as in \eqref{eq:MainOperator} will inevitably involve some form of \emph{discretisation}.
This is the process in which continuum function data are represented by finite dimensional arrays or tensors.
It also involves replacing mappings of such continuum function data with suitably chosen finite dimensional transformations that act on these arrays or tensors. 
It is in fact fair to claim that much of numerical analysis is devoted to developing and analysing computationally efficient discretisations.

It may therefore be tempting to adopt a `discretise then compute' approach where one first discretises the computing task, then considers computational models.
As we already indicated in the preceding section, there are several advantages that come with a `compute then discretise' approach where computing tasks are formulated in the continuum setting before discretisation is applied.
For this reason, much of the development and analysis of computational  methods in numerical analysis is pursued within a `compute then discretise' approach. 
The corresponding notion for operator learning would be `learn then discretise'.
A motivation for promoting this is that this opens up for the possibility to have a structured way to adapt the learning method to changes in the discretisation, e.g., when the discretisation is refined.
Another is the desire to single out intrinsic approximation properties, which do not depend on discretisation, from those that relate to choice of discretisation. 
Results related to intrinsic approximation properties are independent of the choice of discretisation. 
Additionally, as noted earlier, the analysis of properties of computational tasks, such as stability, is most effectively carried out in an infinite-dimensional continuum setting, where the choice of topology plays a crucial role. 
This perspective is central to the development of regularisation methods for ill-posed inverse problems.

Adopting a `learn then discretise' approach in operator learning means formulating statistical learning in a functional analytic setting.
A minimal requirement for this is that the sets containing data can be endowed with probability measures. 
In addition, these sets need to be topological spaces that are preferably Banach spaces, as the latter enables the use of probability theory on Banach spaces where Borel sets are measurable.

The advantage of formulating operator learning within such a functional analytic framework is that it provides a natural setting for theoretical analysis. 
Analogous to contemporary numerical analysis, which often follows a `compute-then-discretise' paradigm, the theoretical analysis of operator learning is most naturally conducted within a “learn-then-discretise” framework.
Section~\ref{sec:TheoryFoundations} provides some examples of theoretical results that have been derived for operator learning.

\subsection{Major parts of operator learning}\label{sec:OLparts}
A central part of operator learning consists of choices made within the following two topics:
\begin{description}
\item[Learning tasks:]
\emph{Operator learning} relies on principles from statistical decision theory to select the neural operator that `best' approximates the target operator \eqref{eq:MainOperator} given training data. 
This requires one to specify what `best' means along with the statistical properties for training data and how they relate to the target operator.

The typical setup is to define learning as a minimiser to a loss that involves the training data. The various options (leaving out variants based on reinforcement) are outlined in Section~\ref{sec:LearningProbs}. 
Defining the optimisation problem for the learning is still a conceptual step in that it specifies `what' one seeks to compute during operator learning. 
This needs to be complemented with selecting an algorithm (training algorithm) that preferably provably converges to a solution to the optimisation.

\item[Neural operator architecture:]
The \emph{architecture} of the neural operator is a precise parametrisation of the family of mappings one searches over during statistical learning, i.e., it is a parametrisation of the hypothesis space. Stated formally, if the aim is to approximate the target operator \eqref{eq:MainOperator}, then the architecture is a specification of a family of mappings $\{ \RecOp_{\NNparam} \}_{\NNparam\in \NNparamSet}$ where $\RecOp_{\NNparam} \colon \DataSpace \to \RecSpace$ and the parameter set $\NNparamSet$ is a finite dimensional vector space.

As already mentioned, an important aspect of operator learning is extending common neural network architectures, which usually represent mappings between finite-dimensional spaces, to the functional analytic setting in which the input and output spaces can be infinite-dimensional. The choice of architecture strongly influences approximation properties of the learned neural operator. Consequently, much focus has been spent on stating and proving universal approximation theorems for specific architectures. This, along with stability properties, is important when neural operators are used to solve ill-posed inverse problems (learned reconstruction) \cite{arridge2019solving,hauptmann2024convergent,Mukherjee:2023aa}.
\end{description}

In the following, we consider operator learning where the target operator is defined as the solution operator for an inverse problem or an optimisation problem.
Such target operators are often implicitly defined by an iterative scheme, and as previously mentioned, this will be a key consideration in selecting an appropriate architecture for the neural operator.

\subsubsection{Inverse problem solvers (learned reconstruction)}\label{sec:InvProb}
The task of solving an inverse problem amounts to recovering a hidden model parameter (signal) from indirect observations (data).
The setting we consider is when one has a model for generating data from a model parameter (simulator), so the aforementioned task amounts to running a simulator `backwards'.

To formalise the above, we introduce the sets $\RecSpace$ (reconstruction space) and $\DataSpace$ (data space).
Elements in these sets represent possible signals and data, respectively. 
It is also common to furthermore assume that these sets are Banach (or Hilbert) spaces.
Next, we also assume one has access to a \emph{forward operator}, which is a (possibly non-linear) mapping $\FwdOp \colon \RecSpace\to \DataSpace$ that models how an element in $\RecSpace$ gives rise to an element in $\DataSpace$ in absence of noise.
Now, a common formalisation of solving an inverse problem is to view it as solving an operator equation. 
\begin{definition}[Inverse problem, deterministic formalisation]\label{def:InvProbFuncAna}
Assume \emph{data} $\data \in \DataSpace$ relates to a \emph{signal} $\signaltrue \in \RecSpace$ by 
\begin{equation}\label{eq:InvProb}
 \data = \FwdOp(\signaltrue) + \datanoise
\end{equation}
with $\datanoise \in \DataSpace$ denoting an (unknown) observation error.
The inverse problem of \emph{reconstructing} this signal $\signaltrue$ from such observed data $\data$ amounts to solving the operator equation in \eqref{eq:InvProb}. 
\end{definition}
Solving an inverse problem is often ill-posed meaning that there can be multiple solutions consistent with data and/or solutions are sensitive to (small) variations in data (instability).
Lack of a unique solution can often be dealt with by considering other notions for a solution, like minimum norm least-squares solution. 
In contrast, handling the intrinsic instability in solving ill-posed inverse problems is more intricate and this is a central topic in \emph{regularisation theory} \cite{Tikhonov:1995aa,Engl:2000aa,Scherzer:2009aa,Schuster:2012aa}.

The above definition of an inverse problem is deterministic.
It is therefore not an ideal bases for attempts at setting up statistical learning for solving inverse problems. 
Statistical notions enter naturally with the observation error $\datanoise$. 
It is often possible to characterise the statistical distribution of the observation error, but Definition~\ref{def:InvProbFuncAna} does not explicitly account for this. 
Incorporating it formally requires one to introduce a $\DataSpace$-valued random variable $\datanoiserand$ that generates observation errors like $\datanoise$.
As a consequence, observe data becomes a single sample of a $\DataSpace$-valued random variable.
Furthermore, some learning tasks (Section~\ref{sec:LearningProbs}), like supervised and weakly supervised learning, also require one to introduce a statistical model for the signal one seeks to recover.
These considerations lead to the statistical formalisation that views the task of solving an inverse problem as Bayesian inference.
\begin{definition}[Inverse problem, statistical formalisation]\label{def:InvProbStat}
Let $\signalrand$ denote a $\RecSpace$-valued random variable generating possible signals and $\datarand$ denotes the $\DataSpace$-valued random variable generating possible data.
The task of solving the inverse problem is now defined as recovering the (posterior) distribution of the $\RecSpace$-valued random variable $(\signalrand \mid \datarand = \data)$ conditioned on observed data $\data \in \DataSpace$ that is a single sample of the $\DataSpace$-valued random variable $(\datarand \mid \signalrand=\signaltrue)$ (with $\signaltrue \in \RecSpace$ unknown) where 
\begin{equation}\label{eq:InvProbStat}
 \datarand = \FwdOp(\signalrand) +  \datanoiserand
\end{equation}
with $\datanoiserand$ denoting a $\DataSpace$-valued random variable modelling observation noise.
Alternatively, one can define the task of solving the inverse problem as recovering a statistical point estimator of the posterior distribution that approximates the true unknown signal $\signaltrue$.
Examples are the posterior mean, posterior median, maximum likelihood, and maximum a posteriori estimators.
\end{definition}
The posterior distribution in Definition~\ref{def:InvProbStat} is along with the posterior mean often continuous w.r.t.\@ data even when the inverse problem is ill-posed in the context of Definition~\ref{def:InvProbFuncAna} \cite{Latz:2023aa}.
The notion of ill-posedness in the context of Definition~\ref{def:InvProbStat} could mean that the posterior distribution is not continuous w.r.t.\@ data, but a common alternative definition is that the maximum likelihood point estimator is not continuous w.r.t.\@ data.
\begin{remark}
Rigorously phrasing Definition~\ref{def:InvProbStat} in its full generality requires introducing technical conditions on the topologies for the reconstruction and data spaces to ensure the existence of conditional distributions. 
Furthermore, the forward operator must also satisfy some basic regularity conditions, like being measurable and preferably also Fréchet differentiable.
We will refrain from further pursuing the narrative along this direction, the interested reader may consult \cite{Nelsen:2025aa} for more about the statistical formalisation of an inverse problem.
\end{remark}

In many applications one either has a forward operator that is linear, or one can recast the inverse problem by pre-processing data into a form where the forward operator is linear.
The latter is e.g. the case with \ac{CT} image reconstruction where data can after pre-processing be interpreted as samples of the ray transform, i.e., the corresponding forward operator is linear. 
There are however inverse problems, like electrical impedance tomography (EIT), that are intrinsically non-linear, meaning that there are no pre-processing steps or linearisations that allow one to assume a linear forward operator. 

A solution method for the inverse problem in Definition~\ref{def:InvProbFuncAna} is represented by a \emph{reconstruction operator} $\RecOp \colon \DataSpace \to \RecSpace$ that approximates some form of (pseudo) inverse of the forward operator. 
Likewise, for Definition~\ref{def:InvProbStat} it is either an operator that maps a point in $\DataSpace$ to a probability distribution on $\RecSpace$, or in case of a point-estimator, it is an operator of the form $\RecOp \colon \DataSpace \to \RecSpace$.

The reconstruction operator must be well-defined (existence and uniqueness) and ideally it is also stable w.r.t.\@ (small) variations in data (well-posed). 
It must also provide a good approximation to the original hidden signal $\signal$ that gives rise to observed data $\data$, i.e., $\RecOp(\data)\approx \signal$ whenever $\data \approx \FwdOp(\signal)$. 
Classically, such a reconstruction operator is handcrafted from the knowledge about the forward operator and its inverse, or formulated in the variational framework as an optimisation problem (maximum a posteriori estimator). 
However, it is also possible to learn a reconstruction operator from example data.
A formal definition in the setting of operator learning is given as follows.
\begin{definition}[Learned reconstruction operator]
A family $\{ \RecOp_{\NNparam} \}_\NNparam$ of parametrised mappings  
\[ \RecOp_{\NNparam} \colon \DataSpace \to \RecSpace 
  \quad\text{where $\NNparam\in\NNparamSet$}
\]
along with a specification of the parameter set $\NNparamSet$ is called a \emph{learned reconstruction operator} for the inverse problem in Definition~\ref{def:InvProbFuncAna} if the parameters $\NNparam$ are determined (learned) from example data (training data) that is generated in a way that is consistent with \eqref{eq:InvProb}.
\end{definition}
One can `lift' the above definition to the statistical inverse problem in Definition~\ref{def:InvProbStat} by defining considering a family $\{ \RecOp_{\NNparam} \}_\NNparam$ of parametrised mappings 
 \[ \RecOp_{\NNparam} \colon \DataSpace \to \mathscr{P}_\RecSpace 
  \quad\text{where $\NNparam\in\NNparamSet$}
 \]
where $\mathscr{P}_\RecSpace$ is the space of probability measures on $\RecSpace$.
The parameters $\NNparam$ are now determined (learned) from example data (training data) that is generated in a way that is consistent with \eqref{eq:InvProbStat}.
Alternatively, one can also require that $\RecOp_{\NNparam} \colon \DataSpace \to \RecSpace$ is a statistical point estimator for the conditional distribution of the $\RecSpace$-valued random variable $(\signalrand \mid \datarand=\data)$ in Definition~\ref{def:InvProbStat}.

\subsubsection{Optimisation solvers (learned optimisation)}\label{sec:OptimSolver}
Operator learning is here used for optimisation. 
More precisely, the aim is to train a neural operator so that it approximates a target operator $\RecOp \colon \DataSpace \to \RecSpace$ that minimises a parametrised convex objective $\signal \mapsto \CostFunc_{\data}(\signal)$, i.e., 
\begin{equation}\label{eq:learning2Opt_reconOp}
 \RecOp(\data) := \argmin_{\signal \in \RecSpace} \CostFunc_{\data}(\signal)
 \quad\text{for $\data \in \DataSpace$.}
\end{equation}
The resulting trained solver is specifically designed to compute minimisers to objectives in a class $\{ \CostFunc_{\data}  \}_{\data \in \DataSpace}$, i.e., minimisers to functionals $\CostFunc_{\data} \colon \RecSpace \to \Real$ that are parametrised by $\data\in \DataSpace$.

The next step is to set-up an appropriate learning problem so that training the neural operator $\RecOp_{\NNparam} \colon \DataSpace \to \RecSpace$ corresponds to approximating $\RecOp$ (learning-to-optimise).
Any learning task that requires access to values of $\RecOp(\data)$ in \eqref{eq:learning2Opt_reconOp} will quickly become unfeasible.
This also means that a classical supervised approach as in \eqref{eq:SupDataX-Loss} is unfeasible as its empirical counterpart \eqref{eq:SupDataX-Loss_emp} assumes access to pairs $(\signal_i,\data_i)$ where $\signal_i \in \RecSpace$ minimises $\signal \mapsto \CostFunc_{\data_i}(\signal)$. 
Obtaining sufficient amount of such paired data $(\signal_i,\data_i)$ may be impossible.
An alternative could be \eqref{eq:UnsupDataY-Loss}, but this learning task results in a bi-level optimisation problem as the target operator in \eqref{eq:learning2Opt_reconOp} is itself given as the solution of an optimisation problem.
Hence, it is non-trivial to evaluate it and also to compute its derivative, thus rendering approaches based on \eqref{eq:UnsupDataY-Loss} as unfeasible. 

Given the above considerations, a better way to set-up the learning task for learning to optimise is to consider an unsupervised approach.
In particular, one can use the objective function as the loss as in \cite{Banert:2020aa,Banert:2024aa}, which is similar to the \acp{PINN} approach for solving a \ac{PDE} where the loss-function consists of the \ac{PDE} and its boundary conditions.
Other similar approaches are nicely surveyed in \cite[Sec.~4]{Carioni:2025aa}. 
There are systematic comparisons of how such learning-to-optimise approaches perform compared to handcrafted methods for minimising $\signal \mapsto \CostFunc_{\data}(\signal)$ for some given $\data$.
These include systematic tests against several classes of large-scale smooth and non-smooth convex optimisation problems taken from inverse problems in imaging. 
Results show superior performance of learning-to-optimise approach compared to established accelerated solvers, the interested reader may consult \cite[Sec.~5.6]{Banert:2024aa}.

\subsubsection{Concluding remarks}
It is important to note that the same neural operator, with an architecture given by a learned iterative network, can be used to either solve an inverse problem (learned reconstruction) or to compute a minimiser (learned optimisation).
The choice of architecture has some influence on \emph{how well} one can approximate a target operator as this partly depends on expressivity of the hypothesis space.
However, \emph{what} target operator one seeks to approximate is determined by the learning task as outlined in Section~\ref{sec:LearningProbs}, see also Remark~\ref{rem:RoleOfUnrolling} that further discusses this in the context of learned iterative networks.
This clarification of roles of for learning task vs.\@ choice of neural operator architecture can at times become confusing, and especially so when one seeks to solve an inverse problem by computing a maximum a posterior estimator as the estimator is then itself defined by an optimisation problem.

\subsection{Learning tasks}\label{sec:LearningProbs}
The idea in operator learning is to set-up an explicitly parametrised family of $\RecSpace$-valued mappings defined on $\DataSpace$, and then obtain the `best' parameter $\NNparam$ from training data. 
The type of training data and choice of method for quantifying performance during training (loss function) influences how to set up the training procedure for the operator learning, which we henceforth refer to as the \emph{learning problem}. 

As we will show, the same neural operator can be trained in different ways depending on how one formulates the learning problem. 
These choices correspond to usage of different concepts from machine learning.
Our focus will henceforth be on solving inverse problems. 
To prepare for discussing the various different ways of setting up the learning problem, we distinguish between two primary cases for training data in inverse problems: fully \emph{supervised} and \emph{unsupervised} data. 

The first setting, the supervised case, assumes access to example data (training data) of the form
\begin{equation}\label{eq:SupervisedData}
(\signal_1,\data_1), \ldots, (\signal_n,\data_n) 
  \in \RecSpace \times \DataSpace
\end{equation}
that are \ac{IID} random draws from an $(\RecSpace \times \DataSpace)$-valued random variable $(\signalrand,\datarand)$ that satisfies the operator equation $\signalrand=\RecOp(\datarand)$ where $\RecOp \colon \DataSpace \to \RecSpace$ is the target operator one seeks to approximate.
For solving inverse problem (Section~\ref{sec:InvProb}), an equivalent assumption is that \eqref{eq:InvProb} holds. 
While this is the desired case from a learning perspective, e.g., it is not the usual case for inverse problems to have access to a large set of ground-truth samples.
It is also not a suitable setting for learning to optimise as explained in Section~\ref{sec:OptimSolver}.

The second case, the unsupervised setting, is therefore the most natural one for both solving inverse problems and for learning to optimise.
Here we only have training data that consists of the observed data, i.e., samples $\data_1, \ldots, \data_n \in \DataSpace$ that are generated by a $\DataSpace$-valued random variable $\datarand$. Alternatively, we also consider the case where samples $\signal_1, \ldots, \signal_n \in \RecSpace$ by a $\RecSpace$-valued random variable $\signalrand$ are given. 

\subsubsection{Supervised operator learning}
\label{sec:Supervised}
The most common setting for supervised learning is with a loss function in $\RecSpace$, i.e., $\Loss_{\RecSpace} \colon \RecSpace \times \RecSpace \to \Real$, like a $p$-norm. 
The learned operator $\RecOp_\NNparam$ can then be defined as the associated Bayes estimator, which is given as $\RecOp_{\est{\NNparam}} \colon \DataSpace \to \RecSpace$ where $\est{\NNparam} \in \NNparamSet$ solves the following learning problem:
\begin{equation}\label{eq:SupDataX-Loss}
  \est{\NNparam} \in \argmin_{\NNparam \in \NNparamSet} 
   \Expect_{(\signalrand,\datarand)}\Bigl[ 
    \Loss_{\RecSpace}\bigl( \RecOp_{\NNparam}(\datarand), \signalrand \bigr)
   \Bigr].
\end{equation}
In practice we only have finite number of samples $(\signal_1,\data_1), \ldots , (\signal_n,\data_n) \in \RecSpace \times \DataSpace$ from $(\signalrand,\datarand)$. 
Hence, instead of solving the learning problem in \eqref{eq:SupDataX-Loss}, we compute the corresponding empirical risk estimator by solving the training problem
\begin{equation}\label{eq:SupDataX-Loss_emp}
  \est{\NNparam} \in \argmin_{\NNparam \in \NNparamSet}
    \frac{1}{n}\sum_{i=1}^n 
    \Loss_{\RecSpace}\bigl( \RecOp_{\NNparam}(\data_i), \signal_i \bigr).
\end{equation}

The above formalisation of learning is a straightforward generalisation of statistical learning into a setting where $\RecSpace$ and $\DataSpace$ can be infinite dimensional vector spaces.
The full statistical interpretation requires some additional structure, like assuming that $\RecSpace$ and $\DataSpace$ are separable Hilbert spaces.
Then Bayes theorem holds and if the output variable follows a white noise model, then the Bayes estimator can also be interpreted as a maximum likelihood estimator over the of neural operators specified by the architecture, see \cite{Reinhardt:2024aa} for further details.

Supervised learning is the most often considered setting in the literature for solving inverse problem, since it is in fact the most desirable case for learned reconstruction.
This is simply because the learning involves training data that provide access to ground-truth/measurement pairs which are directly linked by the underlying operator equation and hence carry the most information of the inverse problem. 
But, this is also its main drawback, namely that each output data in the training data needs to have a corresponding ground-truth input data.
This can be problematic to obtain in practice.

\subsubsection{Unsupervised operator learning}\label{sec:unsuperlearn}
It is often difficult to gain access to paired training examples that are drawn from the joint distribution of the $(\RecSpace \times \DataSpace)$-valued random variable $(\signalrand,\datarand)$.

To illustrate the difficulty, consider learned reconstructions for low-dose \ac{CT} image reconstruction.
The latter refers to image reconstruction when x-ray \ac{CT} data is acquired with dose reduction techniques \cite{Aloud:2025aa}. 
One such technique commonly used in clinical \ac{CT} reduces radiation intensity by adjusting the tube current and voltage, but this also leads to noisier data. 
Another is to reduce number of exposures by sub-sampling data (sparse view \ac{CT}), but this leads to less data.
Supervised data would in this context consist of pairs $(\signal_i,\data_i) \in \RecSpace \times \DataSpace$ where $\signal_i$ is the true image and $\data_i$ is the low-dose \ac{CT} data generated by $\signal_i$ (see Section~\ref{sec:Supervised}).
Experimentally acquiring such data is challenging since it is difficult to get access to the true image $\signal_i$ that generated the low-dose data $\data_i$. A typical approach would be to scan the object twice, one high-dose scan and one low-dose scan.
The former is used for computing the `true' image $\signal_i$. One also needs to align the `true' image to the image reconstructed from corresponding low-dose data to ensure exact correspondence (one could avoid this step by using a loss that is invariant to rigid body transformations \cite{Adler:2017ab}).
Supervised training in \ac{CT} image reconstruction is for these reasons commonly based on simulated data.

Consequently, a more practical setting in inverse problems is when training data represent \ac{IID} samples from the marginal distribution of $\signalrand$ or $\datarand$.
More precisely, one typically encounters on of the following three scenarios (or, some combinations thereof) that we now outline. 
\begin{remark}
We only present a subset of possible learning problems to provide an intuition on potential formulations, rather than a comprehensive survey of loss functions.
\end{remark}

\paragraph{Fully unsupervised training data}
This is the setting where one only has access to \ac{IID} samples $\data_i \in \DataSpace$ from the marginal distribution of $\datarand$.
Such a setting for operator learning without access to ground-truth is the most natural setting in inverse problems since data is the only entity that is always observed. Nevertheless, formulating a fully unsupervised learning problem is not straight-forward.

One possibility in this setting is to consider `learning-to-optimise'. 
Here, the training aims to learn how to accelerate an optimisation solver. 
Hence, the reconstruction operator $\RecOp\colon\DataSpace\to\RecSpace$ is given by  handcrafted variational model as in \eqref{eq:learning2Opt_reconOp} and the aim of the learning is to speed up this solution method. The learning problem is then 
\begin{equation}\label{eq:UnsupDataY-Loss}
 \est{\NNparam} \in \argmin_{\NNparam}
  \Expect_{\datarand}\Bigl[
    \DataLoss\bigl(
     \RecOp_{\NNparam}(\datarand),
     \RecOp(\datarand)
    \bigr)
   \Bigr].
\end{equation}
Clearly, such training can be done without any access to ground-truth solutions.

Another possible learning problem is to consider self-supervised learning.
This applies specifically to solving an inverse problem (Section~\ref{sec:InvProb}) as it makes use of a forward operator $\FwdOp \colon \RecSpace \to \DataSpace$. 
The idea is to use the parametrisation of $\RecOp_{\NNparam} \colon \DataSpace \to \RecSpace$ as an implicit regulariser by setting up the learning problem
\begin{equation}\label{eq:DIP_loss}
 \est{\NNparam} \in \argmin_{\NNparam}
  \Expect_{\datarand}\Bigl[
    \DataLoss\bigl(
     (\FwdOp \circ \RecOp_{\NNparam})(\datarand),
     \datarand
    \bigr)
   \Bigr]
\end{equation}
for some loss function $\DataLoss \colon \DataSpace \times \DataSpace \to \Real$. For the 2-norm this loss would correspond to minimising the data-fidelity and hence care needs to be taken with respect to overfitting noise and considerations of the range of $\FwdOp$.

\begin{remark}
Note that deep image prior applied to inverse problems often coincides with the above self-supervised learning formulation \eqref{eq:DIP_loss}, when only one data sample $\data$ is considered. In fact, the learned reconstruction operator $\RecOp_\NNparam$ is expected to fit low-frequency components first, i.e., implicitly regularising, but still requires early stopping or the use of additional regularisers to avoid overfitting noise  \cite{baguer2020computed,barbano2022educated,dittmer2020regularisation}.
\end{remark}

\paragraph{Ground truth examples as training data}
This is the setting where one only has access to \ac{IID} samples $\signal_i \in \RecSpace$ from the distribution of the ground-truth solutions to the inverse problem, i.e., samples from the true prior.
In this setting, one could of course use the forward operator to generate corresponding noisy data $\data_i \in \DataSpace$, thus recasting the learning problem to a supervised setting, which may represent real measurement data to varying degrees.

A natural approach for setting up a learned reconstruction method that avoids simulated supervised data is to base it on a regulariser that is derived from the original training data, treating $\signal_i \in \RecSpace$ as \ac{IID} samples from the true prior.
Many approaches have been based on this typically leveraging deep generative models, e.g., constructing a projection on the range of the pre-trained generator and approximating its inverse. 
One of the first approaches along these lines was \cite{Lunz:2018aa} (see also \cite{Lunz:2023aa}) and many other variants have since then been developed \cite{Asim:2020aa,Bora:2017aa,Daras:2021aa,Shah:2018aa,Xia:2023aa}.

\paragraph{Weakly supervised training data}
This is the setting when training data consists of unpaired \ac{IID} samples $\signal_i \in \RecSpace$ and $\data_i \in \DataSpace$ from the $\RecSpace$- and $\DataSpace$-marginals $\pi_{\RecSpace}$ and $\pi_{\DataSpace}$ of the joint measure for the $(\RecSpace \times \DataSpace)$-valued random variable $(\signalrand,\datarand)$, respectively.
One can then set up a learning problem that is based on conditional variants of common generative models. 

One formulation of the learning problem that applies to solving an inverse problem (Section~\ref{sec:InvProb}) as it makes use of a forward operator $\FwdOp \colon \RecSpace \to \DataSpace$ is based on training the learned reconstruction operator $\RecOp_{\est{\NNparam}} \colon \DataSpace \to \RecSpace$ where $\est{\NNparam}$ is obtained as
\begin{multline}\label{eq:UnpairedLearning}
\est{\NNparam} \in \argmin_{\NNparam}
 \biggl\{
   \Expect_{\substack{\signalrand \sim \pi_{\RecSpace} \\ \datarand \sim  \pi_{\DataSpace}}}\biggl[ 
     \DataLoss\Bigl( \FwdOp\bigl(\RecOp_{\NNparam}(\datarand)\bigr), \datarand \Bigr)
     +  
     \SignalLoss\bigl( \RecOp_{\NNparam}(\datarand), \signalrand \bigr)
   \biggr] 
\\ 
 + \lambda 
    \Loss_{\PClass{\RecSpace}}\bigl( (\RecOp_{\NNparam})_{\#}(\pi_{\DataSpace}),\pi_{\RecSpace} \bigr) 
  \biggr\}.        
\end{multline}
In the above, $\SignalLoss \colon \RecSpace \times \RecSpace \to \Real$ and $\DataLoss \colon \DataSpace \times \DataSpace \to \Real$ are loss functions on $\RecSpace$ and $\DataSpace$, respectively.
Next, $\PClass{\RecSpace}$ and $\PClass{\DataSpace}$ denote the class of probability measures over $\RecSpace$ and $\DataSpace$,  $\Loss_{\PClass{\RecSpace}} \colon \PClass{\RecSpace} \times \PClass{\RecSpace} \to \Real$ is a distance notion between probability measures on $\RecSpace$, and $(\RecOp_{\NNparam})_{\#}(\pi_{\DataSpace}) \in \PClass{\RecSpace}$ denotes the push-forward of the probability measure $\pi_{\DataSpace} \in \PClass{\DataSpace}$ by $\RecOp_{\NNparam} \colon \DataSpace \to \RecSpace$.
It is common to evaluate $\Loss_{\PClass{\RecSpace}}$ using techniques from \acp{GAN}, which introduce a separate  neural operator as discriminator.
Finally, the parameter $\lambda$ controls the balance between the distributional consistency, noise suppression and data consistency.
See \cite[Sec.~3]{Carioni:2025aa} for a survey of various special cases of the above learning problem. 
One may also consider variants when there is access to a large sample of unpaired (weakly supervised) data combined with a small amount of paired data, or when parts of the probability distributions involved are known.

On a final note, another common setting in inverse problems is when one has access to a large amount of unsupervised data and a smaller amount of supervised data. 
One could then consider fine tuning, which is commonly used for re-training a generic foundation model (often a large language model) for a specific task \cite{Engelen:2020aa,Luo:2024aa,Wang:2025aa}.

\subsection{Neural operator architectures}\label{sec:NeuralOperatorArchi}
Key considerations in approximating a target operator $\RecOp \colon \DataSpace \to \RecSpace$ with a neural operator $\RecOp_{\NNparam} \colon \DataSpace \to \RecSpace$ is to ensure the learned approximation is robust against noise and accurate also for unseen examples.
As argued for in Section~\ref{sec:MotivationStructuredLearning}, this depends to large extent on the neural operator architecture, i.e., on how the neural operators in the hypothesis space $\{ \RecOp_{\NNparam} \}_{\NNparam}$ are parametrised.
Both accuracy and robustness rely on using an architecture with sufficient model capacity that generalises well and is numerically stable. 

For the above reasons, there has been much interest in extending various neural network architectures to a functional analytic setting. 
The approaches differ in how one defines the layer, i.e., how one extend the combination of affine transformations and non-linear activations to a functional analytic setting.

One of the earliest examples of a neural operator architecture is the one introduced in \cite{Chen:1995aa}, which later served as a foundation for encoder-decoder architectures, like \ac{DeepONet} (Section~\ref{sec:OtherNeuralOpArch}). 
A more direct construction was the basic neural operator architecture proposed in \cite{Anandkumar:2020aa,Kovachki:2023aa}, where neural operators are defined in close analogy with conventional neural networks by replacing weight matrices in the hidden layers with integral operators.
Subsequent developments have largely focused on incorporating domain adaptation through inductive bias (structured learning) that is primarily motivated by attempts at solving inverse problems and \acp{PDE} with neural operators (Section~\ref{sec:MotivationStructuredLearning}).
These approaches range from relatively simple modifications of domain-agnostic architectures (Section~\ref{sec:PosPreArch}) to more advanced designs, like learned iterative networks that are defined through an underlying iterative scheme, to architectures where the layers are given by integral operators that have structures tailored to the target operator $\RecOp$ being approximated (Section~\ref{sec:OtherNeuralOpArch}), e.g., as in \acp{FNO}.

As already mentioned, the aforementioned learned iterative networks are the central topic of this survey. 
As outlined in Section~\ref{sec:MotivationStructuredLearning}, and more properly described in Section~\ref{sec:LearnedIterNetwork}, these architectures are obtained by unrolling an appropriate underlying iterative scheme.
Learned iterative networks are in the scientific literature typically formulated in the finite dimensional setting.
The analogous formulation in infinite dimensional setting involves two steps. 
First is to formulate the iterative scheme in such a setting, and second, to replace the neural networks used in the unrolled scheme with corresponding neural operators (neural updating operators).
The first step is often straightforward, so formulating learned iterative networks in infinite dimensional setting almost always reduces to specifying the neural updating operators, which are the infinite dimensional analogues of the neural networks used in the unrolled scheme.
An example that illustrates this is \cite{Andrade-Loarca:2022aa}, which considers the functional analytic version of the learned iterative network in \cite{Adler:2018aa} (a primal-dual architecture) by replacing the \acp{CNN} with differential operators. 
Likewise, \cite{Runkel:2024aa} views a neural \ac{ODE} as the functional analytic version of the same architecture.

In conclusion, the key part in formulating neural operator variants of learned iterative networks lies in finding infinite dimensional analogues of basic neural network components, which will be discussed next.

\subsubsection{Basic domain agnostic architectures}\label{sec:Component}
Focus here is on domain-agnostic neural operator architectures.
As already mentioned, a direct construction is the \emph{feed-forward neural operator architecture} in \cite{Anandkumar:2020aa,Kovachki:2023aa} that is obtained by generalising the compositional feed-forward structure of standard deep neural networks to function spaces.
This yields an architecture of the form in \cite[eq.~(3.13)]{Nelsen:2025aa}, see also \cite[Sec.~1.1]{Boulle:2024aa}. 
A key step here is to replace the weight matrices in the hidden layers of the feed-forward network with integral operators.
A popular development is to introduce domain adaptation by choosing kernels with specific structure.
Such domain adapted integral kernel network architectures are surveyed in Section~\ref{sec:OtherNeuralOpArch}. 
\begin{remark}
    Feed-forward neural operator architectures are often only referred to as neural operators. 
    This can be somewhat confusing as we in this survey reserve the term neural operator to mappings that generalise neural networks to infinite dimensional setting (Section~\ref{sec:NeuralOperatorArchi}).
\end{remark}
Another line of development is to extend convolutional and attention mechanisms to a functional-analytic setting, thus opening up for formulating neural operator versions of basic \acp{CNN} and transformer networks. 
We here focus on these as such components also serve as building blocks for more advanced domain-adapted architectures, like learned iterative networks (Section~\ref{sec:LearnedIterNetw}).

\paragraph{Convolutional neural operators}
One of the earliest examples of viewing the discrete convolutional layer in a \ac{CNN} as a discretisation of an operator is \cite{Ruthotto:2017aa}. 
Here, they consider \ac{CNN} based \acp{ResNet} and the discrete convolutional layers are viewed as discretisations of a differential operator.
The \ac{ResNet} can thereby be seen as a numerical integration method, so a stable numerical integration could serve as a blue print for setting up a stable \ac{CNN} based \ac{ResNet} architecture. 
The approach is further developed in \cite{Ruthotto:2020aa}.
This shows that one can extend a \ac{CNN} to function spaces by viewing convolutional layers as differential operators.
Such an approach was adopted in \cite{Andrade-Loarca:2022aa} where the aim was to characterise the microlocal canonical relation of the learned iterative network in \cite{Adler:2018aa} (\acl{LPD} architecture, see Section~\ref{sec:LPD}). Specialised architectures for certain inverse problems have also been developed by discretising the involved differential operators in the governing \ac{PDE} \cite{arridge2020networks}.

The need to extend \ac{CNN} type of architectures to a functional analytic setting was independently considered in the context of solving \acp{PDE}.
One of the earliest systematic approaches is presented in \cite{Karras:2021aa} that introduced the convolutional neural operator for image generation.
Here, one replaces the discrete convolutions in a \ac{CNN} with convolution operators.
The activation function can be any non-linear function that is applied pointwise. 
This generates aliasing errors as it does not necessarily respect any band-limits of the underlying function space.  
In particular, non-linear activations can generate features at arbitrarily high frequencies.
To address this drawback, \cite{Raonic:2023aa,Raonic:2023ab} modulated the application of the activation function so that the resulting outputs fall within desired band-limits. 
This results in a convolutional neural operator that maps band-limited functions to band-limited functions, thus respecting the continuous-discrete equivalence which is important when solving \acp{PDE}. 

A functional analytic version of the U-Net architecture was proposed in \cite{Cheng:2023aa}, showing increased performance over its discrete counterpart in many imaging applications and wrapped with theoretical guarantees around its performance and robustness. 
Another approach views a U-Net as a one-step unrolling an operator-splitting algorithm that solves some control problem \cite{Tai:2025aa}.

On a final note, we mention that the \ac{FNO} introduced in \cite{Li:2021aa} can be seen as a neural operator variant of a \ac{CNN} that uses the Fourier space parametrisation of a convolution.
Furthermore, one can also view the \ac{FNO} in its functional analytic formulation as a \ac{DeepONet} with a specific architecture of the branch and trunk networks represented by a trigonometric basis \cite{Kovachki:2021aa}.

\paragraph{Transformers}
The attention mechanism in the transformer architecture has, since its introduction in \cite{Bahdanau:2014aa}, found widespread use on various applications.
A key feature has been its ability to significantly enhance model performance and generalisation by offering a mechanism for capturing long range dependencies. 

One of the first extensions of transformer architectures to operator learning came in \cite{Cao:2021aa} with the introduction of the GK-Transformer. 
The attention mechanism is here viewed as an integral transform with a learnable
kernel (which differs from the \ac{FNO} that uses a fixed kernel).
Input and output features of the GK-Transformer need to have the same discretisation, thus preventing its use in problems where discretisations may vary.
Addressing this drawback led to the development of the general neural operator transformer (GNOT) \cite{Hao:2023aa}. This is a neural operator architecture capable of accommodating multiple input functions, irregular grids, and multi-scale problems with high flexibility, allowing input and output at any location. 
See also \cite{Shih:2025aa} for similar approach for defining a neural operator based on Fourier type-linear attention.

One can also consider neural operator architectures in which the attention is represented by a kernel integral operator \cite{Boya:2025aa,Kovachki:2023aa}.
Finally, it is also possible to define a operator formulations of self-attention that retains its role as a sequence-to-sequence model \cite[Definition~5]{Calvello:2025aa}.
This functional analytic operator formulation of self-attention connects to the standard definition in the finite dimensional setting by applying Monte Carlo approximation of the relevant integrals \cite[Thm.~6]{Calvello:2025aa}.

\subsubsection{Need for domain adapted architectures}\label{sec:MotivationStructuredLearning}
Machine learning methods, and in particular those based on deep learning, are commonly adapted to a specific domain by choosing appropriate training and test data, whereas architectures are chosen to be fairly general. 

In case of operator learning, domain agnostic architectures, like those based on \acp{CNN} or transformers that make up foundation models and \acp{LLM}, quickly become unfeasible to set-up and train due to the large-scale nature of operator approximation.
This \emph{scalability} challenge is especially prominent when deep learning is used to solve inverse problems arising in imaging.
In addition, approaches based on generic architectures tend to \emph{generalise poorly} even if trained against very large data sets. 

\paragraph{Example from tomographic image reconstruction}
We here illustrate the aforementioned challenges that comes with using a neural operator with domain agnostic architecture for image reconstruction in tomography. 
\begin{itemize}
\item \emph{Scalability:}
There has been a few attempts at training neural networks with task agnostic architectures, like AUTOMAP \cite{Zhu:2018aa}, which involves fully connected layers to approximate the inverse operator. 
An issue with such architectures is that they scale poorly.
This is evident already for small 2D tomographic imaging problems, like recovering a $512 \times 512$ pixel image from parallel beam tomographic data gathered from 128 different projection angles with a 512 pixel line detector.
The first two layers of the AUTOMAP has for this problem $128 \times 512^3 + 512^4 \approx 8.5 \cdot 10^{10}$ parameters, which implies $>$340~Gigabytes of GPU memory \cite{Bazrafkan:2019aa}.
ETER-net \cite{Oh:2018aa}, which is a further development of AUTOMAP, does reduce the required parameters by over 80\% (by replacing the fully connected/convolutional auto-encoder architecture in AUTOMAP with a recurrent neural network architecture). However, this still results in a network with about $6.8 \cdot 10^{10}$ parameters that need to be learned from training data.

A similar `back of the envelope' calculation for clinical 3D tomography indicates that using AUTOMAP type of architectures, or some other foundation model based on a generic \ac{LLM}, results in a network with about $10^{16}$ learnable parameters. 
There is not enough training data for learning that many parameters in a reliable way. 

\item \emph{generalisation:}
A deep learning based approach for tomographic image reconstruction needs to generalise well across variations in instrumentation and data acquisition protocols.
Domain agonistic architectures generalise poorly as shown in \cite{Antun:2020aa} for the case of AUTOMAP.
\end{itemize}

The above clearly outlines the drawbacks that come with using generic neural operator architectures for tomographic image reconstruction.
In contrast, the same task can be performed reliably with a learned iterative network that has only about $10^{5}$ (in 2D) and $10^{8}$ (in 3D) learnable parameters \cite{Rudzusika:2024aa}. 
Despite having far less parameters, this learned iterative network also yields state-of-the-art results, is numerically stable, and it generalises well across reasonable variations in data noise level, instrumentation, and data acquisition protocols even when it is trained against a relatively small data set (data from $\sim 200$ patients).

\paragraph{Structured learning}
As outlined above, operator learning can quickly become challenging due to difficulties in ensuring sufficient generalisation and handling scalability issues.
This applies to both solving ill-posed inverse problems as well as solving optimisation problems.

Inverse problems are often ill-posed meaning that solution methods that only seek to fit observed data will amplify observation errors. 
Additionally, despite clever discretisations, many inverse problems lead to large-scale numerical computations since elements in the domain and range of the forward operator need to be represented with very high-dimensional arrays.
Likewise, solving optimisation problems that arise in high dimensional signal and image processing quickly lead to large-scale numerical computations.
This is especially the case in imaging applications.
Adopting a naive `discretise first, then train' approach means using a trained neural network for solving very large-scale ill-posed inverse problems.
Doing this with reasonable accuracy and generalisation properties is challenging. 
In particular, using generic neural network architectures and then relying entirely on the scaling law is not a feasible approach as we briefly argue for in Section~\ref{sec:MotivationStructuredLearning}.

The solution to the above scaling challenge is to use \emph{structured learning}.
This refers to domain adaptation that goes beyond choosing appropriate training data.
The key component is to introduce inductive bias in the form of additional (handcrafted) `structure', e.g., by using neural operator with domain adapted architectures.

In many cases there is (partial) knowledge about how the target operator $\RecOp$ in \eqref{eq:MainOperator} is defined.
A typical example is that it is defined implicitly through some iterative scheme.
It is then natural to use a neural operator architecture for approximating such a target operator that accounts for this iterative scheme. 
For inverse problems, this also corresponds to incorporating explicit operator(s) associated with simulating and modelling data (simulator informed architectures).
As an example, when solving inverse problems it is natural to account for the forward operator (simulator) and its adjoint (for linear forward operators) or adjoint of its Fréchet derivative (for non-linear forward operators). 
Such neural operator architectures have been essential for learning based approaches to solve high-dimensional ill-posed inverse problems (Section~\ref{sec:InvProb}) or optimisation problems (Section~\ref{sec:OptimSolver}).
The resulting data-driven methods have for both these tasks had considerable success in matching, or outperforming, state-of-the-art while at the same time offering a significant computational speed-up. 
On a final note, as outlined in Section~\ref{sec:TheoryFoundations}, we still lack formal mathematical guarantees for the majority of structured data-driven approaches to operator learning, like sample complexity results, accuracy guarantees, or useful estimates of the generalisation gap.
We cannot address this shortcoming here, so our aim is instead to provide a survey of learned iterative networks. 
These are a class of domain adapted architectures that have been very successful in solving inverse problems and optimisation problems.

\subsubsection{Pre/post-processing architectures}\label{sec:PosPreArch}
The simplest example in how to adapt a neural operator is to make use of an initial approximation $\RecOp_0 \colon \DataSpace \to \RecSpace$ of the target operator $\RecOp$ in \eqref{eq:MainOperator}.
This can be used to define a learned reconstruction operator as \emph{pre/post-processing architectures}, which are neural operator architectures of the form 
\begin{equation}\label{eq:PrePostReco} 
\RecOp_{\NNparam} := \NNOp^{\text{post}}_{\NNparam''}
\circ \RecOp_0 \circ 
\,\NNOp^{\text{pre}}_{\NNparam'}
\quad\text{with $\NNparam=(\NNparam',\NNparam'')$}
\end{equation}
where $\NNOp^{\text{pre}}_{\NNparam'} \colon \DataSpace \to \DataSpace$ and $\NNOp^{\text{post}}_{\NNparam''} \colon \RecSpace \to \RecSpace$ represent learned pre- and post-processing operators.

The above is quite a common situation for solving inverse problems where the initial approximation $\RecOp_0$ implicitly accounts for how data is generated.
The pre- and post-processing neural operators represent some form of learned restoration and denoising networks, which can be trained separately. Post-processing networks would be trained to improve the output from $\RecOp_0$ \cite{Jin:2017aa,kang2017deep}. For pre-processing networks these would remove measurement noise or fill missing parts of data. Finally, a joint training of both is possible, which is sometimes also referred to as dual-domain approach \cite{lin2019dudonet}.

\subsubsection{Learned iterative networks}\label{sec:LearnedIterNetw}
Pre/post-processing architectures are popular for learning based approaches to solve inverse problems, and especially so for solving mildly ill-posed inverse problems arising in imaging. 
This is due to their simplicity and relatively good performance.  

Learned iterative networks, which are the main focus of this survey, represent a further refinement of pre/post-processing architectures.
These architectures are essentially built by stacking several neural operators of the type in \eqref{eq:PrePostReco}, i.e., neural operators with pre/post-processing architectures.
An example is the operator recurrent network that is outlined in \cite[eq.~(4.15)--(4-16)]{Nelsen:2025aa}.

Equivalently one can also obtain such an architecture by unrolling a suitable iterative scheme.
This is in fact the way these architectures were originally introduced (Section~\ref{sec:UnrollHistory}) and it is also the reason for their name: `learned iterative networks' or `unrolled networks'.
Such unrolling can be done in different ways, thus resulting in different neural operator architectures. 
Exploring these architectures is the central theme of this survey and Section~\ref{sec:LearnedIterNetwork} is the starting point for this.

\subsubsection{Other neural operator architectures}\label{sec:OtherNeuralOpArch}
We here mention encoder-decoder networks and integral kernel networks, which are classes of neural operator architectures that have enjoyed much attention. 
They have been primarily developed in the context of solving \acp{PDE}, but they can also be used for solving inverse problems.  

These architectures enjoy the most comprehensive theoretical analysis (Section~\ref{sec:TheoryFoundations}), like universal approximation \cite{Kovachki:2024aa}, sharp approximate rates \cite{Lanthaler:2023aa}, and sample complexity bounds \cite{Lanthaler:2023aa,Liu:2024ab,Kovachki:2024ab}.
A more detailed account of them along with a survey of their theoretical properties is provided in \cite{Goswami:2023aa,Boulle:2024aa,Shin:2024aa,Kovachki:2024aa,Liu:2025aa} that mainly focuses on their use in solving \acp{PDE}, whereas \cite{Nelsen:2025aa} is a survey that focuses on their use in solving inverse problems.

\paragraph{Encoder-decoder networks}
Following \cite[Sec.~3.2.1]{Nelsen:2025aa}, we say that a neural operator $\RecOp_{\NNparam} \colon \DataSpace \to \RecSpace$ has an encoder-decoder architecture if it is of the form
\begin{equation}\label{eq:EncDecNN}
\RecOp_{\NNparam} = \operator{D}_{\NNparam'''} \circ \operator{A}_{\NNparam''} \circ \operator{E}_{\NNparam'}
\quad\text{with $\NNparam=(\NNparam',\NNparam'',\NNparam''')$,}
\end{equation}
where $\operator{E}_{\NNparam'} \colon \DataSpace \to \Real^{d_{\text{enc}}}$ (encoder), $\operator{A}_{\NNparam''} \colon \Real^{d_{\text{enc}}} \to \Real^{d_{\text{dec}}}$ (approximator), and $\operator{D}_{\NNparam'''} \colon  \Real^{d_{\text{dec}}} \to \RecSpace$ (decoder) are all assumed to be continuous maps.

The \ac{DeepONet} architecture \cite{Lu:2021aa}, which builds on \cite{Chen:1995aa}, is a specific type of encoder-decoder architecture where $\RecOp_{\NNparam} \colon \DataSpace \to \RecSpace$ is of the form 
\begin{equation}\label{eq:DeepONet}
\RecOp_{\NNparam}(\data) = 
  \sum_{j=1}^{d_{\text{dec}}} (a^j_{\NNparam''} \circ \operator{E})(\data) \phi^j_{\NNparam'''}
  \quad\text{for $\data \in \DataSpace$ with $\NNparam=(\NNparam'',\NNparam''')$.}
\end{equation}
Note here that the encoder $\operator{E} \colon \DataSpace \to \Real^{d_{\text{enc}}}$ is not learned, it is handcrafted, whereas $a^j_{\NNparam''} \colon \Real^{d_{\text{enc}}} \to \Real$ (branch net) and $\phi^j_{\NNparam'''} \in \RecSpace$ (trunk net) are neural networks with the latter defining the (linear) decoder $\operator{D}_{\NNparam'''} \colon  \Real^{d_{\text{dec}}} \to \RecSpace$ through
\[ \operator{D}_{\NNparam'''}(z) = \sum_{j=1}^{d_{\text{dec}}} z_j \phi^j_{\NNparam'''}
  \quad\text{for $z \in \Real^{d_{\text{dec}}}$.}
\]
The surveys \cite{Shin:2024aa,Boulle:2024aa,Nelsen:2025aa} provide a more detailed overview of \acp{DeepONet} that also includes subsequent extensions of this architectures. 

PCA-Net \cite{Bhattacharya:2021aa} is a related encoder-decoder architecture of the form in \cite[eq.~(3.10)]{Nelsen:2025aa}.
Like \ac{DeepONet}, it also uses a linear decoder defined through a basis expansion.
However, unlike \ac{DeepONet} that uses learned basis functions, in PCA-Net they are handcrafted through principal component analysis. 

\paragraph{Integral kernel networks}
The feed-forward neural operator architecture mentioned earlier in Section~\ref{sec:NeuralOperatorArchi} is obtained by replacing weight matrices in the hidden layers of a basic feed-forward neural network with their functional-analytic counterparts.
Integral kernel networks refer to the case when integral operators are used as for this purpose. 

One can introduce injective bias into a integral kernel network by considering specialised integral operators whose structures are tailored to the class of target operators being approximated.
This represents a form of domain adaptation that could improve upon scalability and generalisation, and especially so for solving linear \acp{PDE}.
Here, the target operator can often be represented as an integral operator with a kernel (Green's function) whose structure is induced by the properties of the linear \ac{PDE}. 
Examples of structured kernels are translation-invariant, circulant, or off-diagonal low-rank kernels, see \cite[Table~1]{Boulle:2024aa} for a nice summary.
Discretising the action of these integral operators corresponds to a matrix-vector product. 
Hence, structured matrix recovery becomes the discrete analogue of operator learning and this correspondence is nicely summarised in \cite[Table~2]{Boulle:2024aa}.

The most popular integral kernel network is the \ac{FNO} \cite{Li:2021aa}.
Here, the kernels have a structure that is closely related to circulant matrix recovery.
Hence, the \ac{FNO} architecture is especially suitable for approximating target operators that are given as integral transforms with a periodic and translation invariant kernel.
Since its introduction, theoretical properties like universal approximation and error bounds have also been derived for \acp{FNO} \cite{Kovachki:2021aa}, see also Section~\ref{sec:TheoryFoundations}.
The basic \ac{FNO} architecture has also undergone many changes.
See the survey in \cite{Boulle:2024aa} for more on \acp{FNO}, including the multitude of variants that have been developed since their introduction.
A particular generalisation that is of interest in solving inverse problems is Fourier neural mappings \cite[Sec.~4.2.1]{Nelsen:2025aa}. 
These generalise \acp{FNO}, which only map functions to functions, by accommodating a vector-to-function or function-to-vector maps. 
The basic structure of a Fourier neural mapping is given in \cite[eq. (4.13)]{Nelsen:2025aa}.
Finally, we also note that the \ac{FNO} can in its functional analytic formulation be viewed conceptually as a \ac{DeepONet} with a specific architecture for the branch and trunk neural networks represented by a trigonometric basis \cite{Kovachki:2021aa}.

Next, banded matrix recovery serves as blueprint for the \acp{GNO} architecture \cite{Anandkumar:2020aa}, see also \cite[Sec.~3.4]{Boulle:2024aa}.
\Acp{GNO} uses graphs to represent pairwise local relationships and it disregards distant causalities. This allows \acp{GNO} to efficiently represent target operators
that are local.
This bears similarities to the fast multipole method, which has also served as a blue print in the design of neural network architectures \cite{Sushnikova:2022aa,Kang:2025aa}.

Another kernel structure that often appears when discretising solution operators for elliptic and parabolic \acp{PDE} is hierarchical off-diagonal low rank matrix recovery, which is a multi-scale version of banded matrix recovery.
Using this structure as blue print leads to the \ac{MGNO} architecture \cite{Li:2020ab}, see also \cite[Sec.~3.5]{Boulle:2024aa} which represents operators at multiple scales. 
\Acp{MGNO} are based on a hierarchical graph, each representing pairwise interactions at some fixed scale/level. 
This ability to incorporate local (near-field) and global (far-field) interactions allows \acp{MGNO} to effectively learn complex multi-scale patterns. 
Other similar neural network architectures that are designed around a hierarchical operator decomposition are \cite{Li:2018aa,Fan:2019ab,Fan:2019aa,Fan:2019ac,Feliu-Faba:2020aa} that have been used for approximating pseudodifferential or Fourier integral operators.
There are also transformer based multi-scale architectures of the above type as in \cite{Liu:2024aa,Zhdanov:2025aa}.
This aim to represent multi-scale structure is also an example of a active area of research towards architectures that encode/represent a handcrafted hierarchical arrangement of higher order relations (topological deep learning) \cite{Hajij:2023aa,Hajij:2024aa,Hajij:2024ab,Papillon:2025aa}.

\paragraph{Concluding remarks}
Much of the above development mentioned in this section on has been driven by the need to solve \acp{PDE}, and in particular multi-scale \acp{PDE} where the solution operator can be expressed through a hierarchical operator decomposition \cite{Migus:2022aa}.
These architectures can for this reason also be useful for solving inverse problems.
More precisely, in many inverse problem the formal inverse to the forward operators is a pseudodifferential or a Fourier integral operator.
Such operators also admit a similar hierarchical decomposition, so the aforementioned architectures designed for solving \acp{PDE} can therefore also be useful for solving inverse problems as shown in \cite{Fu:2022aa} for the case of tomographic image reconstruction (inverting the ray transform, which is a Fourier operator).

An overwhelming majority of the architectures discussed in this section and defined in the finite dimensional setting, i.e., they are neural network architectures rather than neural operator architectures.
They are however made up of basic domain agnostic components, like convolutional or attention layers, and replacing these with their neural operator versions (Section~\ref{sec:Component}) should result in a neural operator version.

Finally, integral kernel networks lets the structure of the target operator guide the design of architecture.
On a high level, this design principle bears similarities to learned iterative networks introduced in Section \ref{sec:LearnedIterNetw}, where one uses an underlying iterative scheme as blue print in designing the architecture. 
Both approaches leverage on numerical analysis in the design of neural operator architecture.

\section{Learned iterative networks}\label{sec:LearnedIterNetwork}
A learned iterative network is a neural operator with an architecture that is specifically designed to approximate a target operator $\RecOp \colon \DataSpace \to \RecSpace$ that is defined implicitly through an iterative scheme, as  already shortly mentioned in Section~\ref{sec:LearnedIterNetw}.
That is, we consider a target operator of the form 
\begin{equation}\label{eq:OpIterScheme}
\RecOp(\data) := \lim_{k\to \infty} \signal^k
\quad\text{where}\quad
\signal^k := \NNOp^{\!k}(\data,\signal^0,\ldots,\signal^{k-1})
\quad\text{for any $\data \in \DataSpace$,}
\end{equation}
where the handcrafted mappings $\NNOp^{\!k} \colon \DataSpace \times \RecSpace^k \to \RecSpace$, referred to as updating operators, and the initialisation $\signal^0 \in \RecSpace$ are assumed to be known.
The initialisation may be chosen in various ways. A simple option is  $\signal^0=0$, while a common alternative when solving a inverse problems is to set $\signal^0 := \RecOp_0(\data)$ where $\RecOp_0 \colon \DataSpace \to \RecSpace$ is some handcrafted reconstruction operator. 
For \ac{CT} this would be the backprojection (adjoint of the ray transform) or filtered backprojection.

A learned iterative network obtained by unrolling an iterative scheme of the form in \eqref{eq:OpIterScheme} means replacing the updating operators, or parts of it, with neural operators.
These are then learned from training data using a suitable learning task (Section~\ref{sec:LearningProbs}).

This can be done in different ways, which in turn leads to different neural operator architectures. 
The common ones are:
\begin{itemize}
\item Learned gradient networks (Section~\ref{sec:LearnedGradient}) 
\item \Acl{LPD} networks (Section~\ref{sec:LPD})
\item Unrolling higher order iterative schemes for non-linear problems (Section~\ref{sec:HighOrderNonLin})
\end{itemize}
In the following section~\ref{sec:UnrollAbstract}, a description of the idea of unrolling is presented in an abstract setting followed by a historical account in Section~\ref{sec:UnrollHistory}. The reader may also consult the surveys \cite{Adler:2023aa,Chen:2023aa,Gui:2023aa,Monga:2021aa} for additional information.

\subsection{Unrolling an iterative scheme}\label{sec:UnrollAbstract}
To describe unrolling in an abstract setting, consider a target operator that is defined by an iterative scheme as in \eqref{eq:OpIterScheme}. 
The first step in unrolling this scheme is to truncate iterates by replacing the limit $k \to \infty$ with $k=\UnrollIter$.
The next step is to replace the handcrafted updating operators $\NNOp^{\!k} \colon \DataSpace \times \RecSpace^k \to \RecSpace$ in the truncated scheme with neural operators (\emph{neural updating operators}).
This results in a learned reconstruction operator $\RecOp_{\NNparam} \colon \DataSpace \to \RecSpace$ that is defined by the following finite recursive scheme:
\begin{equation}\label{eq:UnrolledRecursive}
\RecOp_{\NNparam}(\data) := \signal^{\UnrollIter}
\quad\text{where}\quad
\begin{cases}
 \signal^0 := \RecOp_0(\data) &
 \\[0.5em]
 \signal^k := \NNOp^{\!k}_{\NNparam_k}(\data,\signal^0,\ldots,\signal^{k-1}) 
 & \text{for $k=1,\ldots,\UnrollIter$.}  
\end{cases}
\end{equation}
Here $\NNparam :=(\NNparam_1,\ldots,\NNparam_\UnrollIter)$ are the network parameters, $\RecOp_0 \colon \DataSpace \to \RecSpace$ is the initialisation, and each neural updating operator
\[ 
\NNOp^{\!k}_{\NNparam_k} \colon \DataSpace \times \RecSpace^k \to \RecSpace
\quad\text{for $k=1,\ldots,\UnrollIter$}
\]
comes with its own weights (no a priori assumption of weight sharing).
A natural special case is to use neural updating operators with fixed architectures.
In this case we will omit the upper index and write $\NNOp^{\!k}_{\NNparam_k}=\NNOp_{\NNparam_k}$. 
Additionally, one can consider weight sharing, i.e., each network also has the same weights $\NNOp_{\NNparam_k}=\NNOp_{\NNparam}$, but as mentioned this is not assumed beforehand.
Finally, it is also common to further mimic the structure of an iterative scheme by considering residual architectures for computing $\signal^k$ in \eqref{eq:UnrolledRecursive}, i.e., 
\begin{equation}\label{eq:UnrolledResidual}
 \signal^{k} := \signal^{k-1} + \NNOp^{\!k}_{\NNparam_k}(\data,\signal^0,\ldots,\signal^{k-1}) 
 \quad\text{for $k=1,\ldots,\UnrollIter$.}  
\end{equation}

\begin{remark}\label{rem:RoleOfUnrolling}
The definition of unrolling as in \eqref{eq:UnrolledRecursive} may suggest that the target operator one seeks to approximate must be defined by the iterative scheme being unrolled, as in \eqref{eq:OpIterScheme}. 
Formally, however, this is not required; indeed, the underlying iterative scheme need not even be convergent. 
Rather, unrolling should be understood primarily as a means of constructing a neural operator architecture, with the truncated iterative scheme serving as a blueprint. 
What target operator one seeks to approximate is instead determined by the learning task as shown in Section~\ref{sec:LearningProbs}.

That said, it is also reasonable to argue that defining a learned iterative network by unrolling a scheme \eqref{eq:UnrolledRecursive} that also characterises the target operator, as in \eqref{eq:OpIterScheme}, yields a domain-adapted neural operator architecture that is particularly well suited to settings where scalability and generalisation pose significant challenges (Section~\ref{sec:MotivationStructuredLearning}).
\end{remark}

\subsection{Historical account of learned iterative networks}\label{sec:UnrollHistory}
The idea of using an iterative scheme as a blue print for a network architecture can be traced back to \cite{Gregor:2010aa}, which considers the case of learning-to-optimise, i.e., training a neural network so that it approximates an optimisation solver that maps an objective to its minimiser (Section~\ref{sec:OptimSolver}).
More precisely, the focus in \cite{Gregor:2010aa} is on objective functions that arise in sparse coding.  
It is therefore natural to consider a proximal gradient method as the iterative scheme, like the \ac{ISTA}.
Using \ac{ISTA} as a blue print for a \ac{CNN}-type of neural network is based on the observation that each \ac{ISTA} iteration consists of a linear operation followed by a non-linear soft-thresholding (proximal to the $\ell_1$-functional).
In fact, the latter can be represented by composing a 2-layer neural network with \ac{ReLU} activation,
so one can assemble a neural network by stacking a finite number of layers where each layer corresponds to an \ac{ISTA} iterate in which the soft-thresholding has been replaced appropriately by \ac{ReLU} activation. 
Additionally, deploying weight sharing across the layers results in a \ac{RNN} type of architecture henceforth called \ac{LISTA}.

Learned iterative networks that were obtained by unrolling (unfolding) a truncated iterative scheme can also be trained in supervised manner to solve an ill-posed inverse problem (Section~\ref{sec:InvProb}). 
One of the earliest examples of this was for blind deconvolution \cite{Schuler:2016aa}.
Another inverse problem is 2D tomographic image reconstruction. 
The possibility to use learned iterative networks, like \ac{LISTA}, for this inverse problem is mentioned in \cite{Jin:2017aa}. 
However, the paper does not implement such an architecture. 
Instead, it moves on to observe that the normal operator $\FwdOp^*\FwdOp$ in this setting is a convolution type operator, i.e., it is smoothing of order~1, which follows from the fact that $\FwdOp$ is the ray transform.
Hence, a single \ac{LISTA} layer has the form of a filtering followed by pointwise non-linearity. 
This in turn serves as a motivation for a post-processing based approach in which a U-Net is trained to improve upon \ac{FBP} reconstructions (FBPConvNet), i.e., the \ac{LISTA} architecture is merely used as a way to motivate the use of U-Net as a post-processing step. 
First examples of using a learned iterative network for solving more general inverse problems, appeared for deconvolution \cite{Kobler:2017aa} and tomographic image reconstruction in \cite{Adler:2017aa}. 
The architectures here are given by unrolling a gradient descent scheme for minimising a differentiable objective representing a penalised (regularised) least-squares fidelity term. 
The forward operator is here replaceable in a plug-and-play manner. 
Such networks were then trained end-to-end for 2D tomographic image reconstruction.

It is also worth mentioning early work on recurrent inference machines that dates back to early work from 2015, which later appeared as a pre-print two years later \cite{Putzky:2017aa}. 
This survey essentially considers a learned iterative network for performing various linear image processing tasks. 
The examples given are image denoising and super-resolution.
The reader may also consult \cite[Table~1]{Monga:2021aa} that lists publications where learned iterative networks are used for performing various signal and image processing tasks.

In conclusion, it is important to note that a learned iterative network is merely a choice of neural network architecture. 
Which task one seeks to perform with such a neural network, i.e., which target operator one seeks to approximate, depends to large extent on how it is trained.
Hence, the choice of learned iterative networks is more about \emph{how to compute} rather than \emph{what to compute}.
Many learned reconstruction methods with a learned iterative network are trained end-to-end against supervised data. 
The unrolling is often based on an iterative scheme that in the limit yields a minimiser to a variational formulation that represents a solution to the inverse problem.
An unfortunate consequence of this is that many papers erroneously claim that more unrolled iterates yield a better approximation to the minimiser, or that it relates to a minimiser of said variational problem.
Without further assumptions, the correct interpretation is that unrolling more iterates only increases the model capacity of the network.

\section{Learned Gradient Networks}\label{sec:LearnedGradient}
A learned gradient network is a learned reconstruction operator with an architecture derived by unrolling some gradient based optimisation scheme.
The starting point in constructing a learned gradient network for approximating $\RecOp \colon \DataSpace \to \RecSpace$ is to specify a \emph{variational model} that implicitly defines $\RecOp$. 
Hence, we assume there is a differentiable cost functional $\CostFunc_{\data} \colon \RecSpace \to \Real$ such that 
\begin{equation}\label{eq:VarModLearnedGrad}
 \RecOp(\data) := \argmin_{\signal \in \RecSpace} \CostFunc_{\data}(\signal)
 \quad\text{for $\data \in \DataSpace$.}
\end{equation}
The target operator in learning-to-optimise (Section~\ref{sec:OptimSolver}) is directly defined by such a variational model whereas in learned reconstruction (Section~\ref{sec:InvProb}) the target operator is defined by the learning objective and does not automatically approximate \eqref{eq:VarModLearnedGrad}.
Next, consider a gradient descent scheme for \eqref{eq:VarModLearnedGrad}:
\begin{equation}\label{eq:Gradient}
\begin{cases}
 \signal^0 := \RecOp_0(\data) &
 \\[0.5em]
 \signal^k := \signal^{k-1} - \omega_k \grad\CostFunc_{\data}(\signal^{k-1}) 
 & \text{for some $\omega_k>0$ and $k=1,2,\ldots$.}  
\end{cases}
\end{equation}
If the cost functional $\signal \mapsto \CostFunc_{\data}(\signal)$ is convex, then the sequence $(\signal^k )_k \subset \RecSpace$ in \eqref{eq:Gradient} that depends on $\data$, can be used to evaluate $\RecOp$ as $\RecOp(\data)=\lim_{k\to \infty} \signal^k$.

\emph{Learned gradient networks} are obtained by unrolling the gradient descent scheme in \eqref{eq:Gradient} as in \eqref{eq:UnrolledRecursive}.
This yields a neural operator $\RecOp_{\NNparam} \colon \DataSpace \to \RecSpace$ with $\NNparam := (\NNparam_1,\ldots, \NNparam_\UnrollIter)$ that is defined as 
\begin{equation}\label{eq:LearnedGradient}
\RecOp_{\NNparam}(\data) := \signal^{\UnrollIter}
\quad\text{where}\quad
\begin{cases}
 \signal^0 := \RecOp_0(\data) &
 \\[0.5em]
 \signal^k := \NNOp_{\NNparam_k}\bigl(
  \signal^{k-1},
  \grad\CostFunc_{\data}(\signal^{k-1})
 \bigr) 
 & \text{for $k=1,\ldots,\UnrollIter$.}
\end{cases}
\end{equation}
In addition, it is common to use residual architectures for the neural update operators in \eqref{eq:LearnedGradient}, which yields a residual update structure as in \eqref{eq:UnrolledResidual}:
\begin{equation}
  \label{eq:residual_gradientNet}
\signal^k := \signal^{k-1} + \NNOp_{\NNparam_k}\bigl(\signal^{k-1},
  \grad\CostFunc_{\data}(\signal^{k-1})
 \bigr).
\end{equation}
Section~\ref{sec:ClassesOfLearnedGradient} discusses learned gradient networks obtained by making specific choices that introduce more structure into the updates \eqref{eq:LearnedGradient}.
Below are some remarks that specifically relate to adaptation of learned gradient networks for solving inverse problems.

\paragraph{Adaptation to inverse problems}
One natural adaption of the gradient scheme in \eqref{eq:Gradient} to the domain of inverse problems is to ensure minimisers to \eqref{eq:VarModLearnedGrad} correspond to (regularised) solutions of the operator equation \eqref{eq:InvProb}. 
It is here natural to also restrict attention to linear forward operators $\FwdOp\colon\RecSpace\to\DataSpace$, the case when $\FwdOp$ is non-linear is discussed in Section~\ref{sec:HighOrderNonLin}. 

The original use case for solving ill-posed inverse problems was based on reconstruction operators that are defined as minimisers to objectives $\CostFunc_{\data} \colon \RecSpace \to \Real$ of the form 
\begin{equation}\label{eq:CostFunc}
  \CostFunc_{\data}(\signal) := \operator{Q}_{\data}(\signal) + \RegOp(\signal)
  \quad\text{for $\signal \in \RecSpace$.}
\end{equation}
Here,  $\operator{Q}_{\data} \colon \RecSpace \to \Real$ depends on $\data$  and $\RegOp \colon \RecSpace \to \Real$ is the regulariser. 
The former is commonly given by a loss in $\DataSpace$ using the  knowledge of $\FwdOp$ as
\begin{equation}\label{eq:DataFidelity}
  \operator{Q}_{\data}(\signal) := \DataLoss\bigl( \FwdOp\signal, \data \bigr)
  \quad\text{for $\signal \in \RecSpace$.}
 \end{equation}
where $\DataLoss\colon \DataSpace \times \DataSpace \to \Real$ is a data-fidelity term, often given as suitable norm. The explicit regulariser $\RegOp\colon\RecSpace\to\Real$ is usually omitted in learned iterative networks and replaced by the learned components. We will get back to this in the remainder of this section.

Thus, to illustrate the concept of learned iterative networks, let us consider a common special case, which is when the objective $\CostFunc_{\data} \colon \RecSpace \to \Real$ in \eqref{eq:CostFunc} only consists of the data-fidelity: $\CostFunc_{\data}(\signal) := \operator{Q}_{\data}(\signal)$.
As discussed earlier, if $\signal \mapsto \operator{Q}_{\data}(\signal)$ is differentiable, then one can find a local minima by a gradient descent scheme:
\begin{equation}\label{eq:GradDescent}
\signal^k := \signal^{k-1} - \NNparam_k \grad\operator{Q}_{\data}(\signal^{k-1})
\quad\text{for $k=0,1,2\ldots$,}
\end{equation}
where $\grad$ is the $\RecSpace$-gradient.
A common setting is when the data-fidelity $\DataLoss$ is the squared $\DataSpace$-norm, i.e., $\DataLoss(\dataother,\data):= \frac{1}{2}\Vert \dataother - \data \Vert_\DataSpace^2$. 
Then 
\begin{equation}\label{eq:DataFidelityL2} 
\operator{Q}_{\data}(\signal) = \frac{1}{2}\bigl\Vert \FwdOp\signal - \data \bigr\Vert_\DataSpace^2,
\end{equation}
so the gradient $\grad\operator{Q}_{\data} \colon \RecSpace \to \RecSpace$ is given as
\begin{equation}\label{eq:DataFidelityL2Grad} 
\grad\operator{Q}_{\data}(\signal) = \FwdOp^{\ast}\bigl(\FwdOp\signal - \data\bigr)
\quad\text{for $\signal \in \RecSpace$.}
\end{equation} 

The iterates in \eqref{eq:GradDescent}, which coincide with Landweber iterations, will then represent a special case of \eqref{eq:LearnedGradient} with an updating rule of the form 
\[
\signal^{k}=\NNOp_{\NNparam_k}( \signal^{k-1}, \grad\operator{Q}_{\data}(\signal^{k-1})) := \signal^{k-1} - \NNparam_k \grad\operator{Q}_{\data}(\signal^{k-1}).
\]
The above can be seen as a gradient descent (Landweber) step with a learnable step-size $\NNparam_k$. 
Now, this can readily be generalised to the general form \eqref{eq:LearnedGradient} by replacing $\NNOp_{\NNparam_k}$ within the above update rule with neural updating operators. 
The parameter $\NNparam_k$ then represents some high-dimensional array of network parameters instead of a scalar step-size.

Using suitably chosen neural updating operators $\NNOp_{\NNparam_k} \colon \RecSpace \times \RecSpace \to \RecSpace$ provides the standard form of a \emph{learned gradient network} as in \cite{Adler:2017aa}.
This yields the learned reconstruction operator $\RecOp_{\NNparam} \colon \DataSpace \to \RecSpace$ with $\NNparam := (\NNparam_1,\ldots, \NNparam_\UnrollIter)$ that is defined as 
\begin{equation}\label{eq:LearnedGradientMethod}
\RecOp_{\NNparam}(\data) := \signal^{\UnrollIter}
\quad\text{where}\quad
\signal^{k} = \NNOp_{\NNparam_k}\bigl( \signal^{k-1}, \grad\operator{Q}_{\data}(\signal^{k-1}) \bigr)
\text{ for $k=1,\ldots, N$.}
\end{equation}
Here, $\signal^0 \in \RecSpace$ is an initialisation, either given by $\signal^0 = \RecOp_0(\data)$ where $\RecOp_0$ is some handcrafted reconstruction method, $\signal^0 = \FwdOp^{\ast}\data$, or $\signal^0=0$.  

The learned reconstruction operator in \eqref{eq:LearnedGradientMethod} is a neural operator with an architecture of the form \eqref{eq:UnrolledRecursive} without memory.
The neural updating operator $\NNOp_{\NNparam_k} \colon \RecSpace \times \RecSpace \to \RecSpace$ and its architecture does not involve data or the forward operator (or its adjoint). 
These components only enter into the evaluation of the gradient direction $\grad\operator{Q}_{\data}(\signal)$. 
In particular, one obtains different  learned gradient networks depending on the choice of data fidelity $\operator{Q}_{\data} \colon \RecSpace \to \Real$ in \eqref{eq:LearnedGradientMethod} 
and choice of neural updating operators in \eqref{eq:LearnedGradientMethod}.
We will next consider some of these, like the network one gets by having a least-squares data-fidelity  (Section~\ref{sec:LearnedLSQ}). 

\subsection{Classes of learned gradient networks}\label{sec:ClassesOfLearnedGradient}
The way the neural updating operators enter into  \eqref{eq:LearnedGradientMethod} determines how learning is intertwined with handcrafted components. 
In fact, many learned iterative networks are variants of \eqref{eq:LearnedGradientMethod} where one applies specific handcrafted updates on $\signalother:=\grad\operator{Q}_{\data}(\signal)$. Simplified these can be summarised as:
\begin{alignat}{2}
\label{eqn:VarNet_proto}
\NNOp_{\NNparam}(\signal,\signalother)
 &:= \signal -\signalother + \NNOpOther_{\NNparam}(\signal)
 &\quad \text{(variational networks)} &
\\ 
\label{eqn:proxNet_proto}
\NNOp_{\NNparam}(\signal,\signalother) 
 &:= \NNOpOther_{\NNparam}(\signal - \signalother) 
 & \quad \text{(learned proximal gradient)}. &
\end{alignat}
Here, $\NNOpOther_{\NNparam} \colon \RecSpace \to \RecSpace$ is a neural operator representing the learned component of the updates, and remaining parts, like $\signalother:=\grad\operator{Q}_{\data}(\signal)$, are handcrafted. 

When applying the above methods to inverse problems in imaging, note that in particular all neural operators are mappings from $\RecSpace$ to $\RecSpace$, e.g., images to images.
Hence, learned iterative networks stack a sequence of such image-to-image transforms using couplings that involve handcrafted components through the update direction $\grad\operator{Q}_{\data}(\signal)$. Additionally, the neural update operators $\NNOp_{\NNparam_k}$ are usually not very expressive as we will discuss in the following subsections. This can be intuitively understood by realising that the update directions $\grad\operator{Q}_{\data}(\signal)$ encode non-local relations through the involvement of the forward operator (and its adjoint), thus the learned network component primarily learns local interactions. Consequently, very deep and expressive networks that can learn long range relations may not be needed. 
Nevertheless, this statement needs to be taken with care and is highly dependent on the application as well as the ability of the update direction $\grad\operator{Q}_{\data}(\signal) \in \RecSpace$ to encode the needed long range relations.

\subsubsection{Learned least squares networks}\label{sec:LearnedLSQ}
Let us first discuss the standard learned least squares network \eqref{eq:LearnedGradientMethod} further, with respect to \cite{Adler:2017aa}. Here the gradient updates are given by the gradient of the squared data-fidelity \eqref{eq:DataFidelityL2Grad}. The learned reconstruction operator is then given as $\RecOp_{\NNparam}(\data) := \signal^{\UnrollIter}$ where $\NNparam = (\NNparam_1,\ldots, \NNparam_\UnrollIter)$ and
\begin{equation}\label{eq:LearnedGradient_withL2}
\begin{cases}
 \signal^0 := \RecOp_0(\data) &
 \\[0.5em]
 \signal^k := \NNOp_{\NNparam_k}\Bigl(\signal^{k-1},
  \FwdOp^{\ast}\bigl(\FwdOp\signal^{k-1}-\data
 \bigr) \Bigr)
 & \text{for $k=1,\ldots,\UnrollIter$.}
\end{cases}
\end{equation}

In the above variant the neural update operator $\NNOp_{\NNparam_k} \colon \RecSpace \times \RecSpace \to \RecSpace$ receives only two inputs, the current iterate $\signal^{k-1}$ as well as the gradient information. 
The original implementation in \cite{Adler:2017aa} additionally used a memory of $m$ previous iterates and the gradient of a handcrafted regulariser as the input to the network, that is the update becomes
\begin{equation}\label{eqn:classicLGD}
\begin{cases}
 \signal^{k} :=0 \quad\text{if $k<0$,} &
 \\[0.5em]
 \signal^k := \NNOp_{\NNparam_k}\Bigl(\signal^{k-1-m},\dots,\signal^{k-1},
  \FwdOp^{\ast}\bigl(\FwdOp \signal^{k-1}-\data
 \bigr),\grad \RegOp(\signal^{k-1}) \Bigr)
 \quad\text{if $k \geq 0$,} &
\end{cases} 
\end{equation}
where $\RegOp \colon \RecSpace \to \Real$ was the Dirichlet energy as in classical Tikhonov regularisation. 
However, it should, be noted that further experiments in the supervised setting showed that using an explicit regulariser in the updates is not necessary for improving the reconstruction quality.
In fact, an additional explicit regulariser will modify the neural network architecture, but it is unclear in which way this may impact the regularising properties of the learned gradient network for solving ill-posed inverse problems.
Instead, the common approach is to assume that the prior  information and regularising effect is implicitly contained in the training data. 
On the other hand, the memory in \eqref{eqn:classicLGD} can improve reconstructions slightly and helps to stabilise the training.

\paragraph{Implementation}
The formulation presented in \eqref{eq:LearnedGradient_withL2} leaves the freedom to choose suitable architectures for the neural updating operators $\NNOp_\NNparam$ that are $\RecSpace$-valued mappings defined on $\RecSpace\times\RecSpace$. 

This choice largely depends also on the discretisation of $\RecSpace$, that means, if we treat $\RecSpace$ as a function space, then a neural operator such as \acp{FNO} can be a suitable choice. 
Whereas, for classic discretisations in $\Real^{n\times n}$ it is common to apply a \ac{CNN} or related architectures.
For example, the original implementation presented in \cite{Adler:2017aa} unrolls $N=10$ iterations and uses \acp{CNN} as neural updating networks.
The updates follow a residual formulation as in \eqref{eq:residual_gradientNet}, where each neural network $\NNOp_{\NNparam_k}$ is a simple 3-layer \ac{CNN} with 32 channels. 
It should be noted, that this architectural choice is similar to \ac{ResNet} architectures \cite{He2016ResNet}.

Finally, the use of the forward operator $\FwdOp$ is essential for domain adaptation in the context of solving inverse problems.
This allows for the use of smaller networks and improves generalisation, but it also introduces a large computational overhead to the training. Additionally, the software for the evaluation of $\FwdOp$ also needs to be seamlessly integrated with whatever framework one uses for deep learning.
The notion of seamless means in particular that differentiable programming works as expected and this can in itself become challenging. 
We will discuss these aspects further in Section~\ref{sec:implementation}.

\subsubsection{Learned proximal networks}\label{sec:LearnedProximal}
The learned least squares network in the previous section is based on the variational model in \eqref{eq:VarModLearnedGrad} with a cost functional of form \eqref{eq:CostFunc}, where the neural updating operator does not make use of further structure apart from the use of gradient information of the cost functional $\CostFunc_{\data}$. Next we will introduce such further structure with \emph{learned proximal networks}. 

Here, the motivation is to improve upon interpretability and generalisation properties.
The guiding principle in learned proximal networks is to make use of the structure that is offered by  forward-backward splitting, also called proximal gradient descent schemes. Such networks are specifically designed for approximating operators $\RecOp \colon \DataSpace \to \RecSpace$ that are defined by a variational model \eqref{eq:VarModLearnedGrad} in which the cost functional $\CostFunc_{\data} \colon \RecSpace \to \Real$ for data $\data \in \DataSpace$ has the form in \eqref{eq:CostFunc}:
\begin{equation}\label{eq:Costfunc_as_DataDiscAndReg}
  \CostFunc_{\data}(\signal)=\operator{Q}_{\data}(\signal) + \RegOp(\signal) 
  \quad\text{with}\quad
  \operator{Q}_{\data}(\signal):=\DataLoss\bigl(\FwdOp\signal,\data\bigr).
\end{equation}
Here, $\DataLoss \colon\DataSpace\times\DataSpace\to\Real$ (data-fidelity), $\FwdOp \colon \RecSpace \to \DataSpace$ (forward operator), and  $\RegOp\colon\RecSpace\to\Real$ (regulariser) are all handcrafted.

If $\operator{Q}_{\data}(\signal) \colon \RecSpace \to \Real$ is convex and differentiable and $\RegOp$ is convex, but not necessarily differentiable, then one can evaluate $\RecOp \colon \DataSpace \to \RecSpace$ using the mentioned proximal gradient descent scheme. Then we get $\RecOp(\data)=\lim_{k\to \infty} \signal^k$ where the sequence $(\signal^k )_k \subset \RecSpace$ is generated as 
\begin{equation}\label{eq:ProxGadient}
\begin{cases}
 \signal^0 := \RecOp_0(\data), &
 \\[0.5em]
 \signal^{k} := \prox_{\omega_k\RegOp}\bigl(\signal^{k-1} - \omega_k \grad \operator{Q}_\data(\signal^{k-1})\bigr), 
 & \text{for $k=1,2,\ldots$}  
\end{cases}
\end{equation}
for some suitable choice $\omega_k>0$ and initialisation $\RecOp_0\colon\DataSpace\to\RecSpace$. 
In the above, the functional $\prox_{\RegOp} \colon \RecSpace \to \Real$ is the proximal operator defined as
\begin{equation}\label{eq:proxOperator}
\prox_{\RegOp}(\signal):=\argmin_{\signalother\in\RecSpace}\left\{ \frac{1}{2}\Vert\signalother-\signal\Vert_2^2 + \omega_k\RegOp(\signalother)\right\}
\quad\text{for $\signal \in \RecSpace$.}
\end{equation}
This can be interpreted as denoising of the input $\signal \in \RecSpace$ with respect to $\RegOp$ and thereby projecting it onto the admissible set defined by the regulariser $\RegOp$. 

The scheme in \eqref{eq:ProxGadient} separates the iterative procedure into two steps, one explicit gradient step with respect to the data-fidelity followed by the proximal operator. 
Since the proximal operator enforces the regulariser, which contains prior knowledge on the reconstructions, it is now straight-forward to replace it by a learned component. 
This leads to the more structured \emph{learned proximal network} where the neural operator $\RecOp_{\NNparam} \colon \DataSpace \to \RecSpace$ with $\NNparam = (\NNparam_1,\ldots, \NNparam_{\UnrollIter})$ is given as $\RecOp_{\NNparam}(\data) := \signal^{\UnrollIter}$ where
\begin{equation}\label{eq:LearnedProximal}
\begin{cases}
 \signal^0 := \RecOp_0(\data) &
 \\[0.5em]
 \signal^k := \NNOp_{\NNparam_k}\bigl(
 \signal^{k-1} - \omega_k \grad\operator{Q}_{\data}(\signal^{k-1})
 \bigr)
 & \text{for $k=1,\ldots,\UnrollIter$}.
\end{cases}
\end{equation}
Further structure from \eqref{eq:ProxGadient} can be included into the above architecture by requiring to use same weights for all unrolled iterates \eqref{eq:LearnedGradient} (weight-sharing). Then, the updating operator $\NNOp_\NNparam$ has the same parameters for each iterate, so the architecture of the resulting learned proximal network with weight-sharing is given by
\begin{equation}
\label{eq:proximalNet_weightShare}
\signal^k :=
 \NNOp_{\NNparam}\bigl(
 \signal^{k-1} - \omega_k \grad \operator{Q}_{\data}(\signal^{k-1})
 \bigr).
\end{equation}
Weight-sharing reduces expressive power of the architecture, but it may improve upon interpretability. 
In fact, learned proximal networks with weight-sharing can be interpreted as including one more layer of structure as it allows to make connections to a fixed proximal operator for the regulariser $\RegOp$. Nevertheless, if no constraints on the parameter space $\NNparamSet$ are made, then there is no guarantee that the learned updating operator $\NNOp_\NNparam$ corresponds to a proximal operator for some regulariser. 

Finally, weight-sharing allows to perform more iterations than the fixed amount $N$ being unrolled. 
It should be noted though, that if the reconstruction operator is trained for $N$ iterates, iterating longer usually does not improved results without further constraints on the network.
More specifically, we can exploit the structure in \eqref{eq:proximalNet_weightShare} by introducing constraints on the updating operator $\NNOp_\NNparam$ and the parameter space $\NNparamSet$. This will be the basis for some theoretical results that we will discuss in the following. 
In particular, this allows us to introduce two popular approaches from the literature, \ac{PnP} schemes and \ac{DEQ} networks. 
We note, that these two approaches can be formulated with the same reconstruction operator $\RecOp_\NNparam$ in \eqref{eq:proximalNet_weightShare}, where both offer the possibility to establish a convergence analysis. In our context, the primary difference will be in distinctive training procedure to achieve said convergence, as it require a notation of limit.

Let us note to this end, that if the reconstruction operator in \eqref{eq:LearnedProximal} is trained end-to-end without restrictions, the same considerations regarding implementation as for learned least-squares networks apply as discussed in Section \ref{sec:LearnedLSQ}.

\paragraph{Relation to \acf{PnP} schemes}
\Ac{PnP} schemes are considered a separate research direction to unrolling schemes, but one can make the connection from the viewpoint of learned iterative schemes and a learned reconstruction operator. 
As will be shown, the difference between \ac{PnP} and unrolling is in the choice of learning problem and not in the formulation of the learned reconstruction operator. 

Unrolling and \ac{PnP} schemes have two main primary distinctions. First, \ac{PnP} methods separate the training of the network component $\NNOp_\NNparam:\RecSpace\to\RecSpace$ from the iterative algorithm, i.e., a different learning problem. Second, conventionally unrolling schemes require truncated iterations, whereas \ac{PnP} schemes base the learned version \eqref{eq:proximalNet_weightShare} on the formulation in \eqref{eq:ProxGadient}. This enables one to perform a convergence analysis. It also means unrolling schemes consider $\RecOp_\NNparam(\data)=\signal^N$ instead of  $\RecOp_\NNparam(\data)=\signal^\infty$ as in \ac{PnP}. Note, that in practice one still has to truncate also in the latter case, but this is supposedly done at an assumed point of convergence. Additionally, in general the learned reconstruction operator does not approximate the variational problem.
In the following, we summarise shortly some underlying theory and refer to \cite{hauptmann2024convergent,kamilov2023plug,Mukherjee:2023aa} for further details. 

The viewpoint of \ac{PnP} schemes originates from the insight that the proximal operator \eqref{eq:proxOperator} in the updates \eqref{eq:ProxGadient} essentially serves as a denoiser, so it could be replaced by a generic denoiser \cite{venkatakrishnan2013plug}. More precisely, the proximal operator represents a Gaussian denoising problem, and a such the denoiser should ideally be designed to remove Gaussian noise.  
This relates to the first distinction mentioned above, which now allows for training of the neural operator $\NNOp_\NNparam$ as a  denoiser. The (supervised) training data is then given as pairs in $\RecSpace$ only, i.e., $(\signal_i,\widetilde{\signal}_i)_i \in \RecSpace \times \RecSpace$, with $\widetilde{\signal}_i=\signal_i + \signalnoise_i$ where $\signalnoise_i$ represents the observation error/noise that is assumed to be zero-mean Gaussian noise. 

In inverse problems, the noisy realisation of $\signal$ may be also obtained from observed measurements $\data$ with a handcrafted reconstruction operator, such that $\widetilde{\signal}=\RecOp(\data)$, if it preserves Gaussian noise. Nevertheless, it should be noted that for many inverse problems the reconstructed image $\widetilde{\signal}$ does contain artefacts and noise that is not purely zero-mean Gaussian, especially so if an undersampled problem is considered, e.g., sparse or limited-angle X-ray tomography. This is a primary limitation of \ac{PnP} schemes for inverse problems and hence \ac{PnP} schemes are mostly successful for fully-sampled inverse problems and have limited success for severely ill-posed or undersampled cases. We note, that some recent advances have been made to extend \ac{PnP} schemes to Poisson data \cite{hurault2023convergent,klatzer2025efficient}. 

On the other hand, and with respect to the second distinction, since the proximal operator in the classic iterative scheme has been replaced, one can examine the limiting case and establish convergence results of varying degree. 
For instance, we observe that if the iterative update given by the mapping $\NNOp_\NNparam\colon\signal^{k-1}\mapsto \signal^k$ is contractive, then the iterative process of applying $\NNOp_\NNparam$ repeatedly will converge to a fixed-point $\signal^\infty$ with
\begin{equation}\label{eqn:fixedPoint}
 \lim_{k\to\infty}\NNOp_\NNparam(\signal^k) = \NNOp_\NNparam(\signal^\infty)=\signal^\infty.
\end{equation}
This requires, that the networks $\NNOp_\NNparam$ in composition with the iterative procedure $\signal^{k-1} - \omega \grad \operator{Q}_{\data}(\signal^{k-1})$ are contractive. Conditions on the step-length, networks, and operators have been established in various settings \cite{chan2016plug,ryu2019plug} and generally require to constrain the Lipschitz constant of the network $\NNOp_\NNparam$ together with an appropriate step-size choice for $\omega$. Finally, stronger results can be achieved by further restricting the network. For instance, objective convergence \cite{hurault2022gradient,nair2021fixed}, or a convergent regularisation method \cite{ebner2024plug,hauptmann2024convergent}. Finally, it should be said that while strong theoretical results can be achieved by restricting the networks and as such the parameter space, this will lead to less expressive networks and most commonly will result in deteriorated quantitative performance. 

To summarise, for a learned reconstruction operator $\RecOp_\NNparam(\data)=\signal^N$ the \ac{PnP} framework offers an alternative learning problem to effectively train the neural updating operator $\NNOp_\NNparam$ with only data on $\RecSpace$. 
Nevertheless, the \ac{PnP} theory only holds, when a reconstruction operator $\RecOp_\NNparam(\data)=\signal^\infty$ is considered.

\paragraph{Relation to \acf{DEQ} networks}
We are starting from \eqref{eq:proximalNet_weightShare} and consider again a learned reconstruction operator in the limit $\RecOp_\NNparam(\data)=\signal^\infty$, and assuming that the iterates converge to a fixed point \eqref{eqn:fixedPoint}. The difference for \ac{DEQ} networks to the previous \ac{PnP}, is that in fact an unrolled algorithm is trained, which means here that the forward pass of the reconstruction operator \eqref{eq:LearnedProximal} is evaluated in the loss function. Clearly, while we cannot evaluate the limit we can make use of the fixed point formulation to compute the gradients for training as we will outline next shortly. 

We assume that the learned reconstruction operator $\RecOp_\NNparam\colon\DataSpace\to\RecSpace$ iteratively applies the neural updating operator $\NNOp_\NNparam\colon\RecSpace\to\RecSpace$ and that $\RecOp_\NNparam(\data)=\signal^\infty$ is a fixed point with $\NNOp_\NNparam(\signal^\infty)=\signal^\infty$ as in \eqref{eqn:fixedPoint}. Given the loss function $\SignalLoss\colon\RecSpace\times\RecSpace\to \Real$ with 
\[
\SignalLoss\left(\RecOp_\NNparam(\data),\signal\right)=\SignalLoss\left(\signal^\infty,\signal\right),
\]
we now want to compute the gradient with respect to the network parameters $\theta$. Differentiating and applying the chain rule gives
\begin{equation}\label{eqn:DEQ_lossTotal}
\frac{\partial\!\SignalLoss\left(\RecOp_\NNparam(\data),\signal\right)}{\partial\NNparam}
=
\frac{\partial\!\SignalLoss\left(\RecOp_\NNparam(\data),\signal\right)}{\partial\!\RecOp_\NNparam(\data)}
\frac{\partial\!\RecOp_\NNparam(\data)}{\partial\NNparam}.
\end{equation}
The first term is simply the loss function differentiated with respect to the reconstruction. For the standard least squares loss and using the fixed point $\RecOp_\NNparam(\data)=\signal^\infty$ we obtain
\[
\frac{\partial\!\SignalLoss\left(\signal^\infty,\signal\right)}{\partial\!\signal^\infty} = \frac{\partial \frac{1}{2}\| \signal^\infty -\signal\|_2^2}{\partial\!\signal^\infty}
=
\signal^\infty-\signal.
\]
The second term in \eqref{eqn:DEQ_lossTotal} can be simplified by also using the fixed point formulation $\RecOp_\NNparam(\data)=\signal^\infty = \Lambda_\theta(\signal^\infty)$
and differentiating both sides of $\signal^\infty = \Lambda_\theta(\signal^\infty)$ using implicit differentiation gives
\[
\frac{\partial\! \signal^\infty}{\partial\NNparam} = \frac{\partial\! \NNOp_\NNparam(\signal^\infty)}{\partial\NNparam} + 
\frac{\partial\! \NNOp_\NNparam(\signal^\infty)}{\partial\!\signal^\infty}\frac{\signal^\infty}{\partial\NNparam}
\]
and solving for $\partial\!\signal^\infty/\partial\NNparam$ to obtain
\[
\frac{\partial\! \signal^\infty}{\partial\NNparam} = \left( \Id - 
\frac{\partial\! \NNOp_\NNparam(\signal^\infty)}{\partial\!\signal^\infty}\right)^{-1} \frac{\partial\! \NNOp_\NNparam(\signal^\infty)}{\partial\NNparam}.
\]
Substituting this into \eqref{eqn:DEQ_lossTotal} we get a short form of the gradient
\begin{equation}\label{eqn:DEQ_lossTotal_2nd}
\frac{\partial\!\SignalLoss\left(\RecOp_\NNparam(\data),\signal\right)}{\partial\NNparam}
= 
\frac{\partial\!\SignalLoss(\signal^\infty),\signal}{\partial\!\signal^\infty}
\left( \Id - \frac{\partial\! \NNOp_\NNparam(\signal^\infty)}{\partial\!\signal^\infty}\right)^{-1} \frac{\partial\! \NNOp_\NNparam(\signal^\infty)}{\partial\NNparam} .
\end{equation}
As we see, only the first term is loss-function dependent. The second term involves differentiating the network with respect to its input 
$\frac{\partial\! \NNOp_\NNparam(\signal^\infty)}{\partial\!\signal^\infty}$, which amounts to calculating the network Jacobian in practice and can be done via automatic differentiation.
Most notably, to compute the gradient of the loss \eqref{eqn:DEQ_lossTotal_2nd} we do not need to perform backpropagation through all iterates, which reduces memory consumption to train the networks. Nevertheless, the forward pass needs to be evaluated. Thus, this technique primarily allows to reduce the high memory consumption needed for efficient backpropagation, but computational limitations remain for expensive forward operators.

In practice, the fixed point is computed using acceleration techniques to reduce the amount of iterations in the forward pass and hence a finite number of iterations is considered in training, see \cite{bai2019deep,gilton2021deep} for further details. The expensive calculation of the Jacobians in \eqref{eqn:DEQ_lossTotal} can in fact be simplified as well, most notably using Jacobian free updates \cite{DAVY2025315,fung2022jfb}.

Finally, let us note that \ac{PnP} and \ac{DEQ} models share a close connection. Both are based on considering the limiting case $\RecOp_\NNparam(\data)=\signal^\infty$ for the theory and not a finite amount of iterates. In both cases, the networks are considered to be contractive. In fact, the convergence results for \ac{DEQ} follow closely the classic \ac{PnP} theory for contractive updates and require constraints on the step-length parameters $\omega_k$ and Lipschitz constant of the network $\NNOp_\NNparam$ and forward operator $\FwdOp$. A crucial difference to \ac{PnP} networks is that \ac{PnP} networks are additionally parametrised as denoiser, while \ac{DEQ} are not further constrained. The former does allow to establish stronger results than fixed point convergence, as discussed earlier. The latter offers computational advantages for applications in limited-angle and or very sparse tomographic settings where artefacts can not be easily described as common noise distributions. In such cases \ac{DEQ} models show better performance that is closer matched to conventional unrolling schemes \cite{hauptmann2023model}.

\subsubsection{Variational networks and total deep variation}\label{sec:VarNets}
The starting point for \emph{variational networks} follows a similar setup as for learned proximal networks. In particular, also {variational networks} aim at parametrising the regulariser in a gradient descent scheme. Rather than through the use of a proximal operator, variational networks parametrise the gradient of the regulariser directly.

The underlying optimisation algorithm for variational networks is a standard gradient descent scheme for minimising $\CostFunc_{\data} \colon \RecSpace \to \Real$ in \eqref{eq:Costfunc_as_DataDiscAndReg} for given data $\data \in \DataSpace$.
The data-fidelity functional $\operator{Q}_{\data}\colon \RecSpace \to \Real$, 
forward operator $\FwdOp \colon \RecSpace \to \DataSpace$, and regulariser $\RegOp\colon\RecSpace\to\Real$ are here handcrafted.
For differentiable $\operator{Q}_{\data} \colon \RecSpace \to \Real$ and $\RegOp\colon \RecSpace \to \Real$ we can write the standard gradient descent scheme as in \eqref{eq:Gradient} by
\begin{equation}\label{eq:GradientVarNet}
\begin{cases}
 \signal^0 := \RecOp_0(\data) &
 \\[0.5em]
 \signal^k := \signal^{k-1} - \omega_k \bigl( \grad\operator{Q}_{\data}(\signal^{k-1}) + \grad\RegOp(\signal^{k-1}) \bigr)
 & \text{for $\omega_k>0$ and $k = 1,2,\ldots$.}  
\end{cases}
\end{equation}
Variational networks keep the gradient descent structure and aim at parametrising the gradient of the regulariser $\grad\RegOp\colon \RecSpace\to\RecSpace$ directly. Note that the gradient of the regulariser only acts on the image space $\RecSpace$ and can be replaced by a neural network. While, in the original publication \cite{Kobler:2017aa} that introduced variational networks a \ac{FoE} model has been used to create an interpretable representation, we will follow here a general formulation using a neural operator for consistency with the previous sections. Thus, let us introduce the neural operator $ \NNOp_{\NNparam_k}\colon\RecSpace\to\RecSpace$ with iteration dependent parameters, which is replacing the gradient of the regulariser in each iteration. The learned reconstruction operator is then given as $\RecOp_{\NNparam}(\data) := \signal^{\UnrollIter}$ where
\begin{equation}\label{eq:VarNet}
\begin{cases}
 \signal^0 := \RecOp_0(\data) &
 \\[0.5em]
 \signal^k := 
 \signal^{k-1} - \omega_k \grad \operator{Q}_\data(\signal^{k-1}) + \NNOp_{\NNparam_k}(\signal^{k-1})
 & \text{for $k=1,\ldots,\UnrollIter$}.
\end{cases}
\end{equation}
Without introducing further structure in the neural operator $\NNOp_{\NNparam_k}$, there are no theoretical guarantees available for the above learned reconstruction operator. 

The paper \cite{Kobler:2017aa} considered the \ac{FoE} parametrisation, which allowed to learn activation functions. While it provided some insights into what kind of non-linear features the network learns, this original parametrisation did not provide any convergence or stability guarantees. Nevertheless, more recent work addressed this by introducing an extended framework as the so-called Total Deep Variation \cite{kobler2020total,Kobler:2022aa} which does provide a stability analysis. We will discuss both approaches in the following.
\begin{remark}
If the neural updating operator $\NNOp_\NNparam$ is fixed for each iteration being unrolled, then one can interpret \eqref{eq:VarNet} as minimising $\CostFunc_{\data}$ in \eqref{eq:Costfunc_as_DataDiscAndReg} by learning-to-optimise (Section~\ref{sec:OptimSolver}). 
\end{remark}

\paragraph{Implementation and extensions}
Variational networks are among the first learned iterative networks and for the time the connection to variational regularisation was the most apparent. Even though, there is only a minor theoretical difference to other reconstruction operators based on unrolling schemes.

For this purpose, let us shortly discuss the original variational networks as proposed by Kobler et al. \cite{hammernik2018learning,Kobler:2017aa}, which take their inspiration from variational regularisation models where the regularisation functional $\RegOp$ is equipped with parametrisations that go beyond handcrafted regularisers such \ac{TV}.
More concretely, variational networks take their inspiration from the \ac{FoE} model 
described in \cite{chen2014insights,roth2009fields,roth2005fields}. Let us consider here a general \ac{FoE} parametrisation of the regulariser as
\begin{equation}\label{eq:FoE_linOp}
\RegOp_{\NNparam}(\signal) 
 := \sum_{i=1}^m \Bigl[ \sum_{k=1}^n \rho_i \bigl( (J_{i} \signal)_k \bigr) \Bigr]
 \quad\text{with $\NNparam= (\rho_1, \ldots, \rho_m, J_1, \ldots, J_m)$.}
\end{equation}
Here, $J_i\colon \RecSpace\to\RecSpace^n$ are linear transforms, usually given by a convolution with $n$ filters, and 
$m$ is the number of regularisation functions, with parameterisable non-linear functions $\rho_i\colon \Real\to \Real$. Note that the non-linear function $\rho_i\colon \Real\to \Real$ as well as its derivative $\rho'_i\colon \Real\to \Real$ acts pointwise.

The updates in \eqref{eq:GradientVarNet} require the gradient of the regulariser $\RegOp_{\NNparam}$, which can be computed by
\begin{equation}\label{eq:chenFoEbilevel}
\grad \RegOp_{\NNparam}(\signal) 
 := \sum_{i=1}^m J_i^{\ast} \rho'_i \bigl( J_{i} \signal \bigr),
\end{equation}
with the adjoint operator $J^{\ast}\colon \RecSpace^n \to \RecSpace$. Note, that the linear operator $J$ expands the image space and the adjoint $J^{\ast}$ collapses it again. This is similar to a two layer neural network, where the filters are tied together. 

From here, we obtain the associated reconstruction operator for variational networks \cite{hammernik2018learning,Kobler:2017aa} as
$\RecOp_{\NNparam}(\data) := \signal^{\UnrollIter}$ with
\begin{equation}\label{eq:VarNetOrig}
\begin{cases}
 \signal^0 := \RecOp_0(\data) &
 \\[0.5em]
 \signal^k := 
 \signal^{k-1} - \omega_k \grad \operator{Q}_\data(\signal^{k-1}) + \grad \RegOp_{\NNparam_{k-1}}(\signal^{k-1}) 
 & \text{for $k=1,\ldots,\UnrollIter$},
\end{cases}
\end{equation}
where now each iterate has its own set of parameters. The training of the reconstruction operator is performed fully supervised and end-to-end. The first paper by Kobler et al. \cite{Kobler:2017aa} also suggests to parameterise the data-fidelity term $\operator{Q}_\data$ in a similar manner to \eqref{eq:FoE_linOp}, later developments \cite{hammernik2018learning} have omitted the learned component on the data-fidelity. 

Despite the name variational networks, the connection to variational methods is restricted to parametrising the learned update from a \ac{FoE} perspective. Training the reconstruction operator supervised and with changing weights in each iteration removes the possibility to interpret $\RecOp_\NNparam$ as a minimiser of a variational cost functional. 

Finally, since the connection to a minimiser of a variational cost functional is not satisfied, it is easy to replace the learned component by a general neural operator as in our formulation \eqref{eq:VarNet}. This offers to use larger and more expressive networks for $\NNOp_\NNparam\colon\RecSpace\to\RecSpace$, such as U-Nets.

\paragraph{Total Deep Variation}
The framework of variational networks has been extended to include multi-scale networks, likely inspired by the popular U-Net architecture, introduced as \ac{TDV} in \cite{kobler2020total,Kobler:2022aa}. Additionally, the work establishes robustness and stability results with respect to input as well as learned parameters based on a mean-field optimal control interpretation. That is, the formulation of the reconstruction operator is based on a continuous-in-time gradient flow rather than a discrete iterative process. One can then formulate the reconstruction operator $\RecOp_\theta\colon\DataSpace\to \RecSpace$ that assigns the reconstruction at finite time $T>0$, i.e., $\RecOp_\theta(\data)=\signal(T)$, where the gradient flow is defined  to minimise a variational energy \eqref{eq:Costfunc_as_DataDiscAndReg}. The training problem is then stated as a mean-field optimal control problem to determine the stopping time $T$, or time discretisation after time normalisation, as well as the network parameters. The continuous formulation as gradient flows allows for the subsequent analysis of stability. We refer to the original papers for the details \cite{kobler2020total,Kobler:2022aa}.

\section{\Acl{LPD} architectures}\label{sec:LPD}
The previous learned gradient architectures can be extended by also including neural operators in the data space $\DataSpace$. The resulting learned iterative network does not fall into the larger classification of a learned gradient network and thus forms its own class. Such \emph{primal-dual} neural operator architectures, which are also called dual domain architectures \cite{lin2019dudonet}, were first introduced in \cite{Adler:2018aa}.

A \acf{LPD} architecture is, as the name suggests, obtained by unrolling a suitable primal-dual proximal splitting scheme.
Splitting schemes are designed to minimise convex objectives that are given as a sum of a differentiable functional and a possibly non-smooth functional.
Primal-dual schemes were introduced to exploit structure in a wider range of minimisation problems than those handled by splitting methods and involve setting up two iterative schemes for explicitly updating the primal and the dual variable. 

Such primal-dual proximal splitting schemes are particularly well suited to account for specific structure in minimising objectives that typically arise in variational regularisation of ill-posed inverse problems. 
The objective is given as a functional of the form in \eqref{eq:Costfunc_as_DataDiscAndReg}, i.e.
\begin{equation}\label{eq:Costfunc_as_DataDiscAndReg2}
  \CostFunc_{\data}(\signal) := \RegOp(\signal) + \operator{G}_{\data}\bigl(\FwdOp\signal\bigr)
\end{equation}
where $\operator{G}_\data \colon \DataSpace \to \Real$ is given as $\operator{G}_{\data}(\dataother):= \DataLoss(\dataother,\data\bigr)$, so $\operator{G}_{\data}\bigl(\FwdOp\signal\bigr)=\operator{Q}_{\data}(\signal)$ in the earlier notation.
The primal variable in a primal-dual scheme will then represent the signal in $\RecSpace$-space, and the dual variable (in the Hilbert space setting) can be identified with the data variable in $\DataSpace$-space.

An early example of a primal-dual proximal splitting scheme is the Arrow-Hurwicz algorithm \cite{Arrow:1958aa}.
When used for minimising $\CostFunc_{\data}(\signal)$ in \eqref{eq:Costfunc_as_DataDiscAndReg2}, it takes the form
\begin{equation}\label{algo:Arrow-HurwiczBasic}
\begin{cases}
 \data^{k+1} := \prox_{\sigma \operator{G}^{\#}_\data}\bigl(\data^{k} + \sigma\FwdOp\signal^{k}\bigr) &  
 \\[0.5em]
 \signal^{k+1} := \prox_{\sigma\RegOp}\bigl(\signal^{k} - \sigma\FwdOp^{\ast}\data^{k+1}\bigr) & 
\end{cases}
\quad\text{for $k=0, 1,2,\ldots$.}
\end{equation}
Here, the step length $\sigma >0$  is chosen appropriately to ensure convergence and $\operator{G}^{\#}_{\data} \colon \DataSpace \to \Real$ is the convex conjugate (Fenchel conjugate) of $\operator{G}_{\data}$, which in the Hilbert space setting can be defined as 
\[
 \operator{G}^{\#}_{\data}(\dataother) = \sup_{\dataother'\in \DataSpace}\, \langle \dataother',\dataother \rangle_{\DataSpace} - \operator{G}_{\data}(\dataother')
 \quad\text{for $\dataother \in \DataSpace$.}
\]
The primal-dual hybrid gradient algorithm \cite{Chambolle:2011aa} introduces the possibility to have separate step lengths $\tau, \sigma>0$ for the primal and dual variables. It also includes a momentum governed by $0<\rho<1$ in the update of the primal variable to accelerate the convergence rate, thus resulting in an iterative scheme of the following form: 
\begin{equation}\label{algo:PDHG}
\begin{cases}
 \data^{k+1} := \prox_{\sigma \operator{G}^{\#}_\data}\bigl(\data^{k} + \sigma\FwdOp\bar{\signal}^{k}\bigr) &  
 \\[0.5em]
 \signal^{k+1} := \prox_{\tau\RegOp}\bigl(\signal^{k} - \tau\FwdOp^{\ast}\data^{k+1}\bigr) & 
 \\[0.5em]
 \bar{\signal}^{k+1} := \signal^{k+1} + \rho(\signal^{k+1}-\signal^{k}) & 
\end{cases}
\quad\text{for $k=0,1,2,\ldots$.}
\end{equation}
The convergence speed and versatility of such schemes has made them a popular choice for non-smooth minimisation, due to sparsity regularisation, that typically arises in variational regularisation of inverse problems.

A primal-dual type of iterative scheme, like \eqref{algo:Arrow-HurwiczBasic} or \eqref{algo:PDHG}, can serve as a blueprint for a learned iterative network architecture via unrolling.
One could replace only the proximal related to the regulariser $\RegOp \colon \RecSpace \to \Real$ in \eqref{eq:Costfunc_as_DataDiscAndReg2} with a neural operator, or both proximal operator with neural operators.
The \ac{LPD} architecture presented in \cite{Adler:2018aa} adopts the latter by unrolling the iterative scheme in \eqref{algo:PDHG} and replacing the proximal operators in both $\RecSpace$- and $\DataSpace$-space with neural operators. 
This gives rise to the following learned reconstruction operator for $\RecOp_{\NNparam} \colon \DataSpace \to \RecSpace$ with $\NNparam := (\NNparam_0,\NNparamOther_0\ldots, \NNparam_{N-1},\NNparamOther_{N-1})$ where $\RecOp_{\NNparam}(\data)=\signal^N$ is given as 
\begin{equation}\label{eq:LPDArchitecture}
\begin{cases}
\data^{k+1} := \NNOpOther_{\NNparamOther_{k}}\bigl(\data^{k}, \FwdOp\signal^{k},\data \bigr)
& \\[0.5em]
\signal^{k+1} := \NNOp_{\NNparam_{k}}\bigl(\signal^k , \FwdOp^{\ast}\data^{k+1} \bigr)
&
\end{cases}
\quad\text{for $k=0,1,\ldots, \UnrollIter-1$.}
\end{equation}
In the above, we assume there is an initialisation $(\signal^0,\data^0)\in \RecSpace \times \DataSpace$, typically chosen as $\data^0 := \data$ and $\signal^0 := \RecOp_0(\data)$ as initial reconstruction from a handcrafted reconstruction operator.
The two neural updating operators $\NNOp_{\NNparam_k} \colon \RecSpace\times\RecSpace\to\RecSpace$ and $\NNOpOther_{\NNparamOther_k} \colon \DataSpace\times\DataSpace\times\DataSpace \to \DataSpace$ replace the proximal operators in \eqref{algo:Arrow-HurwiczBasic}. We note, that similar architectural choices as for learned gradient networks are possible, such as weight-sharing, memory for primal and dual variables, as well as residual formulations of the update.

The use of the two spaces and associated neural updating operators instead of one improves expressivity of the learned reconstruction operator due two primary factors. 
First, and most obvious, the use of two neural operators doubles the number of learnable parameters and hence provides a stronger approximation capacity. Second, the added information in the dual space encodes crucial information on topology of the data and enables direct suppression of noise. 
\begin{remark}
\Ac{LPD} architectures in \eqref{eq:LPDArchitecture} extends the learned gradient scheme in \eqref{eq:LearnedGradientMethod}.
To see this, note that setting $\NNOpOther_{\NNparamOther_k}(\dataother,\dataother',\dataother'') = \dataother'-\dataother''$ (independent of $\dataother$) in \eqref{eq:LPDArchitecture} yields the learned least squares architecture in \eqref{eq:LearnedGradient_withL2}. 
\end{remark}

To summarise, \ac{LPD} architectures extend learned gradient networks by introducing a second neural updating operator in the data space. This allows to increase expressivity for a fixed amount of iterations, and hence under the same associated cost with evaluating the forward $\FwdOp$ and adjoint $\FwdOp^\ast$.

\paragraph{Implementation}
The original implementation in \cite{Adler:2018aa} is largely similar to the implementation of the learned least squares network \cite{Adler:2017aa} as described in Section~\ref{sec:LearnedLSQ}. Specifically, both neural updating operator $\NNOp_{\NNparam_k} \colon \RecSpace\times\RecSpace\to\RecSpace$ and $\NNOpOther_{\NNparamOther_k} \colon \DataSpace\times\DataSpace\times\DataSpace \to \DataSpace$ are parametrised as a ResNet style \ac{CNN}. Additionally, a memory in both primal and dual variable has been used. 

As before, the resulting learned reconstruction operator can be trained depending on available data and possible loss functions. Even though supervised and end-to-end training is the most successful, some studies have utilised unsupervised data \cite{mukherjee2021end,AnttiThesis2024}. Additionally, the same considerations concerning computational complexity of including $\FwdOp$ and $\FwdOp^*$ in the training remain, which will be discussed further down in Section~\ref{sec:implementation}.

\section{Non-linear inverse problems and higher order methods}\label{sec:HighOrderNonLin}
The previous section have been concentrated on learned reconstruction operators with linear forward operators, in the following we will now consider the more general setting where the forward operator $\FwdOp \colon \RecSpace \to \DataSpace$ in the inverse problem \eqref{eq:InvProb} may be non-linear and is assumed to be Fréchet differentiable.
To simplify the setting somewhat, we will still restrict our attention to the case when $\RecSpace$ and $\DataSpace$ are Hilbert spaces. In fact, we will first see that the previous two sections on gradient networks and primal dual networks can be readily extended to the non-linear case.

\subsection{Learned gradient networks for non-linear problems}\label{sec:non-linearGrad}
Since $\FwdOp$ is assumed to be Fréchet differentiable, we get that the functional $\operator{Q}_{\data} \colon \RecSpace \to \Real$ with $\operator{Q}_{\data}(\signal) = \frac{1}{2}\bigl\Vert \FwdOp(\signal) - \data \bigr\Vert_\DataSpace^2$ as in \eqref{eq:DataFidelityL2} is also Fréchet differentiable and its derivative is a bounded linear operator $\partial\!\operator{Q}_{\data}(\signal) \colon \RecSpace \to \RecSpace$.
Next, $\RecSpace$ is a Hilbert space so from Riesz representation theorem we know there exists a unique element $\grad \operator{Q}_{\data}(\signal) \in \RecSpace$, the gradient of $\operator{Q}_{\data}$, such that 
\[ \partial\!\operator{Q}_{\data}(\signal)(\signalother) =
 \bigl\langle \grad \operator{Q}_{\data}(\signal),\signalother \bigr\rangle_{\RecSpace}
 \quad\text{for all $\signalother \in \RecSpace$.}
\]

One can now use a gradient descent scheme as in \eqref{eq:GradDescent} to compute a minimiser to the functional $\operator{Q}_{\data} \colon \RecSpace \to \Real$. 
This yields an iterative scheme of the form 
\[ 
\signal^{k+1} = \signal^{k} - 
\omega_k \triangle\signal^{k}
\quad\text{with}\quad
\triangle\signal^{k} := \grad\operator{Q}_{\data}(\signal^k).
\]
where the step length $\omega_k>0$ is chosen according to some pre-defined rule.
A particular case of interest is when $\operator{Q}_{\data}$ is given as in \eqref{eq:DataFidelityL2}. Then 
\begin{equation}\label{eq:DataFidelityL2GradNonLin} 
\grad\operator{Q}_{\data}(\signal) = \bigl[\partial\!\FwdOp(\signal)\bigr]^{\ast}\bigl(\FwdOp(\signal) - \data\bigr)
\quad\text{for $\signal \in \RecSpace$}
\end{equation} 
where the linear mapping $\bigl[\partial\!\FwdOp(\signal)\bigr]^{\ast} \colon \DataSpace \to \RecSpace$ is the adjoint of the derivative of the forward operator at $\signal \in \RecSpace$.
The above iterative scheme becomes 
\[ 
\signal^{k+1} = \signal^{k} - 
\omega_k \bigl[\partial\!\FwdOp(\signal^{k})\bigr]^{\ast}\bigl(\FwdOp(\signal^{k}) - \data\bigr).
\]

Unrolling the above scheme in the same way as in section~\ref{sec:LearnedGradient} yields a learned operator architecture for solving ill-posed non-linear inverse problems. Analogous, we can replace the linear forward operator for the primal-dual schemes in Section~\ref{sec:LPD} with the non-linear forward $\FwdOp\colon\RecSpace\to\DataSpace$ and the adjoint with the adjoint of its derivate $\bigl[\partial\!\FwdOp(\signal)\bigr]^{\ast} \colon \DataSpace \to \RecSpace$ to obtain a learned-primal dual architecture for non-linear inverse problems. 

\subsection{Newton and Newton-type methods}
For non-linear inverse problems the computation of the Fréchet derivative is usually computationally more demanding, as it depends on the current solution and needs to be re-computed every time. Thus, optimisation algorithms with faster convergence rates are often preferred to gradient-based and first order methods. Therefore, we will now discuss Newton methods and Newton-type methods as well as the corresponding learned reconstruction operators that result from unrolling them.

\subsubsection{Newton iterates}
When computing least-squares solutions to the inverse problem in \eqref{eq:InvProb}  with $\RecSpace, \DataSpace$ Hilbert spaces, the solution is a minimiser to $\operator{Q}_{\data} \colon \RecSpace \to \Real$ in \eqref{eq:DataFidelityL2}.
Therefore, also a critical point to $\operator{Q}_{\data}$ and a natural approach for computing it is based on solving $\grad \operator{Q}_{\data}(\signal)=0$ in $\RecSpace$.
If $\operator{Q}_{\data}$ is twice Fréchet differentiable, then it is natural to consider (relaxed) Newton iterates for computing a minimiser by solving $\grad\operator{Q}_{\data}(\signal)=0$. 
These yield a sequence $(\signal^k)_k \subset \RecSpace$ generated by an iterative scheme of the form
\begin{equation}\label{eq:NewtonIterates}
  \signal^{k+1} = \signal^{k} + \omega_k \triangleSignal^k
  \quad\text{where}\quad
  \triangleSignal^k := - \bigr[\HessianOp{\operator{Q}_{\data}}(\signal^{k})\bigl]^{-1} 
  \bigl( \grad \operator{Q}_{\data}(\signal^k) \bigr).
\end{equation}
Here, $\HessianOp{\operator{Q}_{\data}} \colon \RecSpace \to \LinOp(\RecSpace,\RecSpace)$ is the Hessian of $\operator{Q}_{\data}$ (Section~\ref{sec:HessianCalc}) and the step-length  $0<\omega_k \leq 1$ is chosen to ensure the objective function $\operator{Q}_{\data}$ decreases `sufficiently' much along the Newton direction $\triangleSignal^k$. The corresponding reconstruction operator  
$\RecOp \colon \DataSpace \to \RecSpace$ is then given by $\RecOp_{\NNparam}(\data) := \signal^{\UnrollIter}$ with
\begin{equation}\label{eq:ReconOperatorNewton}
\begin{cases}
 \signal^0 := \RecOp_0(\data) &
 \\[0.5em]
 \signal^k := \signal^{k-1} + \omega_k \triangleSignal^{k-1}
 & \text{for $k=1,\ldots$.}
\end{cases}
\end{equation}
Computing the Newton direction $\triangleSignal^k$ requires inverting the Hessian operator of $\operator{Q}_{\data}$ at each iterate $\signal^k$ and this quickly becomes computationally unfeasible.
If the main computational bottle neck lies in the inversion, then one could consider computing the Newton direction by solving the linear system 
\[ 
 \bigl[\HessianOp{\operator{Q}_{\data}}(\signal^k)\bigr](\triangleSignal^k) = -\grad \CostFunc(\signal^{k})
\]
When the Hessian is a positive definite operator, then one can consider various iterative approaches for solving the above linear problem. 
However, the mere task of forming the Hessian is often computationally unfeasible in its own right.
Much effort has therefore been devoted to formulating various computationally feasible approximations to the Hessian or its inverse.
We will consider next two of these, namely Gauss-Newton methods (Section~\ref{sec:GaussNewton}) and quasi-Newton methods (Section~\ref{sec:QuasiNewton}), which basically provide a different update direction $\triangleSignal$ for the reconstruction operator \eqref{eq:ReconOperatorNewton}.

\subsubsection{Gauss-Newton methods}\label{sec:GaussNewton}
The starting point is to consider Newton iterates \eqref{eq:NewtonIterates} for computing a least squares solution to the inverse problem in \eqref{eq:InvProb}.
This requires computing the inverse of the Hessian of $\operator{Q}_{\data}$ in \eqref{eq:DataFidelityL2} at each iterate. Due to the aforementioned cost of computing the Hessian $\HessianOp{\operator{Q}_{\data}}(\signal^k) \in \LinOp(\RecSpace,\RecSpace)$ given in \eqref{eq:QHessian} the idea is now to approximate it by dropping the 2nd term, so that we get 
\[
 \HessianOp{\operator{Q}_{\data}}(\signal^k) \approx 
 \bigl[\partial\!\FwdOp(\signal^k)\bigr]^{\ast} \circ \partial\!\FwdOp(\signal^k).
\]
This resulting scheme is $\signal^{k+1} = \signal^{k} + \omega_k \triangleSignal^k$ where
\begin{equation}\label{eq:Gauss-NewtonIterates}
  \triangleSignal^k := \Bigl( \bigl[\partial\!\FwdOp(\signal^k)\bigr]^{\ast} \circ \partial\!\FwdOp(\signal^k) \Bigr)^{-1} 
  \bigl( \grad \operator{Q}_{\data}(\signal^k) \bigr).
\end{equation}

Computing the Hessian (or its inverse) can however be ill-conditioned, and this is especially the case for ill-posed inverse problems.
Thus, one may consider applying the above iterative scheme not on $\operator{Q}_{\data} \colon \RecSpace \to \Real$, but on a regularised version $\CostFunc_{\data} \colon \RecSpace \to \Real$ of the form
\[
 \CostFunc_{\data}(\signal):=\operator{Q}_{\data}(\signal) + \alpha \RegOp(\signal,\signalother)
\] 
where $\RegOp(\,\cdot\,,\signalother) \colon \RecSpace \to \Real$ for given $\signalother \in \RecSpace$ is assumed to be twice Fréchet differentiable with Hessian $\HessianOp{\RegOp(\,\cdot\,,\signalother)} \colon \RecSpace \to \LinOp(\RecSpace,\RecSpace)$.
This leads to the following Gauss-Newton type of scheme where the Newton direction $\triangleSignal^k \in \RecSpace$ in \eqref{eq:Gauss-NewtonIterates} is replaced with
\begin{equation}\label{eq:Gauss-NewtonIterates2}
 \triangleSignal^k := 
  \Bigl( \bigl[\partial\!\FwdOp(\signal^k)\bigr]^{\ast} \circ \partial\!\FwdOp(\signal^k) 
       + \HessianOp{\RegOp(\,\cdot\,,\signal^{k-1})}(\signal^k) 
  \Bigr)^{-1} 
  \bigl( \grad\CostFunc_{\data}(\signal^{k}) \bigr).
\end{equation}
This was first introduced in \cite{Bakushinskii:1992aa} for regulariser $\RegOp(\signal,\signalother)=\| \signal-\signalother \|^2$ (classical Tikhonov functional). An alternative popular choice for $\RegOp \colon \RecSpace \to \Real$ is given by smoothed total variation \cite{gonzalez2017isotropic} to ensure twice differentiability.

\paragraph{Learned Gauss-Newton}
Unrolling the iterative scheme in \eqref{eq:ReconOperatorNewton} where $\triangleSignal^k$ is given as in \eqref{eq:Gauss-NewtonIterates} or \eqref{eq:Gauss-NewtonIterates2} yields the \emph{learned Gauss-Newton} neural operator for solving the inverse problem in \eqref{eq:InvProb}.
More precisely, this is a learned reconstruction operator $\RecOp_{\NNparam} \colon \DataSpace \to \RecSpace$ with $\NNparam := (\NNparam_1,\ldots, \NNparam_\UnrollIter)$ that is given as $\RecOp_{\NNparam}(\data) := \signal^{\UnrollIter}$ where
\begin{equation}\label{eq:LearnedGaussNewton}
\begin{cases}
 \signal^0 := \RecOp_0(\data) &
 \\[0.5em]
 \signal^k := \NNOp^k_{\NNparam_k}\bigl(
   \signal^{0},\triangle\signal^{0},\signal^{1}, \triangle\signal^{1}, \ldots, \signal^{k-1}, \triangle\signal^{k-1} \bigr) 
 & \text{for $k=1,\ldots,\UnrollIter$.}
\end{cases}
\end{equation}
where $\NNOp^k_{\NNparam_k} \colon (\RecSpace \times \RecSpace)^{k} \to \RecSpace$ are neural operators with suitable architectures. 
The weights $\NNparam_k$ are learned during training whereas the Newton directions $\triangle\signal^{k} \in \RecSpace$ are handcrafted as in \eqref{eq:Gauss-NewtonIterates} or \eqref{eq:Gauss-NewtonIterates2}. 

A special case considered by \cite{herzberg2021graph,Mozumder:2022aa} is \eqref{eq:LearnedGaussNewton} without memory and $\RegOp \colon \RecSpace \to \Real$ set to a Gaussian with co-variance operator $\Sigma_S \colon \RecSpace \to \RecSpace$, i.e., 
\[
 \RegOp(\signal) = \exp\Bigl(-\frac{1}{2} \bigl\langle \Sigma_S(\signal),\signal \bigr\rangle\Bigr)
\]
where $\Sigma_S \colon \RecSpace \to \RecSpace$ is bounded, self-adjoint and non-negative.
Then \eqref{eq:LearnedGaussNewton} becomes
\begin{equation*}
  \signal^{k+1} := 
  \NNOp_{\NNparam_k}\biggl(
   \signal^{k}, 
   \Bigl(\bigl[\partial\!\FwdOp(\signal^k)\bigr]^{\ast} \circ \partial\!\FwdOp(\signal^k) 
   + \Sigma_{\RegOp} 
   \Bigr)^{-1}\bigl( \grad\CostFunc_{\data}(\signal^{k}) \bigr)
  \biggr)
 \text{ for $k=1,\ldots,\UnrollIter$}
\end{equation*}
where $\NNOp_{\NNparam_k} \colon \RecSpace \times \RecSpace \to \RecSpace$ is a neural operator with suitable architecture. The training in \cite{herzberg2021graph,Mozumder:2022aa} has been done following the greedy regime \eqref{eqn:greedyTrain}. Whereas the an end-to-end training of the learned reconstruction operator has been utilised in \cite{Colibazzi:2022aa,manninen2025towards}.

\subsubsection{Quasi-Newton methods}\label{sec:QuasiNewton}
The approach here is to iteratively approximate the Hessian or its inverse in \eqref{eq:NewtonIterates}.
This results in the following scheme: $\signal^{k+1} = \signal^{k} + \omega_k \triangleSignal^k$ where
\begin{equation}\label{eq:Quasi-NewtonIterates}
\triangle\signal^{k} := 
 \begin{cases}
  \opH_k^{-1} \bigl( \grad \operator{Q}_{\data}(\signal^k) \bigr)
  &\text{ with } \opH_k \approx \HessianOp{\operator{Q}_{\data}}(\signal^{k})
  \\[0.75em]
  \opB_k\bigl( \grad \operator{Q}_{\data}(\signal^k) \bigr)
  &\text{ with } \opB_k \approx \bigl[\HessianOp{\operator{Q}_{\data}}(\signal^{k})\bigr]^{-1}.
 \end{cases}
\end{equation}
The approximations $\opB_k, \opH_k \colon \RecSpace \to \RecSpace$ typically need to satisfy a secant condition:
\begin{equation}\label{eq:SecantCond}
\begin{split}
 \grad\operator{Q}_{\data}(\signal^{k+1}) 
   &= \grad\operator{Q}_{\data}(\signal^{k}) + \opH_k (\signal^{k+1}-\signal^{k})
\\
 \signal^{k+1}-\signal^{k}
   &= \opB_k \bigl(\grad\operator{Q}_{\data}(\signal^{k+1})-\grad\operator{Q}_{\data}(\signal^{k})\bigr).
\end{split}
\end{equation}

Various methods have been developed for constructing $\opH_k$ or $\opB_k$ when $\RecSpace = \Real^n$.
In that setting, $\operator{Q}_{\data} \colon \Real^n \to \Real$ so its Hessian is represented by a symmetric $n \times n$ matrix. 
The symmetric linear operators $\opB_k, \opH_k \colon \Real^n \to \Real^n$ are therefore represented by $n \times n$ matrices $\matB_k$ and $\matH_k$, respectively. 
Quasi-Newton methods now provide explicit closed-form expressions for assembling $\matB_k$ and $\matH_k$ from previous iterates that only make use of first order information. 
Examples include the \ac{DFP}, \ac{BFGS}, \ac{SR1} and \ac{PSB} methods that are nowadays widely used in non-linear programming. 
As an illustration, we provide the expressions used within the widely used \ac{BFGS} scheme:
\begin{equation}\label{eq:BFGS}
\begin{split}
\matH_{k+1}
&= \bigl(\matI-\gamma_k (\deltaSignal_k \cdot \deltaCost_k^{\!\top}) \bigr)
   \cdot \matH_k\cdot
   \bigl(\matI-\gamma_k (\deltaCost_k \cdot \deltaSignal_k^{\!\top}) \bigr)
  + \gamma_k (\deltaSignal_k \cdot \deltaSignal_k^{\!\top})
\\[0.25em]
\matB_{k+1}
&= \matB_k - \frac{1}{\langle \deltaSignal_k, \matB_k \cdot \deltaSignal_k \rangle}
  \bigl( (\matB_k \cdot \deltaSignal_k) \cdot (\deltaSignal_k^{\!\top} \cdot \matB_k) \bigr)
  + \gamma_k (\deltaCost_k \cdot \deltaCost_k^{\!\top})
\end{split}
\end{equation}
where $\deltaSignal_k :=\signal^{k+1}-\signal^k$, $\deltaCost_k :=\grad\operator{Q}_{\data}(\signal^{k+1})-\grad\operator{Q}_{\data}(\signal^{k})$, and $\gamma_k:=1/\langle \deltaSignal_k, \deltaCost_k \rangle$.

Last two decades has seen efforts in extending quasi-Newton schemes of the above type to the infinite-dimensional Hilbert space setting.
These are nicely summarised in \cite{Vuchkov:2020aa} that provides Hilbert space versions of the \ac{BFGS}, \ac{DFP}, \ac{SR1}, and \ac{PSB} schemes.
The derivations are based on the observation that a quasi-Newton update formula along with the secant condition can be expressed as a solution to a specific constrained variational problem over the space of symmetric matrices.
Formulating and solving the corresponding variational problem over the space of bounded symmetric operators in Hilbert spaces (see \cite[Corollary~3.3]{Vuchkov:2020aa}) yields the functional analytic version of the quasi-Newton update formula.
The generalisation of the \ac{BFGS} scheme in \eqref{eq:BFGS} to the functional analytic setting reads as follows:
\begin{equation}\label{eq:BFGS-Func}
\begin{split}
\opH_{k+1}
&= \bigl(\opI-\gamma_k (\deltaSignal_k \otimes \deltaCost_k) \bigr)
 \circ \opH_k \circ 
 \bigl( \opI-\gamma_k (\deltaSignal_k \otimes \deltaCost_k) \bigr) 
 + \gamma_k (\deltaSignal_k \otimes \deltaSignal_k)
\\[0.25em]
\opB_{k+1}
&= \opB_k - \frac{1}{\bigl\langle \deltaSignal_k, \opB_k(\deltaSignal_k)\bigr\rangle} 
 \bigl( \opB_k(\deltaSignal_k) \otimes \opB_k(\deltaSignal_k) \bigr)
 + \gamma_k (\deltaCost_k \otimes\,  \deltaCost_k)
\end{split}
\end{equation}
where $\deltaSignal_k, \deltaCost_k$ and $\gamma_k$ are defined as in \eqref{eq:BFGS} and the $\otimes$-operator is defined as follows: $\signal \otimes \signalother \colon \RecSpace \to \RecSpace$ for given $\signal, \signalother \in \RecSpace$ is the linear rank one operator $\signalothernew \mapsto \langle \signalother,\signalothernew \rangle \signal$.
On a final note, one can show that if the curvature condition $\langle \deltaSignal_k, \deltaCost_k \rangle > 0$ holds, then $\opB_{k},\opH_{k} \in \LinOp(\RecSpace,\RecSpace)$ in \eqref{eq:BFGS-Func} are self-adjoint, invertible, and positive definite.

\paragraph{Learned quasi-Newton}
Similar to how learned Gauss-Newton architectures were defined by unrolling a Gauss-Newton scheme (Section~\ref{sec:GaussNewton}), we can also define learned quasi-Newton architectures by unrolling a quasi-Newton scheme that has the form \eqref{eq:Quasi-NewtonIterates}.

This yields a learned reconstruction operator $\RecOp_{\NNparam} \colon \DataSpace \to \RecSpace$ with $\NNparam := (\NNparam_1,\ldots, \NNparam_\UnrollIter)$ that is given as in Learned Gauss-Newton \eqref{eq:LearnedGaussNewton}, but this time with Newton directions $\triangle\signal^{k} \in \RecSpace$ that are handcrafted as in \eqref{eq:Quasi-NewtonIterates}. 
One has in addition the possibility to enforce the secant conditions in \eqref{eq:SecantCond}.
This could be achieved either by introducing additional terms to the loss during training of $\RecOp_{\NNparam}$. One could in addition also modifying the architecture to ensure that it satisfies hard constraints of the above type without sacrificing model capacity \cite{Min:2025aa}. Examples of learned quasi-Newton schemes in the literature are considered in \cite{manninen2025towards} for SR1 updates and in \cite{smyl2021efficient} based in Broyden's method.

\subsubsection{Additional notes and remarks}
Based on the reconstruction operator in \eqref{eq:ReconOperatorNewton} one could construct a large class of learned reconstruction operators by replacing the Newton directions $\signalother$ with suitable choices. One may even consider learned directions instead of handcrafted update directions.

Additionally, one can also construct neural operator architectures by unrolling proximal Newton and proximal quasi-Newton schemes and then replacing the proximal with a neural operator are possible. 
This approach was considered in the learned proximal regularised Gauss-Newton network \cite{Alberti:2025aa}, see also \cite{Colibazzi:2022aa}. 

Another computational challenge exists due to the fact that many non-linear inverse problems are given by a \ac{PDE} that is solved with finite element methods (FEM). That means, the reconstructions and iterates $\signal^k$ are given on usually triangular finite element meshes and hence a neural update operator $\NNOp_\NNparam$ based on \acp{CNN} is not directly applicable. This can be overcome for instance by interpolation to a rectangular pixel mesh \cite{Mozumder:2022aa} or by using rectangular meshes \cite{manninen2025towards} to begin with. Alternatively, some authors \cite{Alberti:2025aa,herzberg2021graph} have utilised graph neural networks to operate on finite element meshes. This is a natural choice, where the adjacency matrix in the graph neural network encodes the neighbouring elements and hence allows for computation over irregular meshes. This has been realised as a promising and flexible alternative to train a learned reconstruction operator that is mesh and dimension agnostic \cite{herzberg2023domain,Toivanen2025}.

\section{Implementation related aspects}\label{sec:implementation}
We have previously discussed in Sections~\ref{sec:LearnedGradient} and \ref
{sec:LPD} some fundamental implementation aspects. 
These refer mainly to architecture specific considerations. 
Here, we provide a discussion on more general and commonly occurring aspects. In particular, at the end of this section we will discuss a major limitation of learned iterative networks given by their computational complexity with respect to the forward operator.

\subsection{Performance and interpretability}
The performance of a learned reconstruction method for solving an ill-posed inverse problem will depend on a number of choices.
The first relates to the choice of neural operator architecture and parametrisation.
This has been the main topic of this survey and it affects the accuracy and generalisation properties (robustness) of the learned operator.
As outlined in Section~\ref{sec:MotivationStructuredLearning}, scalability and generalisation issues mandate usage of architectures that incorporate some level of domain adaptation.
Learned iterative networks are especially well suited to approximate target operators that are often defined implicitly through an iterative scheme. 
This covers operators that regularise ill-posed inverse problems (Section~\ref{sec:InvProb}) and operators representing solutions to convex optimisation problems (Section~\ref{sec:OptimSolver}).

The next choice relates to setting up the learning problem. 
The various options are briefly discussed in Section~\ref{sec:LearningProbs} and it is here important to setup a learning problem that is consistent with the type of training data one has.
The choice of loss function used in training will in particular determine what statistical estimator the learned reconstruction method approximates. 
Hence, in contrast to what is claimed in \cite{Monga:2021aa}, using an unrolled architecture as in learned iterative networks does not in itself improve upon interpretability.
As an example, if one uses a loss that is the squared 2-norm, then a perfectly trained learned  operator will approximate the conditional expectation. 
Likewise, using the 1-norm as loss implies that the learned operator approximates the conditional median. 
The choice of architecture is more related to how good this approximation becomes for some given set of training data.

To summarise, the choice of loss function along with type of training data dictates \emph{what} one seeks to compute.
The choice of neural operator architecture is more related to \emph{how to compute} as it influences the model capacity and generalisation gap. We refer to the surveys \cite{hauptmann2024convergent,kamilov2023plug,Mukherjee:2023aa} for further discussions on interpretability and theoretical results for learned reconstructions.

\subsection{Hyper-parameter tuning}
Besides choices of architecture and learning problem (incl.\@ choice of loss function), it is also well-known that the performance of a neural network depends on hyper-parameter settings  \cite{Boukrouh:2025aa,Probst:2019aa,Weerts:2020aa,Yang:2020aa,Raiaan:2024aa}.
These hyper-parameters typically regulate the optimisation method used during the training. 

Appendix~\ref{app:SGD} provides a very brief introduction to gradient descent schemes that are commonly used for training deep neural networks.
Typical hyper parameters that are varied during the training are the learning rate, choice of batch size (mini-batches are typically selected as random subsets of the training data), and initialisation. 
Empirical experience indicates that it is often enough to use default settings of these hyper-parameters for training many learned iterative networks.

There is, however, no methodology in place for tuning a hyper-parameter if that need rises apart from trial-and-error. For example, with respect to neural network size, it is common to start with a ResNet type architecture as updating operator $\NNOp_\NNparam$ in the learned reconstruction operator \cite{Adler:2018aa,Adler:2017aa}. In case performance is not satisfactory, one may increase expressivity of the updating operator by either adding more convolutional layers or use a more expressive architecture like a U-Net for the updating operator \cite{hauptmann2018approximate}. Usually, one will observe a plateau in performance while increasing expressivity of the neural network, indicating a somewhat optimised choice.

\subsection{Risk of overfitting}
An advantage that learned iterative networks have over domain agnostic neural network architectures is that the former has much fewer learnable parameters (Section~\ref{sec:MotivationStructuredLearning}).
This reduces the risk for overfitting. 
Since training data is finite, empirical risk minimisation will lead to overfitting, e.g., the optimal parameter computed from training data will differ to one computed from all possible training data, and parameters that minimise the empirical risk will not be optimal for the expected risk. 

Classical statistical learning theory dictates that more parameters increases the risk of overfitting, although this has recently been called into question \cite{Cherkassky:2024aa,Lafon:2024aa,Schaeffer:2024aa}. Learned iterative networks  have advantageous properties over domain agnostic architectures since the number of learnable parameters is typically much smaller. In particular, this should also be taken into account when optimising the network architecture for performance as outlined above and hence out-of-distribution tests are necessary to properly judge performance.

\subsection{Complexity of implementation}\label{sec:ComplexityImplementation}
A specific challenge that arises when training a learned iterative network is to ensure the software libraries used for computing the handcrafted components are seamlessly integrated with the deep learning framework used for setting up and training the neural network. 
This means in particular that it must be possible to cast these handcrafted components as layers within the architecture for the learned iterative network. In addition, one must also ensure the chain rule works as expected when the network is differentiated with automatic differentiation.

Deep neural networks are typically trained with gradient methods that seek to compute a (local) minimiser to a highly non-linear and non-convex objective, see Appendix~\ref{app:SGD}.
This objective takes a large number of neural network parameters as input, so a gradient scheme typically involves computing gradients in a high dimensional space. One therefore uses automatic differentiation for this purpose, see \cite{Blondel:2024aa} for more details.
Here we simply note that usage of automatic differentiation has been greatly simplified \cite{Baydin:2018aa} thanks to well-maintained and highly optimised libraries that form the backbone of contemporary deep learning, like TensorFlow \cite{Abadi:2016aa} or PyTorch \cite{Paszke:2019aa}.
These allow for computing gradients of standard neural network components, but they very rarely implement a simulator, e.g., the implementation of a forward operator in inverse problems.
Hence, setting up and training a learned iterative network that incorporate a simulator would require one to seamlessly integrate external domain specific software components for the simulator (and the adjoint of its derivative) with frameworks for differentiable programming.
An option is to re-implement the simulator within a framework that supports differentiable programming, like JAX \cite{jax2018github,Lin:2024aa} or DiffTaichi \cite{Hu:2020aa}.
Examples of such differentiable simulators are \cite{Newbury:2024aa} for rigid-body simulation, \cite{Kakkar:2024aa,Zhao:2020aa} for rendering of complex light transport effects, and Mitsuba~3 \cite{Mitsuba3} for Monte Carlo based simulation of forward and inverse light transport.

However, rewriting highly optimised software libraries for specific simulators is often unpractical. This is especially the case when the library in question is extensive and highly optimised, like ASTRA \cite{vanAarle:2016aa,vanAarle:2015aa} for simulating tomographic imaging with various acquisition geometries and FEniCS/Dolphin \cite{Alnaes:2015aa,Logg:2012aa,Logg:2010aa} for finite element based simulation of systems modelled by (coupled) \acp{PDE}.
A far more pragmatic approach is to use the external libraries unaltered. Which requires one to either manually define the automatic differentiation rules within the learning libraries or a more streamlined approach is possible if one wraps the operators defined through the external libraries using some binding library, like ODL \cite{Adler:2017aa} or DeepInverse \cite{tachella2025deepinverse}.

In summary, the implementation of learned iterative methods involves another layer of computational complexity and hence often requires further domain knowledge. Consequently, it is not as accessible as an out-of-the-box neural network for image processing. This underlines the importance of open access codes accompanying papers which present new learned iterative methods as well as the development of designated software packages for learned reconstructions.

\subsection{Computational feasibility of training}\label{sec:CompFeasTrain}
Let us now discuss one major obstacle for training learned iterative reconstructions. That is, the involvement of the forward and adjoint operator in the training procedure, leading to potential extensive computational complexity. Ironically, as we discussed earlier, the involvement of model components also constitutes a major advantage of learned iterative reconstructions to include domain knowledge.

For the purpose of this section, let us quickly recall the basic setup for training a learned iterative reconstruction operator $\RecOp_\NNparam$.  
While conceptually any loss function in Section~\ref{sec:LearningProbs} is possible, most commonly the learned reconstruction operator $\RecOp_\NNparam$ is trained end-to-end given supervised training data $(\signal_i,\data_i) \in \RecSpace \times \DataSpace$ by minimising the empirical loss in \eqref{eq:SupDataX-Loss_emp}:
\begin{equation}\label{eqn:end2end}
\est{\NNparam}=\argmin_{\NNparam\in\NNparamSet}\sum_i \SignalLoss\bigr(\RecOp_{\NNparam}(\data_i),\signal_i\bigr).
\end{equation}

Computationally, solving \eqref{eqn:end2end} can be demanding, as it requires evaluation of the update directions, including the forward model $\FwdOp$ and computation of $\FwdOp^{\ast}$ (or $\bigl[\partial\!\FwdOp(\signal)\bigr]^{\ast}$ in the non-linear case) in each forward as well as backward pass $N$-times.
Hence, there are $4N$ operator evaluations for each training iteration, which can easily be in the order of ten thousands during a training session.
Thus, training of the learned reconstruction operator $\RecOp_{\NNparam}$ can pose a major computational challenge. Particularly, in case the operator $\FwdOp$ is computationally demanding, e.g., when evaluation times of $\FwdOp$ are larger than a second, then training of $\RecOp_\NNparam$ will easily take several days or even weeks. In particular, for non-linear problems where the Fréchet derivates, or Jacobians in the discrete setting, need to be recomputed for each unrolled as well as training iteration. 
Consequently, different training regimes are necessary to train the reconstruction operator $\RecOp_{\NNparam}$ for computationally expensive forward operator. 

One such option has is given by a greedy training approach to avoid the computational demanding evaluation in each training iteration \cite{Hauptmann:2018aa}, see also \cite{aghabiglou2024r2d2}.
Instead of the end-to-end loss above, only iterate-wise optimality is required, thus the terminology greedy; alternatively this may also be called sequential training. 
For the least-squares gradient network in \eqref{eq:LearnedGradient_withL2}, this leads to $N$ sequential training problems of the form
\begin{equation}\label{eqn:greedyTrain}
\est{\NNparam}_k \in \argmin_{\NNparam\in\NNparamSet}\sum_i \SignalLoss\biggl( \NNOp_{\NNparam_k}\Bigl(\signal^{k-1}_i,
  \FwdOp^{\ast}\bigl(\FwdOp\signal^{k-1}_i-\data_i
 \bigr) \Bigr), \signal_i\biggr),
\end{equation}
to obtain the final trained $\RecOp_{\est{\NNparam}}$ with $\est{\NNparam}=\{\est{\NNparam}_1,\dots,\est{\NNparam}_N\}$. 
For each training in \eqref{eqn:greedyTrain} we assume that the previous neural update operator $\NNOp_{\est{\NNparam}_{k-1}}$ for iteration $k-1$ has been trained. Then one can evaluate $\signal^{k-1}_i$ with the fixed network parameters $\est{\NNparam}_{k-1}$ by 
\begin{align*}
& \signal_i^{k-1} := \NNOp_{\est{\NNparam}_{k-1}}\Bigl(\signal^{k-2}_i,\FwdOp^{\ast}\bigl(\FwdOp\signal^{k-2}_i-\data_i \bigr) \Bigr).
\end{align*}
Additionally, the gradients 
$
\FwdOp^{\ast}\bigl(\FwdOp\signal^{k-1}_i-\data_i\bigr)
$
needed for \eqref{eqn:greedyTrain} 
can be pre-computed for all training pairs. 
Note, that for $k=1$ only the gradient for the initialisation $f^0$ is computed. This reduces the operator evaluations to $2Nn$ (with no backward pass), where $n$ is the number of training pairs.  
This greedy procedure allows decoupling of the network training from the evaluation of the model components and enables efficient training even for computationally demanding inverse problems, such as 3D photoacoustic tomography \cite{Hauptmann:2018aa}, where the evaluation of the forward operator takes more than 10 seconds, or in astronomical imaging \cite{aghabiglou2024r2d2}. Furthermore, it enables training learned iterative networks efficiently for non-linear inverse problems \cite{herzberg2021graph,Mozumder:2022aa,manninen2025towards}. Additionally, it simplifies the implementation with respect to the discussion in Section~\ref{sec:ComplexityImplementation}, as the implementation does not need to support the automatic differentiation. 

One should note, that the iterate-wise optimality in \eqref{eqn:greedyTrain} can be easily adjusted to include other formulations of learned iterative networks discussed in Section~\ref{sec:LearnedGradient} as well as the non-linear networks, as long as the loss function can be stated in the image space $\RecSpace$. That means, the greedy training objective does not easily generalise to primal-dual algorithms, since an optimal loss for the data space $\DataSpace$ is not well defined nor available, as it would  require ground-truth measurement data. Hence, it remains an open problem, how a greedy training could be implemented and in particular if it is useful for primal-dual architectures.

Alternatively, to the greedy training, one can aim to reduce the computational cost of the operator itself to enable feasible training times as well as faster inference. Clearly, one could learn the operator itself \cite{arridge2024inverse,Guan2023FNOPAT,Strom:2022aa}. Nevertheless, conceptually this amounts to introducing another learned component in the network and will remove the handcrafted nature of the domain adaptation, especially if generic neural operators are used. Another option to reduce the computational cost of the forward operator include approximations \cite{hauptmann2023model,hauptmann2018approximate,lunz2021learned,Strom:2022aa}, model-reductions or discretisations \cite{hauptmann2020multi}, as well as optimised numerical implementations for special cases \cite{Grohl2025digitaltwin,hauptmann2025fast}.

Another popular option is to decouple training of the network $\NNOp_\NNparam$ and evaluation of the forward operator by plug-and-play approaches as discussed in Section~\ref{sec:LearnedProximal}, here the updating network is trained as a denoiser independent of the iterative scheme, see \cite{kamilov2023plug} for a survey.

On a final note, one may also consider using gradient checkpointing during training. 
This significantly reduces the memory footprint during backpropagation that can easily become prohibitively large for a learned iterative network.
However, GPU computation is  delayed/paused during checkpoint saving for checkpoint GPU-CPU transfer, resulting in significant training interruptions and reduced training throughput. One can, to some extent, mitigate this slow down in training by considering specialised checkpointing strategies \cite{Zhang:2025ab}.

\section{Numerical examples}\label{sec:Comparison}
Let us illustrate some concepts in this survey with a numerical study on performance differences for various formulations of the reconstruction operator. One goal is to understand to what extend the different formulations in result in performance differences. Naturally, a comprehensive study on all choices is out of scope, but we aim to give a intuition here on the influence of such design choices. 

In the first part we will examine the discussed unrolling schemes for linear inverse problems under largely same training schemes. In the second part we will shortly present recent results for non-linear inverse problems and the influence of update directions for the learned reconstruction, together with a comparison of greedy vs. end-to-end training relevant for expensive non-linear problems.

\subsection{Comparison for linear problem}
We consider a standard reconstruction problem in sparse view and low-dose CT imaging in fan-beam geometry in two dimensions. The forward operator is given by the X-ray transform 
\[
\FwdOp\signal=\int_{\ell(\varphi,r)} \signal(s) \mathrm{d} s = \data,
\]
where the integration is w.r.t.\@ $s$, which is a coordinate on the $\ell(\varphi,r)$ that is parametrised by an angle $\varphi\in [0,2\pi)$ and location on the detector $r\in\Real$.
The geometry for each line is then defined by measurement specific variables describing distances between source, object, and detector. The forward operator is implemented using the Astra toolbox \cite{vanAarle:2016aa} supporting GPU computations.

\begin{figure}[h!]
    \centering
    {\includegraphics[width=0.38\textwidth]{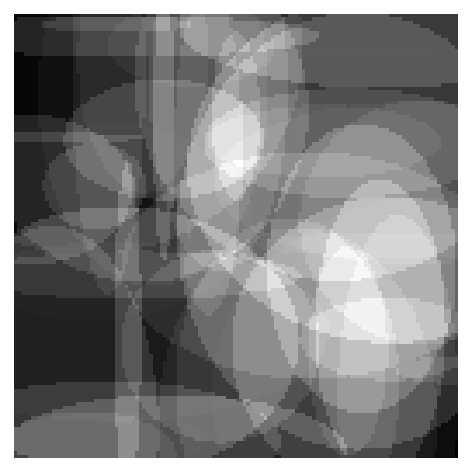}}
    {\includegraphics[width=0.22\textwidth]{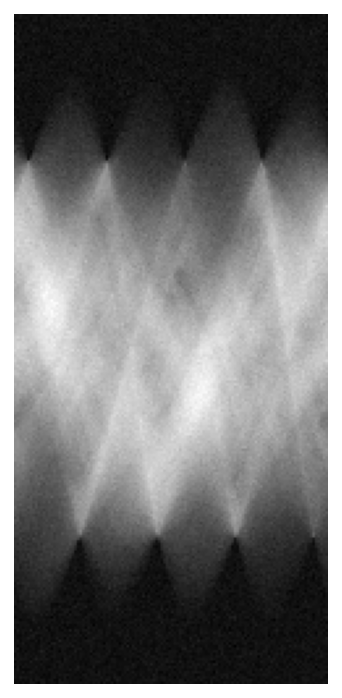}}
    {\includegraphics[width=0.38\textwidth]{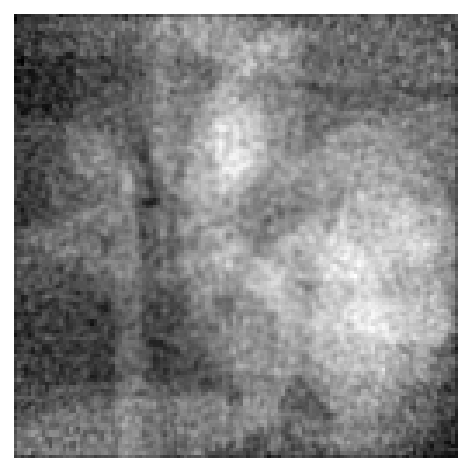}}
    \caption{Illustration of measurement setup. (Left) ground-truth, (Middle) sinogram with 120 angles and 2\% noise, (Right) Reconstruction via filtered backprojection.}
    \label{fig:fanbeam_data}
\end{figure}

For this study we chose 120 equidistant angles, 256 detection points on the detector, i.e., 256 lines per angle, and $2\%$ relative Gaussian noise. The target images are of resolution $128 \times 128$. We sample training data from a distribution of up to 50 random ellipses in the image and create a test set of fixed 50 samples for evaluating the performance after training. The loss functions are chosen to be fully supervised with the mean squared error
\[
\est{\NNparam}=\argmin_{\NNparam\in\NNparamSet}\sum_i \|\RecOp_{\NNparam}(\data_i)-\signal_i\|_\RecSpace^2,
\]
where training samples are drawn from randomly generated ellipses sampled new for each training iteration. An illustration of a ground-truth image $\signal_i$, obtained noisy measurement $\data_i$, and initial  reconstruction by filtered backprojection $\signal^0_i=\RecOp(\data_i)$ are shown in Figure~\ref{fig:fanbeam_data}.

We will consider the previously presented options for learned reconstruction operators $\RecOp_\theta\colon \DataSpace\to\RecSpace$ for the linear inverse problem. That is, the three classes for learned gradient networks, namely learned least squares networks~\ref{sec:LearnedLSQ}, proximal networks~\ref{sec:LearnedProximal}, and variational networks~\ref{sec:VarNets}. That is we have the following update rules with $ \grad\operator{Q}_{\data}(\signal) =\FwdOp^\ast(\FwdOp\signal-\data)$ in respective order: 
\begin{align}
    \signal^k &:= \NNOp_{\NNparam_k}\Bigl(\signal^{k-1},
  \grad\operator{Q}_{\data}(\signal^{k-1}) \Bigr) \\
 \signal^k &:= \NNOp_{\NNparam_k}\bigl(
 \signal^{k-1} - \omega \grad\operator{Q}_{\data}(\signal^{k-1})
 \bigr) \\
 \signal^k &:= 
 \signal^{k-1} - \omega \grad \operator{Q}(\signal^{k-1},\data) + \NNOp_{\NNparam_k}(\signal^{k-1}).
\end{align}
For the two last updates, we also learn the step-size parameter $\omega$ shared for all iterates.
Additionally, we consider the \ac{LPD} as presented in Section~\ref{sec:LPD} with two update networks: 
\begin{equation}
\begin{cases}
\data^{k} := \NNOpOther_{\NNparamOther_{k}}\bigl(\data^{k-1}, \FwdOp\signal^{k-1},\data \bigr)
& \\[0.5em]
\signal^{k} := \NNOp_{\NNparam_{k}}\bigl(\signal^{k-1} , \FwdOp^{\ast}\data^{k} \bigr).
&
\end{cases}
\end{equation}
Each reconstruction operator starts with the initial reconstruction given by the filtered backprojection $\signal^0_i=\RecOp(\data_i)$. The result of the learned reconstruction operator is then given by the final iterate $\RecOp_\theta(\data_i)=\signal_i^N$, depending on the corresponding update equation above. 
We have made an effort to choose the networks and parameter count largely the same, as well as training protocols. We note, that better results may be obtained if training protocols and parameter choices are optimised for each algorithm separately. Additionally, the implementation may differ from the original published versions. Nevertheless, the purpose here is to present a direct comparison under same environmental choices.

The subnetworks $\NNOp_\theta$ and $\NNOpOther_{\NNparamOther}$ for all approaches are chosen as ResNet \cite{He2016ResNet} style with 5 convolutional layers in total, the first layer expands the input to 32 channels, followed by 3 layers with each 32 channels, and a final layer collapsing to one output channel. Each layer has $3\times 3$ convolutional kernels, bias, and ReLU as non-linearity, except the final layer, which does not employ a non-linearity important for the update. All learned reconstruction operators are trained for 25000 iterations, with initial learning rate of $10^{-3}$ and cosine annealing. The performance is then evaluated at the end for the test ellipses. The results of the trained networks are summarised in Table~\ref{tab:recOper_comparison}.
\begin{table}[h!]
\centering
\caption{Comparison of learned reconstruction operators based on different unrolling schemes. \Acf{PSNR} is reported as average over 50 test samples.}
\small{\par\smallskip
\begin{tabular}{l|rrr|rr}
   & \multicolumn{1}{c}{\textbf{VarNet}}  
   & \multicolumn{1}{c}{\textbf{Proximal}} 
   & \multicolumn{1}{c|}{\textbf{Least Sq.}} 
   & \multicolumn{2}{c}{\textbf{LPD}}  
\\
\hline
Iterations & 10 & 10 & 10 & 10 & 5 \\
Parameter  & 283531  & 283531 & 286411 & 575710 & 287855 \\
Training (min.) & 48  & 48  & 49  & 56  & 35  \\
\Acs{PSNR} (dB) & 30.88  & 30.91  & 30.96  & 31.11  & 31.02 
\end{tabular}
\label{tab:recOper_comparison}}
\end{table}

We can see from the quantitative results that the reconstruction quality for all methods is rather close. 
Especially all learned gradient networks perform very similar, whereas \ac{LPD} provided a small increase in performance with the same number of iterations. For the gradient networks it is interesting to observe that the performance increases with decreasing structure, that means the performance decreases, albeit only slightly, with more handcrafted components. In other words, the more freedom the network has to learn, the better it performs. It should be noted though that the test data is in distribution and out-of-distribution generalisation may differ. 

For \ac{LPD} we see a clear improvement in performance, but under the same amount of iterations also the number of parameters doubles due to the two networks, one in data space $\DataSpace$ and one in image space $\RecSpace$. Thus, we have additionally performed a test with half the iterates, so that \ac{LPD} has a similar parameter count as the gradient networks. Interestingly, \ac{LPD} still provides the best reconstructions in terms of \ac{PSNR}, but with a smaller improvement over the learned least squares network.

\begin{figure}[h!]
    \centering
    {\includegraphics[width=0.3\textwidth]{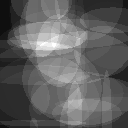}}
    {\includegraphics[width=0.3\textwidth]{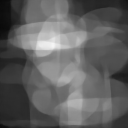}}
    {\includegraphics[width=0.3\textwidth]{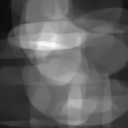}}
    {\includegraphics[width=0.3\textwidth]{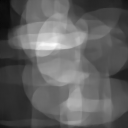}}
    {\includegraphics[width=0.3\textwidth]{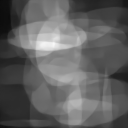}}
    {\includegraphics[width=0.3\textwidth]{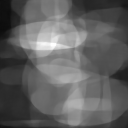}}
    
    \caption{Comparison of reconstructions for the trained networks for one sample in the test data. From left to right and top to bottom: Ground truth, variational network, proximal network, learned least squares, \ac{LPD} (5 iterates), \ac{LPD} (10 iterates).}
    \label{fig:learnedCTrecs}
\end{figure}

Finally, we present the reconstructions obtained with each network in Figure~\ref{fig:learnedCTrecs}. We can see that there is a clear difference in qualitative visual performance between the networks. The three gradient networks produce blurrier images, and \ac{LPD} does provide a sharper reconstruction. The 10 iteration reconstruction with \ac{LPD} is clearly the visually best performing, whereas the Variational Network result fails to recover many fine details. This is also reflected in the quantitative results.

\subsection{Comparison for non-linear problem}

The research on unrolled methods for non-linear inverse problems is much younger with few examples \cite{Alberti:2025aa,Colibazzi:2022aa,herzberg2021graph,Mozumder:2022aa,manninen2025towards}. Where some approaches have considered a greedy training \eqref{eqn:greedyTrain} to overcome the computational burden, as discussed in Section~\ref{sec:LearnedLSQ}. Additionally, in the non-linear case a large class of update directions is available to train the unrolled network. Hence, one major question is how the choice of update directions influence the performance of the learned reconstruction operator and how greedy training compared to end-to-end training influences the performance.

In the following we shortly present results from a recent study \cite{manninen2025towards}, which compares different update directions and training regimes for the non-linear inverse problem of \ac{QPAT}. 
For this purpose let us shortly discuss the basics of \ac{QPAT}, where the goal is to recover the spatially distributed optical parameters of absorption $\mu_a$ and scattering $\mu_s$ by modelling the fluence $\Phi$ in biological tissue after illumination with a short laster pulse. 
The accurate model for light transport in this case is given by the radiative transfer equation, but to allow for efficient computations and training of the unrolled schemes the study \cite{manninen2025towards} considers an elliptic approximation. That is, the underlying model equation in the domain $\Omega$ is then given by the diffusion approximation 
\begin{equation}\label{eq:qpat}
    -\nabla \cdot \kappa(r) \nabla \Phi(r)+\mu_a(r) \Phi(r)=q_0(r), \quad r \in \Omega,
\end{equation}
where $q_0$ is the light source in $\Omega$ and the parameter $\kappa(r) = (n(\mu_a(r) + \mu'_{s}(r)))^{-1}$ in dimension $n=2,3$ is the diffusion coefficient where $\mu'_s =(1-g)\mu_s$ is the reduced scattering coefficient with anisotropy parameter $-1<g<1$. Under suitable boundary conditions \cite{arridge1999optical}. 

The inverse problem in \ac{QPAT} is now to recover $\mu_a,\mu_s$ in $\Omega$ from internal data. That is, the knowledge of the absorbed energy density $h\in\Omega$ given by
\begin{equation}\label{eq:E_density}
    h=\mu_a \Phi(\mu_a,\mu_s),
\end{equation}
where $\Phi(\mu_a(r),\mu_s(r))$ is the solution to the diffusion approximation \eqref{eq:qpat}. This internal data $h$ can be obtained by solving the acoustic problem first, see \cite{beard} for a review. 

The inverse problem \eqref{eq:E_density} is usually solved by second order methods, such as Gauss-Newton and quasi-Newton methods. The paper \cite{manninen2025towards} formulates a learned reconstruction operator to robustly recover the optical parameters considering that the diffusion approximation is not an accurate model. Additionally, it investigates how different choices for update directions and training schemes for the learned reconstruction operator in \eqref{eq:LearnedGaussNewton} effect the reconstruction quality. Specifically, the study compares choices learned gradient descent (Section~\ref{sec:non-linearGrad}), learned Gauss-Newton (Section~\ref{sec:GaussNewton}), and learned quasi-Newton with an SR1 update (Section~\ref{sec:QuasiNewton}) for $\triangleSignal$.
The learned reconstruction operator for all three cases is given by
$\RecOp_{\NNparam}(\data) := \signal^{\UnrollIter}$ where
\begin{equation*}
\begin{cases}
 \signal^0 := \RecOp_0(\data) &
 \\[0.5em]
 \signal^k := \NNOp_{\NNparam_k}\bigl(\signal^{k-1}, \triangle\signal^{k-1} \bigr) 
 & \text{for $k=1,\ldots,\UnrollIter$,}
\end{cases}
\end{equation*}
where $\NNOp_{\NNparam_k}\colon\RecSpace\times\RecSpace\to\RecSpace$ uses a simple \ac{ResNet} style architecture with 4 convolutional layers of 32 channels except the last one which contracts to the singular output channel, similar to the linear inverse problem experiments above.
Training is performed for both, end-to-end and greedy, given supervised training data with increasing amount of iterations. Additionally, a comparison to a simple U-Net post-processing is done as baseline. 
\begin{figure}[ht!]
    \centering
    \includegraphics[width=0.47\textwidth]{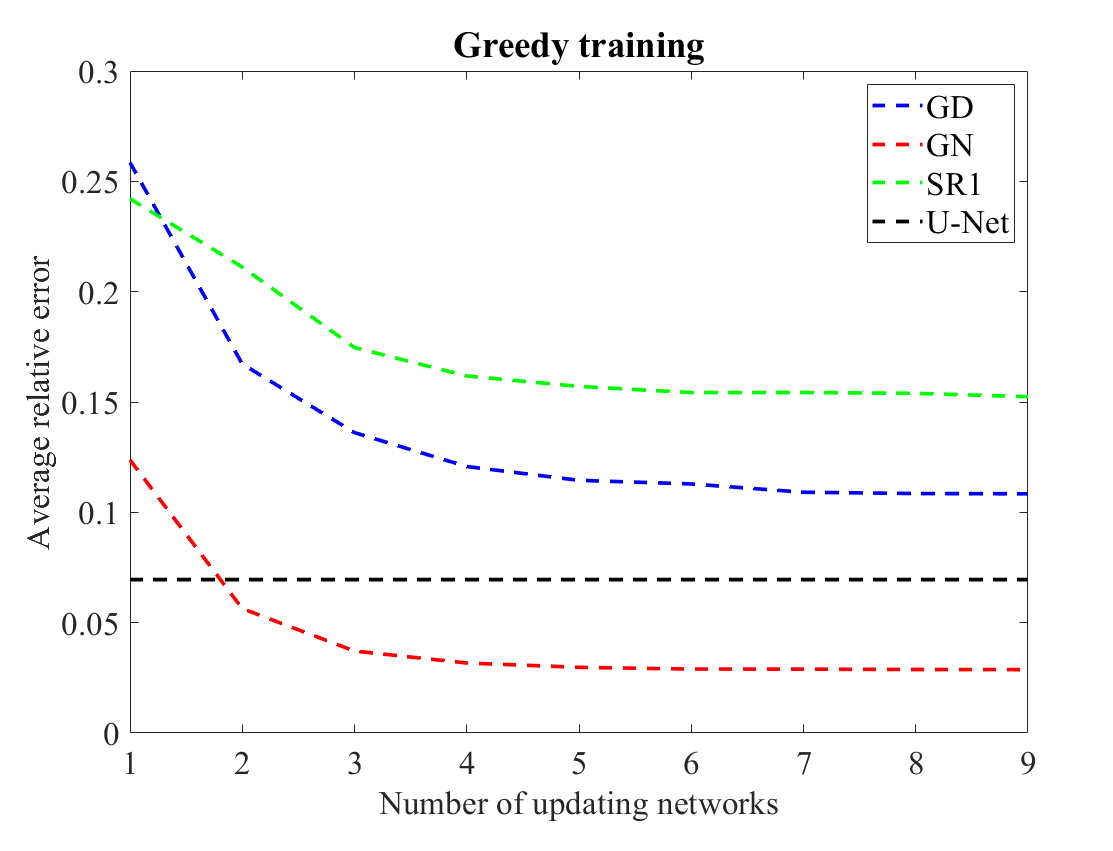}
    \hfill
    \includegraphics[width=0.47\textwidth]{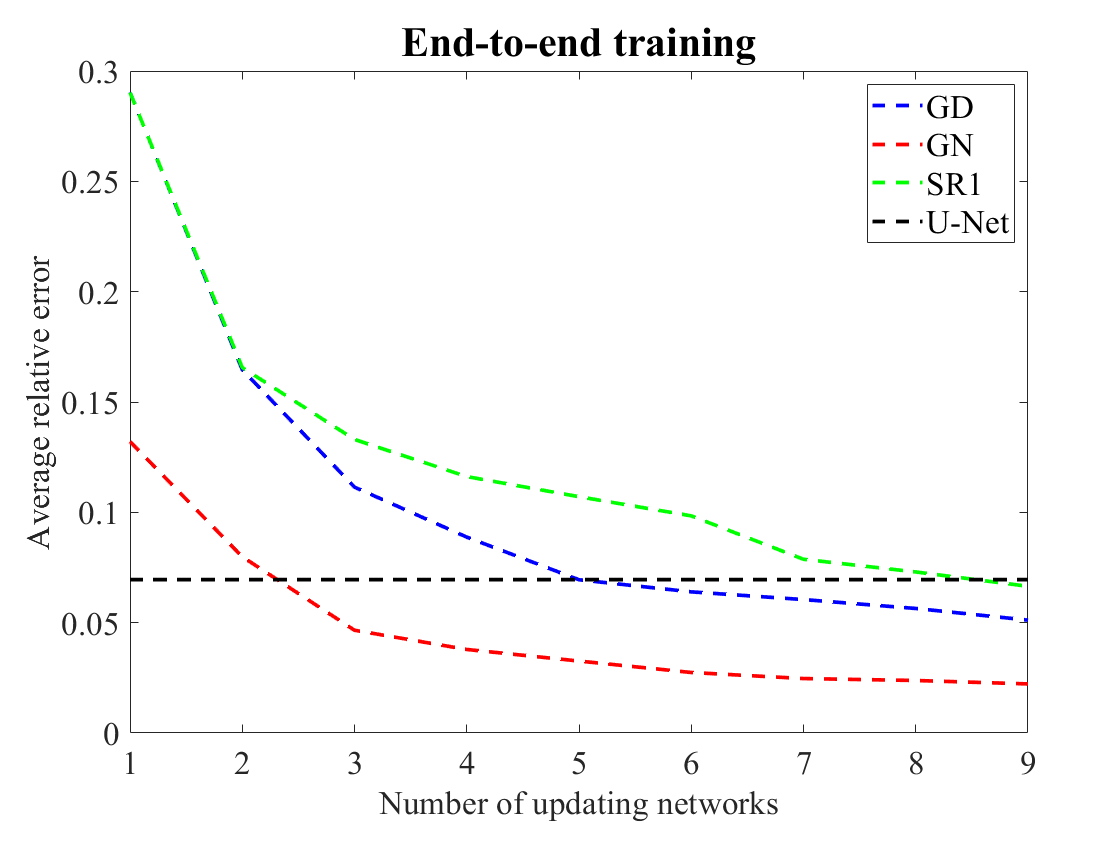}
    \caption{Comparison of training update directions and training schemes for learned reconstructions in \ac{QPAT}. Shown is the relative error in reconstructed absorption coefficients $\mu_a$ for (Left) greedy training schemes and (Right) end-to-end, with three choices for the handcrafted update direction $\triangleSignal$ as gradient descent (GD), Gauss-Newton (GN), and quasi-Newton SR1 update (SR1).}
    \label{fig:non-linearComp}
\end{figure}

The results are presented in Figure~\ref{fig:non-linearComp} for the absorption coefficient $\mu_a$ only, where the scattering behaves. There are a few takeaways from the experiment. First, the learned Gauss-Newton performs well and largely similar for greedy and end-to-end training. Second, gradient descent and the quasi-Newton method does not perform satisfactory in greedy training. This is consistent with previous unreported experiments, that greedy training with gradient descent updates did not perform well for the study in \cite{Mozumder:2022aa}. Third, when trained end-to-end all algorithms are able to beat the U-Net baseline, but need larger amount of iterations. In conclusion this study indicates that learned Gauss-Newton clearly outperforms the other update directions and reinforces the use in previous papers. We refer to \cite{manninen2025towards} for the experiments and an example for a digital twin introducing large modelling error that will be present in biological tissue.

\section{Theoretical analysis}\label{sec:TheoryFoundations}
Many tasks, like those outlined in Sections~\ref{sec:InvProb} and \ref{sec:OptimSolver}, involve continuum function data as inputs or outputs. 
They are therefore mathematically formalised as evaluating a target operator $\RecOp \colon \DataSpace \to \RecSpace$.
The theoretical analysis of methods for learning such a target operator is therefore also naturally carried out within a functional analytic framework. 
The importance of this functional analytic perspective is emphasised in Section~\ref{sec:MotivationFuncAnaly} and it is also a perspective that aligns with contemporary numerical analysis, where theoretical analysis is closely intertwined with algorithmic development. 

In the following, we provide a brief survey of theoretical analyses for operator learning. 
A central theme is the characterisation of the expressiveness of neural operator architectures (through universal approximation theorems) and to derive parameter/error bounds (Section~\ref{sec:ApproxError}). 
Another important line of work concerns trainability and the extent to which learned operators generalise beyond the training data (Section~\ref{sec:GenGap}). 
Next we have sample complexity bounds (Section~\ref{sec:SampComplexity}) and computational complexity (Section~\ref{sec:CompComplexity}).
Role of discretisation invariance is discussed in (Section~\ref{sec:DiscInv}). 
We conclude with stability (Section~\ref{sec:Stability}), which is yet another related topic that is especially important when learning reconstruction operators for ill-posed inverse problems (Section~\ref{sec:InvProb}).
Many of the above topics are closely connected to different sources of error. 
A natural starting point to set the stage is therefore to decompose the total approximation error in operator learning into components that can be analysed separately, which is done in Section~\ref{sec:TypeOfErrors}.

On a final note, as will become clear, a great majority of the results we cite in the subsequent sections concern neural operators for solving \acp{PDE} where the architectures differ from those given by learned iterative networks.
For this reason, they do \emph{not} directly apply to operator learning that is our focus, which is usage of learned iterative networks for solving inverse problems (Section~\ref{sec:InvProb}) or optimisation problems (Section~\ref{sec:OptimSolver}).
Despite this, we decided to include this material as it illustrates the type of results one can state and prove in a theoretical analysis. 
However, this also means we refrain from giving the detailed statements of the results as these typically involve many technical conditions that would require much space to describe.
In contrast we do put specific emphasis on those results that do apply to our setting, which is operator learning with learned iterative networks for solving inverse problems or optimisation problems.

\subsection{Decomposing the total error}\label{sec:TypeOfErrors}
The error in approximating a target operator $\RecOp$ with a neural operator $\{ \RecOp_{\NNparam} \}_{\NNparam}$ is often decomposed into parts. 
In the following we illustrate this decomposition in the case of supervised learning, similar decompositions can be done for the other learning tasks in Section~\ref{sec:LearningProbs}. 

Training data is in supervised learning of the form in \eqref{eq:SupervisedData}, i.e., it is independent random draws $(\signal_i,\data_i) \in \RecSpace \times \DataSpace$ from the $(\RecSpace \times \DataSpace)$-valued random variable $(\signalrand,\datarand)$ where $\signalrand=\RecOp(\datarand)$.
This defines an empirical probability measure $\est{\mu}$ on $\RecSpace \times \DataSpace$ by 
\begin{equation}\label{eq:EmpMeas}
  \est{\mu}(W) = \frac{1}{m} \sum_{i=1}^m \delta_{(\signal_i,\data_i)}(W)
  \quad\text{for any measurable $W \subset \RecSpace \times \DataSpace$.}
\end{equation}
We next consider the approximation error in operator learning for a given empirical measure $\est{\mu}$ that is defined by training data as in \eqref{eq:EmpMeas}:
\begin{equation}\label{eq:TotErrorSupervised}
\ApproxError_{\text{tot}}(\est{\mu}) = \Expect_{(\signalrand,\datarand) \sim \mu}\Bigl[ \bigl\Vert \RecOp_{\LearnAlg(\est{\mu})}(\datarand) - \signalrand \bigr\Vert_{\RecSpace} \Bigr].
\end{equation}
Here, $\mu$ is the probability measure on $\RecSpace \times \DataSpace$ for data generated by the target operator, i.e., it defines the distribution for the $(\RecSpace \times \DataSpace)$-valued random variable $(\signalrand,\datarand)$ where $\RecOp(\datarand)=\signalrand$.
Finally, $\LearnAlg$ is the learning algorithm, which maps the empirical measure (which is a probability measure on $\RecSpace \times \DataSpace$) to a parameter in $\NNparamSet$ that in turn defines a neural operator in the hypothesis space. 

One can now bound and decompose the approximation error in \eqref{eq:TotErrorSupervised} as:
\begin{equation}\label{eq:ErrorDecomp}
\begin{split}
\ApproxError_{\text{tot}}(\est{\mu}) &=
    \Expect_{(\signalrand,\datarand) \sim \mu}\Bigl[ 
        \bigl\Vert \RecOp_{\LearnAlg(\est{\mu})}(\datarand) - \est{\RecOp}(\datarand) 
        + \est{\RecOp}(\datarand) - \signalrand \bigr\Vert_{\RecSpace} 
      \Bigr]
 \\
 &\leq 
    \Expect_{\datarand \sim\mu_{\DataSpace}}\Bigl[ 
      \bigl\Vert \RecOp_{\LearnAlg(\est{\mu})}(\datarand) - \est{\RecOp}(\datarand) \bigr\Vert_{\RecSpace} 
    \Bigr]
    + \Expect_{(\signalrand,\datarand) \sim \mu}\Bigl[ 
          \bigl\Vert \est{\RecOp}(\datarand) - \signalrand \bigr\Vert_{\RecSpace} 
       \Bigr]
 \\
 &= 
      \underbrace{\Expect_{(\signalrand,\datarand) \sim\mu}\Bigl[ \bigl\Vert \est{\RecOp}(\datarand) - \signalrand \bigr\Vert_{\RecSpace} \Bigr]}_{\ApproxError_{\text{intr}}}
   + \underbrace{\Expect_{\datarand \sim\est{\mu}_{\DataSpace}}\Bigl[ 
         \bigl\Vert \RecOp_{\LearnAlg(\est{\mu})}(\datarand) - \est{\RecOp}(\datarand) \bigr\Vert_{\RecSpace} 
       \Bigr]}_{\ApproxError_{\text{approx}}}
 \\
 &\quad\,
   + \underbrace{\Expect_{\datarand \sim\mu_{\DataSpace}}\Bigl[ \bigl\Vert \RecOp_{\LearnAlg(\est{\mu})}(\datarand) - \est{\RecOp}(\datarand) \bigr\Vert_{\RecSpace} \Bigr]
   - \Expect_{\datarand \sim\est{\mu}_{\DataSpace}}\Bigl[ \bigl\Vert \RecOp_{\LearnAlg(\est{\mu})}(\datarand) - \est{\RecOp}(\datarand) \bigr\Vert_{\RecSpace} \Bigr]}_{\ApproxError_{\text{gen}}}.
\end{split}
\end{equation}
Here, $\mu_{\DataSpace}$ and $\est{\mu}_{\DataSpace}$ are the $\DataSpace$-marginals of $\mu$ and $\est{\mu}$, respectively.
Finally, the mapping $\est{\RecOp} \colon \DataSpace \to \RecSpace$ is the optimal statistical estimator (best possible learned operator), which in supervised learning is a solution to \eqref{eq:SupDataX-Loss} when the hypothesis space equals the set of all $\RecSpace$-valued measurable mappings on $\DataSpace$:
\begin{equation}
  \est{\RecOp} \in \argmin_{\RecOpOther \colon \DataSpace \to \RecSpace} 
   \Expect_{(\signalrand,\datarand)\sim \mu}\Bigl[ 
    \Loss_{\RecSpace}\bigl( \RecOpOther(\datarand), \signalrand \bigr)
   \Bigr].
\end{equation}

Each of the terms appearing in the right-hand-side of \eqref{eq:ErrorDecomp} have an interpretation that we now provide: 
\begin{itemize}
\item $\ApproxError_{\text{approx}}$ is the \emph{approximation error}. 
  It depends on the hypothesis space, and in particular on the model capacity of the corresponding neural operator architecture(s).
  
  A closely related and highly relevant analysis is to derive parameter/error bounds and quantitative error rates.
  This is possible for selected architectures and target operators, e.g., in the case of using \acp{DeepONet} or \acp{FNO} (Section~\ref{sec:ApproxError}) to solve \acp{PDE}.
  However, making the connection to $\ApproxError_{\text{approx}}$ requires one to include the approximation error introduced by the training algorithm, and this is often left out.

\item $\ApproxError_{\text{gen}}$ is the \emph{generalisation error} that quantifies how well a trained neural operator generalises to unseen test data. 
  This term is especially important when test and training data come from different statistical distributions.
  
  The generalisation error is closely related to the \emph{generalisation gap}, which in supervised learning is commonly defined as
  \begin{equation}\label{eq:GenGap} 
      \operatorname{GenGap}(\NNparam) = 
     \Expect_{(\signalrand,\datarand) \sim\mu}\Bigl[ 
       \Loss_{\RecSpace}\bigl( \RecOp_{\NNparam}(\datarand),  \signalrand \bigr) 
     \Bigr]
     - \Expect_{(\signalrand,\datarand) \sim \est{\mu}}\Bigl[ 
         \Loss_{\RecSpace}\bigl( \RecOp_{\NNparam}(\datarand), \signalrand \bigr) 
     \Bigr].
  \end{equation}  
  The generalisation error commonly refers to $\operatorname{GenGap}(\est{\NNparam})$ where $\est{\NNparam}$ is assumed to solve \eqref{eq:SupDataX-Loss_emp}, i.e., one disregards the error from the training algorithm.
  Estimates for the generalisation gap has been derived for selected architectures.
  It is therefore tempting to relate $\ApproxError_{\text{gen}}$ to $\operatorname{GenGap}\bigl(\LearnAlg(\est{\mu})\bigr)$.
  As with the approximation error, a difficulty is to handle the additional approximation error from the training algorithm that is introduced by setting $\NNparam = \LearnAlg(\est{\mu})$ in \eqref{eq:GenGap}. 
  Another difficulty is that the loss function $\Loss_{\RecSpace}$ in the learning \eqref{eq:SupDataX-Loss}, and consequently in the generalisation gap in \eqref{eq:GenGap}, does not necessarily coincide with the $\RecSpace$-norm used for defining the error in \eqref{eq:TotErrorSupervised}.

\item $\ApproxError_{\text{intr}}$ is the \emph{intrinsic error} of the operator learning task that is \emph{independent} of choice of \emph{training data}, \emph{neural network architecture}, and \emph{learning algorithm}.

In the context of solving an inverse problem (Section~\ref{sec:InvProb}), this term depends on the degree of ill-posedness. Statistical regularisation theory provides estimates for this term, typically via concentration rates and/or through a characterisation of the posterior’s microscopic fluctuations. These results are often problem-specific, depending in particular inverse problem (target operator); see \cite{Dashti:2017aa} for results on inverse problems with linear forward operators, and \cite{Nickl:2017aa,Nickl:2023aa} survey various results of this type for certain non-linear \ac{PDE} solution operators arising in \ac{PDE} parameter estimation.
\end{itemize}
\begin{remark}
An alternative decomposition is to replace $\est{\RecOp}$ in \eqref{eq:ErrorDecomp} with $\RecOp_{\NNparam^*}$ where $\NNparam^* \in \NNparamSet$ solves \eqref{eq:SupDataX-Loss}. 
Then $\RecOp_{\NNparam^*}$ is the optimal neural operator within the hypothesis space given by the involved neural operator architecture(s).
\end{remark}

An important part of the analysis is the choice of architecture(s) for the neural operators that specify the parametrisation $\{ \RecOp_{\NNparam} \}_{\NNparam \in \NNparamSet}$ of the neural operators $\RecOp_{\NNparam} \colon \DataSpace \to \RecSpace$ (hypothesis space) one searches over during the learning.
Another is the \emph{training algorithm} that is used for finding the `best' parameter in $\NNparamSet$ given the training data.
This is essentially the algorithm (often an iterative scheme) that approximately solves the learning problem.

This is a well-developed research topic in the finite dimensional setting with convergence results as nicely surveyed in \cite[Chap.~6-7]{Jentzen:2025aa}, but much less has been done in the infinite dimensional setting \cite{Kereta:2026aa}.
Formally we can view it as a mapping that takes the (empirical) probability measure associated with training data to a parameter in $\NNparamSet$, which in turn yields the learned neural operator in the hypothesis space.
The probability measure that come into consideration are those that generate training data, so this depends on the learning task one has chosen (Section~\ref{sec:LearningProbs}). 

\subsection{Approximation properties}\label{sec:ApproxError}
Most of the theoretical analysis related to approximation properties do not directly concern the approximation error $\ApproxError_{\text{approx}}$ in \eqref{eq:ErrorDecomp}.
Focus has instead been on the following three closely related topics: universal approximation, parameter/error bounds, and quantitative error rates. 

Universal approximation holds when one can approximate a target operator, or a class of target operators, to any level of accuracy with some neural operator in the hypothesis space.
Parameter/error bounds is a further refined analysis that provides bounds on the number of neural operator parameters (for some architecture) that are needed to achieve a desired approximation accuracy. 
This is complemented by quantitative rates for the decrease in the approximation error as number of model parameters increases (error rates).

Such results are typically formulated for specific architectures and specific target operators. 
The broad picture is that universal approximation holds for several neural operator architectures, like feed-forward neural operators  \cite{Chen:1995aa,Lu:2021aa,Kovachki:2023aa}, \ac{DeepONet} \cite{Chen:1995aa,Lu:2021aa,Lanthaler:2022aa}, PCA-NET \cite{Lanthaler:2023aa}, and \ac{FNO} \cite{Kovachki:2021aa} that where introduced for solving \acp{PDE} (Section~\ref{sec:OtherNeuralOpArch}). 
More recently, there has been an attempt at proving universal approximation results for broader classes of architectures that has \ac{DeepONet} and \ac{FNO} as special cases \cite{Kovachki:2023aa,Lanthaler:2025ab,Godeke:2025aa,Zappala:2025aa}.
Sharp parameter/error bounds are more difficult to derive in general setting, e.g., such results also depend strongly on the type of target operator one seeks to approximate.
In particular, such bounds are strongest when the target operator has additional structure, such as linearity, spectral compressibility, holomorphy, or regularity, the latter in cases when the target operator represents the solution operator for a well posed specific \ac{PDE}. 
As surveyed in \cite{Boulle:2024aa,Shin:2024aa,Luo:2024ab}, such theoretical analysis is especially well-developed in the setting of using \ac{FNO} and \ac{DeepONet} based neural operators for solving well-posed (often linear) \acp{PDE}, as shown in \cite{Kovachki:2021aa,Duruisseaux:2026aa} for \acp{FNO}, \cite{Lanthaler:2023aa} for PCA-NET, and \cite{Deng:2022aa,Ryck:2022aa,Lanthaler:2022aa,Shin:2024aa} for \acp{DeepONet}. 
There are also results in this direction for feed-forward neural operators \cite{Yarotsky:2017aa,Lanthaler:2022aa}.

The references cited above focus mainly on solving specific well-posed \acp{PDE} using either a \ac{DeepONet} or an \ac{FNO}. 
This differs from both the neural operator architectures and the target operators considered in this survey, which is on using learned iterative networks to solve ill-posed inverse problems (Section~\ref{sec:InvProb}) and optimisation problems (Section~\ref{sec:OptimSolver}).
The authors are unaware of any such analysis specifically directed for this setting.
Universal approximation is a natural starting point, as such results typically hold for broad classes of target operators. 
Universal approximation should hold trivially for learned iterative networks when the neural update operators themselves are universal approximators. 
The more interesting case arises when this assumption does not hold.
In that case, an analysis of whether universality holds must incorporate the unrolling.
In the finite dimensional setting, universal approximation holds for operator recurrent networks \cite[Thm.~2.1]{Hoop:2021aa}, which is a specific type of learned iterative network mentioned in Section~\ref{sec:LearnedIterNetw}. 
One can in addition provide parameter/error bounds \cite[Thm.~2.2]{Hoop:2021aa}.

\subsection{Trainability and generalisation}\label{sec:GenGap}
Trainability concerns how effectively the learning problem (Section~\ref{sec:LearningProbs}) can be solved, while generalisation focuses on characterising the generalisation error $\ApproxError_{\text{gen}}$ in \eqref{eq:ErrorDecomp}, i.e., how well a trained neural operator performs on unseen test data.
This is particularly relevant when the test and training data follow different statistical distributions. 

As with the approximation properties (Section~\ref{sec:ApproxError}), most of the theoretical analysis consider supervised learning (Section~\ref{sec:Supervised}). 
Moreover, they do not directly target $\ApproxError_{\text{gen}}$.
Instead, focus is on analysing closely related notions, like the loss landscape and bounding the generalisation gap. 

\paragraph{Loss landscape}
Characterising the (non-convex) landscape for the optimisation for the learning problem in \eqref{eq:SupDataX-Loss} or \eqref{eq:SupDataX-Loss_emp} is a central topic for understanding trainability and generalisation.
This landscape depends to large extent on the choice of loss function and the neural operator architecture. 

In the finite-dimensional setting of supervised neural network training, this landscape plays a fundamental role in determining the trajectory and efficiency of the training process, as well as the generalisation properties of the resulting network.
Surveys such as \cite{Vidal:2022aa,Marchetti:2025aa} demonstrate that exploiting structural properties of the loss landscape can lead to more effective and robust neural network training strategies. 
For instance, the presence of non-differentiable saddle points and suboptimal local minima can significantly complicate training. 
Furthermore, loss landscapes associated with common network architectures and loss functions has a multiscale structure \cite{Ma:2022aa}. 

None of the above applies to learned iterative networks. 
In fact, to the best of the authors’ knowledge, there are no characterisations of the loss landscape for learned iterative networks.
Furthermore, there seems to be no results for characterising the loss landscape in the infinite dimensional setting, i.e., for operator learning.

\paragraph{Bounding the generalisation gap}
Theoretical results on generalisation typically provides bounds for the generalisation gap defined in \eqref{eq:GenGap}, which is closely related to $\ApproxError_{\text{gen}}$ in \eqref{eq:ErrorDecomp}.

One reason is that the generalisation gap is easier to bound, as it does not involve estimating the error introduced by the training algorithm. 
This also means that such bounds isolate how generalisation is influenced by the choice of architecture. 
Another important reason is that the generalisation gap itself provides meaningful insight: it serves as an indicator of both overfitting and predictive reliability.
A small gap suggests that the neural operator performs similarly on both training and unseen data, indicating good generalisation. In contrast, a large positive gap implies that the model fits the training data significantly better than the underlying distribution, signalling overfitting. Finally, a near-zero gap indicates that the empirical learning problem \eqref{eq:SupDataX-Loss_emp} closely approximates the true learning problem \eqref{eq:SupDataX-Loss} where one is assumed to have access to the joint distribution of the random variable $(\signalrand,\datarand)$ generating the training data.

In the finite-dimensional setting, there are a wide range of bounds of the generalisation gap for neural networks \cite{Kawaguchi:2022aa,Berner:2022aa}. 
These bounds typically involve terms that quantify `model complexity', like VC dimension and Rademacher complexity.
While these notions of model complexity were originally developed for neural network classifiers, several of them can also be used for regression problems as they extend to classes of real-valued functions.

Within this finite-dimensional setting, there are bounds of the generalisation gap for operator recurrent networks in the context of solving inverse problems \cite{Hoop:2021aa}. 
These are special cases of learned iterative networks and the target operator $\RecOp$ is here assumed to be continuous non-linear operator function of the form \cite[eq.~(115)]{Hoop:2021aa}.
The bounds are expressed in terms of either Rademacher complexity or metric entropy (that bounds the Rademacher complexity of the operator class in a Hilbert space setting). 
More specifically, the generalisation gap is in \cite[Thm.~5.5]{Hoop:2021aa} bounded as
\[
  \operatorname{GenGap}(\est{\NNparam})
  \leq C \frac{\operatorname{Comp}}{\sqrt{m}}
  \quad\text{with $\est{\NNparam}$ solving \eqref{eq:SupDataX-Loss_emp},}
\]
and where $m$ in the denominator is the number of training data points and Comp in the numerator, which is the complexity of the hypothesis space, scales linearly with the norms of the weights and exponentially with the number of unrolled iterates $L$.
These bound also remain valid as the discretisation is refined, meaning the error does not `blow up' as you increase the resolution of the signal (resolution-independence) \cite[Remark~14]{Hoop:2021aa}.
In addition, it is possible to estimate the expected performance gap between the optimal network and the network obtained from training data \cite[Thm.~5.6 and Remark~14]{Hoop:2021aa}. 
The latter result shows that, with high probability, the trained operator recurrent network performs nearly as well as the optimal network within the same class.

There are also bounds on the generalisation gap via Rademacher complexity for deep iterative recovery networks in the context of solving linear inverse problems \cite[Thm.~3.1]{Schnoor:2023aa}.
Deep iterative recovery networks are also a special case of learned iterative networks. 
Here, one unrolls an iterative scheme that is designed for sparse recovery of a linear inverse problem (sparse-recovery network generalising \ac{LISTA}). 
A key assumption in these networks is weight-sharing between the neural updating networks. 
This enables a unified analysis for very different neural network types, ranging from recurrent ones to networks more similar to standard feed-forward neural networks. 
The dependence of the generalisation gap on the number of layers is at most logarithmic in contrast to many previous results on the generalisation error for deep learning, where the scaling in the number of layers is often exponential.

Finally, there are bounds on the generalisation gap for compound Gaussian networks in the context of solving a linear inverse problem. 
These are a class of learned iterative networks that are obtained by unrolling a scheme designed to compute a maximum a posteriori estimator with a compound Gaussian prior and the architecture is outlined in \cite[Sec.~II.B]{Lyons:2024aa}.
The compound Gaussian networks have shown good performance in solving compressive sensing and tomographic imaging problems. 
The bound on the generalisation gap is formulated by bounding the Rademacher complexity and they show that, at worst, the generalisation gap scales as $O(n\sqrt{\log(n)})$ in the signal dimension $n$ and $O((\text{network size})^3/2)$ in network size.

Analogues results in the infinite dimensional setting (for neural operators) are harder to find.
A difficulty is to adopt an appropriate notion of model complexity bearing in mind tasks relevant to operator learning, like solving inverse problems (Section~\ref{sec:InvProb}) or optimisation problems (Section~\ref{sec:OptimSolver}).
One notion of complexity for neural operators is covering numbers/metric entropy.
Bounds on the generalisation gap in terms of such model complexity are derived \cite{Liu:2024ab} for encoder-decoder networks used to  approximate target operators that are Lipschitz.
There are also more specific results for \ac{DeepONet} \cite{Lanthaler:2022aa}, see also the general infinite-dimensional regression framework of \cite{Reinhardt:2024aa}.
Another is Rademacher complexity, which is the main architecture-specific complexity notion in characterising generalisation properties of \ac{DeepONet} \cite{Gopalani:2024aa} and \ac{FNO} \cite{Kim:2024aa} based architectures.
The proofs exploit the branch-trunk structure of \ac{DeepONet} and the Fourier-layer structure of \ac{FNO}, so the dependence on architecture is more explicit than bounds that are based on covering numbers/metric entropy.
None of the above applies to learned iterative networks. 
In fact, besides the above cited results in the finite dimensional setting, there seems to be no bounds on the generalisation gap that applies to operator learning with learned iterative networks.

\subsection{Sample complexity bounds}\label{sec:SampComplexity}
A further refinement of characterising the approximation  (Section~\ref{sec:ApproxError}) and generalising properties (Section~\ref{sec:GenGap}) is sample complexity bounds.
The aim here is to estimate amount of training data needed to ensure the learned neural operator approximates target operators within some pre-defined class to a prescribed accuracy.
The weak variant considers a fixed input-output distribution for training data, the strong variant takes the worst-case sample complexity over all input-output distributions.
An alternative formulation is to compare the learned neural operator to the best possible neural operator within the hypothesis space.

Such results are highly relevant for the various domain adapted neural operator architectures outlined in Section~\ref{sec:NeuralOperatorArchi}, which were introduced to improve upon scalability and generalisation as discussed in Section~\ref{sec:MotivationStructuredLearning}.
Ample empirical evidence shows that learned iterative networks, encode-decoder networks, like \ac{DeepONet}, and integral kernel networks, like \ac{FNO}, typically require a relatively small amount of training data to approximate target operators associated solving \acp{PDE} \cite{Lu:2021aa,Goswami:2023aa,Kovachki:2023aa,Boulle:2023aa} or inverse problems \cite{Rudzusika:2024aa}.
This contrasts with the vast amount of data that would have been needed if one would attempt at learning these target operator by refining a pre-trained foundation model (which have domain agnostic architecture) \cite{Shen:2024aa,Song:2025aa,Wang:2025aa,Zhou:2025aa}.

The above observations are supported by sample complexity bounds.
An example is \cite{Liu:2024ab,Kovachki:2024ab} that considers  approximating target operators that are Lipschitz with  an encoder-decoder network.
Another is \cite{Boulle:2023ab,Schafer:2024aa}, which focuses on solving certain elliptic \acp{PDE}. 

There are also sample complexity bounds for a broader class of target operators that are not necessarily associated with solution operators of \acp{PDE}. 
An example is \cite{Grohs:2025aa}, which derives lower bounds on achievable convergence rates for learning operators in $L^p$-setting.
Another is \cite[Sec.~5]{Kovachki:2024aa} that derives/surveys sample complexity bounds for various architectures, like \ac{FNO}, \ac{DeepONet} and PCA-NET, when the target operator is a general linear operator, holomorphic operator, and general Lipschitz operator.
We also have \cite{Stepaniants:2023aa,Jin:2023aa} that derive upper and lower bounds on the sample complexity of Hilbert–Schmidt operators between two reproducing kernel Hilbert spaces that depend on the smoothness of the input and output functions.

Target operators mentioned in the preceding paragraph can arise in solving some inverse problems, so these results are of interest for solving inverse problems.
However, the architectures under consideration are not learned iterative networks.
In fact, there are far fewer results that applies to the setting where one solves inverse problems with learned iterative networks.
A notable example is \cite[Thm.~4]{Atchade:2025aa}, which provides a statistical sample complexity result for gradient descent networks in the context of solving linear inverse problems.

\subsection{Computational complexity}\label{sec:CompComplexity}
The aim here is to characterise the computational complexity of the neural operator with a specific architecture.
This affects both training (see also Section~\ref{sec:CompFeasTrain}) and inference and the results are for natural reasons highly architecture specific. The focus is typically on beating the `curse of parametric complexity', which here means identifying conditions for the architecture which ensure that the required size of the neural operator (number of tunable parameters) grows only algebraically with the inverse of the desired accuracy. 
\begin{remark}
The notion of `curse of parametric complexity' was introduced in \cite{Lanthaler:2023aa}. This notion that is specific to operator learning is sometimes referred to as `curse of dimensionality'. 
Still, it is distinct from the conventional usage of curse of dimensionality in the context of numerically solving high-dimensional or multi-scale problems. 
\end{remark}

Numerical experiments are provided in \cite{DeHoop:2022aa} in the context of solving specific \acp{PDE} with some of the neural operator architectures listed in Section~\ref{sec:OtherNeuralOpArch}, like \ac{DeepONet}, PCA-NET and \ac{FNO}.
The paper also lists the parameter complexity and the scaling of the the evaluation cost for these architectures \cite[Tables~4.1 and 4.2]{DeHoop:2022aa}.
A relevant analysis for solving linear \acp{PDE} has been given in \cite{Boulle:2023aa,DeHoop:2023aa}.
Theoretical analysis of the required complexity of neural operators in the context of solving specific \acp{PDE} are studied from a approximation theoretic point of view in \cite{Kovachki:2021aa,Schwab:2023aa,Kutyniok:2022aa,Lanthaler:2022aa,Ryck:2022aa}.
Similar theoretical study for the PCA-NET are provided in \cite{Lanthaler:2023aa}.
For more general classes of target operators, like Lipschitz operators or continuously differentiable operators with bounded Fréchet derivatives, one can show that one suffers from curse of sample complexity \cite{Kovachki:2024ab}, meaning that learning such operators requires an exponential number of samples in a minimax sense that is independent of the reconstruction algorithm. 

\subsection{Discretisation invariance}\label{sec:DiscInv}
Following \cite{Kovachki:2023aa}, we define discretisation invariance in the following way. 
An neural operator architecture is discretisation-invariant if it satisfies the following when it has a fixed number of parameters:
\begin{inparaenum}[(i)]
\item It can act on any discretisation of the input function, i.e. accepts any set of points in the input domain, 
\item it can be evaluated at any point of the output domain, and 
\item it converges to a continuum operator as the discretisation is refined.
\end{inparaenum}

The first two requirements of accepting any input and output points in the domain is a natural requirement for discretisation invariance, while the last one ensures consistency in the limit as the discretisation is refined. 
As an example, families of graph neural networks and transformer models are resolution invariant, i.e., they can receive inputs at any resolution, but they fail to converge to a continuum operator as discretisation is refined. 
Moreover, we require the models to have a fixed number of parameters; otherwise, the number of parameters becomes unbounded in the limit as the discretisation is refined. 
Thus the notion of discretisation invariance allows us to define neural operator models that are consistent in function spaces and can be applied to data given at any resolution and on any mesh. We also establish that standard neural network models are not discretisation invariant. 

To fully adopt a `learn then discretise' approach (Section~\ref{sec:MotivationFuncAnaly}) requires using neural operator architectures that are discretisation invariant \cite{Kovachki:2023aa}.
Such invariance means there is a canonical way to adapt a neural operator so that it can operate on various discretisation of the data. 
This is often a necessary condition if the trained model is to generalise well to unseen data and remain transferable across different discretisations.

Feed-forward neural operator architectures introduced in \cite{Kovachki:2023aa} are universal approximators that are also discretisation-invariant.
In fact, these are the only known class of architectures that guarantee both discretisation-invariance and universal approximation. 
Interestingly, many of the popular architectures in Section~\ref{sec:OtherNeuralOpArch}, like \ac{DeepONet} and \ac{FNO}, are \emph{not} discretisation invariant in their original formulation. 
Some of these can be modified to render them discretisation invariant as outlined in \cite{Boulle:2024aa}, but this may involve elaborate modifications. 
In contrast, the discretisation invariance of learned iterative networks rests upon the ability to formulate discretisation invariant architectures for the neural updating operators.
The architectures of these neural updating operators is often rather simple, thus making it much easier to ensure they are discretisation invariance.
 
\subsection{Stability results}\label{sec:Stability}
Learning an iterative method is different from a learned iterative network. Where the latter is merely an architecture, the former is actually an iterative scheme with learned updates. 
Note that learning to optimise (Section~\ref{sec:OptimSolver}) often reduces to a form of  learning an iterative method, namely the one that solves the optimisation problem. As we have pointed out earlier, this is not the case for learned iterative networks. Thus, the question remains what kind of theoretical guarantees can be given for a learned iterative network. At the time of writing this chapter and to our knowledge, there are only two types of theoretical results specifically for learned iterative networks. Namely, stability with respect to unrolled iterates in the caseof \acf{DEQ} networks and stability with respect to parameters and input in the case of \acf{TDV}. Let us shortly summarise the results in the following.

For \ac{DEQ} networks, we make use of the fixed point formulation. This first necessitates weight-sharing and considering the limit $\RecOp_\NNparam(\data)=\signal^\infty$. The theoretical results concern convergence of the iterates to said fixed point. Note that this does not further characterise the fixed point, e.g., as the minimiser of a cost functional. The basis for convergence is to show that the updates in \eqref{eq:proximalNet_weightShare} are a contraction, see \cite{bai2019deep,gilton2021deep}. For the learned proximal networks in this work, we require that $\NNOp_\NNparam-\Id$ is $\varepsilon$-Lipschitz. Furthermore, in the finite dimensional case, consider the largest $L$ and smallest singular value $\mu$ of the normal $A^TA$. Then a suitable step-size $\omega$ depending on $L,\mu,\varepsilon$ in \eqref{eq:proximalNet_weightShare} exists, such that the iterative updates are a contraction, if $\varepsilon<2\mu/(L-\mu)$. In particular, this has two consequences. Compact forward operators, specifically with rapidly decreasing spectrum, are not well suited for this approach, or in other words \ac{DEQ} networks are better suited for mildly ill-posed inverse problems, which is largely in-line with the literature. Secondly, it is desired that the Lipschitz constant of the network $\NNOp_\NNparam$ is close to 1. This is usually enforced in the training by spectral normalisation, unfortunately this is also known to limit expressivity of the network.

An alternative approach for a stability  analysis is presented by \cite{kobler2020total,Kobler:2022aa}. Based on the variational network parametrisation, a stability result with respect to the input as well as with respect to parameter perturbations is derived under the assumption that test and training data are within the same distribution. The analysis is enabled by stating the learned iterative network in \eqref{eq:VarNet} with a least-squares data-fidelity via an implicit scheme as
\begin{equation}\label{eqn:TDV_implicitupdates}
\signal^{k+1}=g(\signal,\data,T,\NNparam):=\left(\Id+\frac{T}{N}\FwdOp^T A\right)^{-1}\left(\signal+\frac{T}{S}\left(\FwdOp^T\signal-\nabla\NNOp_\NNparam(\signal)\right)\right),
\end{equation}
where $N$ is the number of unrolled iterates and $T$ denotes a termination time. The training is then done as a mean-field optimal control problem over $T$ and $\NNparam$. This formulation allows to provide upper bounds on the trajectories defined by $g$ for different solutions of the operator equation \eqref{eq:InvProb}. That is, given two measurements $\signal$ and $\widetilde{\signal}$ originating from the same ground-truth $\signal$. Then one can bound the difference $\|\signal^{k+1} - \widetilde{\signal}^{k+1}\|_\RecSpace$ of two solutions 
\[
\signal^{k+1}=g(\signal_k,\data,T,\NNparam) \text{ and } \widetilde{\signal}^{k+1}=g(\widetilde\signal_k,\widetilde\data,T,\NNparam),
\]
by the distance of measurements $\|\data-\widetilde\data\|_\DataSpace$, where the bound depends on the local Lipschitz constant of the data distribution, the norm of the normal and of the initial reconstruction operator. A similar estimate can be made for stability with respect to parameters, where the bound is given by the difference in parameters depending on the local Lipschitz constant of the data distribution. We believe that the proposed framework for a stability analysis based on a gradient flow may provide the basis to prove similar stability results for a variety of other learned iterative networks.

\section{Conclusions}
This survey is aimed at developing a unified view on learned iterative reconstructions through the formulation of learned reconstruction operator. Specifically, we separated the \emph{how to compute} via the learned reconstruction operator from the \emph{what to compute} given by the learning problem. This independent treatment allows to identify structural similarities in many established approaches. In fact, we see in Section~\ref{sec:LearnedGradient} that many approaches in the literature are the same in their core and only differ by a specific parametrisation of the neural updating operator. Furthermore, we have presented the framework of learned iterative reconstructions for the usual case of linear inverse problems as well as for non-linear inverse problems in Section~\ref{sec:HighOrderNonLin}. Finally, the computational examples showed that for linear inverse problems and gradient networks, there is only minor difference between the different parametrisations. Whereas in non-linear inverse problems, the choice of update direction, or the underlying algorithm for the unrolled method, has a major effect on the performance.

We expect that this survey will help researchers to view learned iterative reconstructions and learned reconstruction operators on a high level and will open new ideas to design suitable learned reconstructions for challenging inverse problems.

\section*{Acknowledgements}
This survey as been supported in parts by the
Research council of Finland (Project No. 353093, Finnish Centre of Excellence in Inverse Modelling and Imaging; and Project No. 359186, Flagship of Advanced Mathematics for Sensing Imaging and Modelling; Projects No. 338408, Academy Research Fellow (AI-SOL); Project No. 370528, Academy project (AequiLoFi)) and in parts by the 
Swedish Research Council grants 2020-03107,
Swedish Energy Agency P2022-00286,
FORMAS 2022-00469. AH was supported by DigitalFutures Scholar-in-Residence grant KTH-RPROJ-0146472 2

\appendix
\clearpage

\section{Common data loss functionals}\label{sec:DataLoss}
Let $\DataSpace$ be a (real) Hilbert space with inner product $\langle \Cdot,\Cdot \rangle_{\DataSpace}$ and associated norm $\Vert \Cdot \Vert_{\DataSpace}$.
A common special case for the data-fidelity $\operator{Q}_{\data} \colon \RecSpace \to \Real$ in \eqref{eq:DataFidelity}
is when the data loss is given as 
\[ \DataLoss(\data,\dataother) := \DataNorm(\data-\dataother)
\quad\text{for $\data,\dataother \in \DataSpace$}
\]
where $\DataNorm \colon \DataSpace \to \Real$ is some suitable functional.
Then the data-fidelity is of the form 
\begin{equation}\label{eq:Qoperator2} 
\operator{Q}_{\data}(\signal) :=
   \DataNorm\bigl( \FwdOp(\signal) - \data \bigr)
   \quad\text{for $\signal \in \RecSpace$.}
\end{equation}
We next compute the gradient and the Hessian of such a data-fidelity functional. 

\subsection{Gradient calculations}
When $\FwdOp \colon \RecSpace \to \DataSpace$ is Fréchet differentiable, then one can derive an expression for the derivative and gradient of $\operator{Q}_{\data} \colon \RecSpace \to \Real$ in \eqref{eq:Qoperator2}.
To see this, we first note that $\operator{Q}_{\data}$ is given by composing the mapping $\FwdOp(\Cdot)-\data$ with $\DataNorm$. 
Next, the derivative of $\FwdOp(\Cdot)-\data$ at $\signal \in \RecSpace$ is the linear mapping $\signalother \mapsto \partial\! \FwdOp(\signal)(\signalother)$.
Hence, by the chain rule we get that the derivative of $\operator{Q}_{\data}$ in \eqref{eq:Qoperator2} at $\signal \in \RecSpace$ is the linear mapping $\partial\! \operator{Q}_{\data}(\signal) \colon \RecSpace \to \RecSpace$ given as
\begin{equation}\label{eq:QoperatorDiff}
\partial\! \operator{Q}_{\data}(\signal)(\signalother) 
  = \partial\! \DataNorm\bigl( \FwdOp(\signal)-\data \bigr)\bigl( \partial\! \FwdOp(\signal)(\signalother) \bigr)
  \quad\text{for $\signalother \in \RecSpace$.}
\end{equation}
The associated gradient at $\signal \in \RecSpace$ is the element $\grad \operator{Q}_{\data}(\signal) \in \RecSpace$ that is given implicitly by the relation 
\begin{equation}\label{eq:QoperatorGrad}
\partial\! \operator{Q}_{\data}(\signal)(\signalother)
= \bigl\langle \grad \operator{Q}_{\data}(\signal), \signalother \bigr\rangle_{\RecSpace}
\quad\text{for any $\signalother \in \RecSpace$.}
\end{equation}
The existence of a unique such element is guaranteed by the Riesz representation theorem. 
To get more explicit expressions for the derivative/gradient, we need to consider special cases of $\DataNorm \colon \DataSpace \to \Real$ in \eqref{eq:Qoperator2}.
\begin{description}
\item[$\DataNorm(\data) := \Vert \data \Vert_{\DataSpace}^2$:]
This choice corresponds to $\operator{Q}_{\data}(\signal) = \Vert \FwdOp(\signal)-\data \Vert_{\DataSpace}^2$ in \eqref{eq:Qoperator2}.
Next, $\partial\!\DataNorm(\data)(\dataother)
 = 2 \langle \data, \dataother \rangle_{\DataSpace}$
for $\data, \dataother \in \DataSpace$. 
Inserting this into \eqref{eq:QoperatorDiff} yields
\begin{multline}\label{eq:QoperatorDiffCase1}
\partial\! \operator{Q}_{\data}(\signal)(\signalother) 
  = 2 \bigl\langle 
   \FwdOp(\signal)-\data, 
   \partial\! \FwdOp(\signal)(\signalother) 
  \bigr\rangle_{\DataSpace}
  \\
  = 2 \Bigl\langle 
   \bigl(\partial\! \FwdOp(\signal)\bigr)^{\ast}\bigl(\FwdOp(\signal)-\data\bigr), 
   \signalother
  \Bigr\rangle_{\DataSpace}.
\end{multline}
In the above, $\partial\!\FwdOp(\signal)^{\ast} \colon \DataSpace \to \RecSpace$ is the linear mapping representing the adjoint of the derivative of $\FwdOp \colon \RecSpace \to \DataSpace$ at $\signal \in \RecSpace$ (think transpose of the Jacobian).
Finally, comparing the rightmost expression in \eqref{eq:QoperatorDiffCase1} with the equality in \eqref{eq:QoperatorGrad} yields
\begin{equation}\label{eq:L2SquaredGrad}
\grad \operator{Q}_{\data}(\signal)=2 \bigl(\partial\! \FwdOp(\signal)\bigr)^{\ast}\bigl(\FwdOp(\signal)-\data\bigr).
\end{equation}

\item[$\DataNorm(\data) := \Vert \data \Vert_{\DataSpace}$:]
This choice corresponds to $\operator{Q}_{\data}(\signal) = \Vert \FwdOp(\signal)-\data \Vert_{\DataSpace}$ in \eqref{eq:Qoperator2}.
Next, $\partial\!\DataNorm(\data)(\dataother)
 = \frac{1}{\DataNorm(\data)} \langle \data, \dataother \rangle_{\DataSpace}$
for $\data, \dataother \in \DataSpace$. 
Inserting this into \eqref{eq:QoperatorDiff} yields
\begin{multline}\label{eq:QoperatorDiffCase2}
\partial\! \operator{Q}_{\data}(\signal)(\signalother) 
  = \frac{1}{\DataNorm\bigl(\FwdOp(\signal)-\data\bigr)}
   \bigl\langle 
   \FwdOp(\signal)-\data, 
   \partial\! \FwdOp(\signal)(\signalother) 
  \bigr\rangle_{\DataSpace}
  \\
  = \frac{1}{\bigl\Vert \FwdOp(\signal)-\data\bigr\Vert_{\DataSpace}}
   \Bigl\langle 
   \bigl(\partial\! \FwdOp(\signal)\bigr)^{\ast}\bigl(\FwdOp(\signal)-\data\bigr), 
   \signalother
  \Bigr\rangle_{\DataSpace}.
\end{multline}
Finally, comparing the rightmost expression in \eqref{eq:QoperatorDiffCase2} with the equality in \eqref{eq:QoperatorGrad} yields
\begin{equation}\label{eq:L2Grad}
 \grad \operator{Q}_{\data}(\signal)=
\frac{1}{\bigl\Vert \FwdOp(\signal)-\data\bigr\Vert_{\DataSpace}}
\bigl(\partial\! \FwdOp(\signal)\bigr)^{\ast}\bigl(\FwdOp(\signal)-\data\bigr).
\end{equation}
Note that the derivative/gradient of $\operator{Q}_{\data}$ at $\signal$ only exists when $\FwdOp(\signal)-\data \neq 0$. This is to be expected since $\DataNorm$ is not differentiable at $0$.
\end{description}

\subsection{Hessian calculations}\label{sec:HessianCalc}
Assume now that $\FwdOp \colon \RecSpace \to \DataSpace$ is twice Fréchet differentiable.
Then $\operator{Q}_{\data}$ is also twice Fréchet differentiable.
Its 2nd Fréchet derivative at $\signal\in \RecSpace$ is a mapping $\GateuaxSecond{\operator{Q}_{\data}}(\signal) \colon \RecSpace \to \LinOp(\RecSpace, \Real)$ representing a symmetric bilinear form, i.e., 
\[ (\signalother,\signalothernew) \mapsto \GateuaxSecond{\operator{Q}_{\data}}(\signal)(\signalother)(\signalothernew) 
 \quad\text{is symmetric bilinear form.}
\]
Analogous to the case with the first derivative, there exists a unique bounded linear operator $\HessianOp{\operator{Q}_{\data}} \colon \RecSpace \to \LinOp(\RecSpace,\RecSpace)$ (Hessian of $\operator{Q}_{\data}$) such that the following holds: 
\[
\partial^2\!\operator{Q}_{\data}(\signal)(\signalother)(\signalothernew) =
 \bigl\langle \signalothernew, \HessianOp{\operator{Q}_{\data}}(\signal)(\signalother) \bigr\rangle_{\RecSpace}
 \quad\text{for $\signalother,\signalothernew \in \RecSpace$.}
\]
Next, if $\RecSpace$ is a real Hilbert space then $\HessianOp{\operator{Q}_{\data}}(\signal) = \bigl(\HessianOp{\operator{Q}_{\data}}(\signal)\bigr)^{\ast}$ i.e., the Hessian $\HessianOp{\operator{Q}_{\data}}(\signal) \in \LinOp(\RecSpace,\RecSpace)$ is self-adjoint.
This follows from the fact that the 2nd Fréchet derivative represents a symmetric bilinear form.

We conclude with explicit expressions for the gradient $\grad \operator{Q}_{\data}(\signal) \in \RecSpace$ and the Hessian $\HessianOp{\operator{Q}_{\data}}(\signal)(\signalother) \in \RecSpace$ when $\operator{Q}_{\data}$ is given as in \eqref{eq:DataFidelityL2}:
\begin{align}
\grad \operator{Q}_{\data}(\signal) &= [\partial\!\FwdOp(\signal)]^{\ast}\bigl(\FwdOp(\signal)-\data\bigr)
\label{eq:Qgrad}
\\[0.5em]
\HessianOp{\operator{Q}_{\data}}(\signal)(\signalother)
&= \Bigl(\bigl[\partial\!\FwdOp(\signal)\bigr]^{\ast} \circ \partial\!\FwdOp(\signal)\Bigr)(\signalother)
\notag
\\
&\qquad\qquad
+ \bigl[\partial^2\!\!\FwdOp(\signal)(\signalother,\Cdot)\bigr]^{\ast}\bigl(\FwdOp(\signal)-\data\bigr)
\quad \text{for $\signalother\in \RecSpace$.}
\label{eq:QHessian}
\end{align}
In particular, if $\FwdOp \colon \RecSpace \to \DataSpace$ is linear, then $\partial\!\FwdOp(\signal)=\FwdOp$ independent of $\signal \in \RecSpace$, so $\grad\operator{Q}_{\data}(\signal)$ is given as in \eqref{eq:DataFidelityL2Grad} and $\HessianOp{\operator{Q}_{\data}}(\signal)=\FwdOp^{\ast} \circ \FwdOp$.

\section{\Acf{SGD}}\label{app:SGD}
We here provide a brief introduction to optimisation methods used for training deep neural networks. 
The aim is to identify some hyper parameters that are commonly varied  during training. The material is based on \cite{Ruder:2016aa,Tian:2023aa}.

Training a deep neural network typically amounts to computing a (local) minimiser to an objective function $\operator{F} \colon \NNparamSet \to \Real$.
This is a real-valued function defined on the set $\NNparamSet$ of neural network parameters that one seeks to learn during training. 
Section~\ref{sec:LearningProbs} lists various objective functions one may encounter depending on the type of training data one has access to.
As an example, the objective function for supervised data $\Sigma \subset \DataSpace \times \RecSpace$ is given in \eqref{eq:SupDataX-Loss_emp}, i.e., the objective is of the form 
\[
\operator{F}_{\Sigma}(\NNparam) := \sum_{i=1}^n 
    \Loss_{\RecSpace}\bigl( \RecOp_{\NNparam}(\data_i), \signal_i \bigr)
\quad\text{for $\Sigma=\bigl\{(\data_i,\signal_i) \bigr\}_{i=1}^n \subset \DataSpace \times \RecSpace$.}    
\]

Objectives used in training deep neural networks are often chosen so that they are differentiable. 
It therefore makes sense to considers first order gradient methods to compute a minimiser.
The objective is however also highly non-convex and non-linear, still one can prove existence of global minima and a basic convergence analyses for gradient descent methods \cite{Jentzen:2022aa}.
Furthermore, to address the issue with non-convexity, it is common to compute a (local) minimiser by some version of \ac{SGD}. 
The `stochasticity' in \ac{SGD} refers to the fact that the objective is usually not evaluated on the entire training dataset at hand, but instead on randomly sampled subsets of it (mini-batches). The advantages of using
mini-batches are twofold: firstly, it significantly reduces the computational effort required for a parameter update. Secondly, it introduces some randomness into the minimisation procedure, which can be helpful to avoid getting stuck in a local minimum.
At each \ac{SGD} iteration, an unbiased approximation of the gradient $\grad_{\NNparam}\operator{F}_{\Sigma}(\NNparam)$ is computed.
The neural network parameters are subsequently updated via:
\[
\NNparam^{k+1} = \NNparam^k - \gamma_k \grad \operator{F}_{\Sigma_k}(\NNparam).
\]
Here, at each iterate one needs to choose the mini-batch $\Sigma_k \subset \Sigma$ consisting of $n_k \ll n$ elements and the learning rate $\gamma_k>0$.

Numerous modifications of this algorithm exist, which introduce additional features such as momentum or adaptive learning rates. 
\cite{Jentzen:2022aa}. 
A popular variant, which combines both of these extensions, is the Adam (adaptive moment estimation) optimiser \cite{Kingma:2015aa}, see also \cite[Ch.~5]{Bach:2024aa} and \cite[Ch.~11]{Zhang:2025ac} for more in-depth exposition of various \ac{SGD} methods and their properties.

The learning rate, choice of batch size $n_k$ (mini-batches are typically selected as random subsets of the training data), and initialisation are all hyper parameters that are typically varied during the training. 
Appropriately selecting these parameters is part of hyperparameter tuning.

\printbibliography

\end{document}